\pdfoutput=1
\documentclass[%
reprint,
superscriptaddress,
bibnotes,
longbibliography,
amsmath,amssymb,
prb,
floatfix
]{revtex4-2}
\usepackage{graphicx}
\usepackage{dcolumn}
\usepackage{bm}
\usepackage{hyperref}

\usepackage{natbib}
\usepackage{mathrsfs,mathtools,color,wasysym,dsfont,here}
\usepackage{amsthm}
\usepackage{physics}
\usepackage[all]{xy}
\usepackage{tikz}
\usepackage{amsfonts}
\usepackage{array,booktabs}
\usepackage{comment}
\usetikzlibrary{calc}
\usetikzlibrary{arrows}
\usetikzlibrary{decorations.markings,decorations.pathmorphing}
\usepackage{cases}
\usepackage{siunitx}
\usepackage{tablefootnote}

\cmidrulesep=\doublerulesep
\newcommand{\Hamzero}{{\mathcal{H}^{(0)}}}

\newcommand{\Hamtwo}{{\mathcal{H}^{(2)}}}

\begin{document}

\title{Spin Nernst and thermal Hall effects of topological triplons in quantum dimer magnets on the maple-leaf and star lattices}

\author{Nanse Esaki}
\email{esaki-nanse0428@g.ecc.u-tokyo.ac.jp}
\affiliation{Department of Physics, Graduate School of Science, The University of Tokyo, 7-3-1 Hongo, Tokyo 113-0033, Japan}
\author{Yutaka Akagi}
\affiliation{Department of Physics, Graduate School of Science, The University of Tokyo, 7-3-1 Hongo, Tokyo 113-0033, Japan}
\affiliation{Faculty of Core Research Natural Sciences Division, Ochanomizu University, 2-1-1 Ohtsuka, Tokyo 112-8610, Japan}
\author{Karlo Penc}
\affiliation{Institute for Solid State Physics and Optics, Wigner Research Centre for Physics, H-1525 Budapest, P.O.B.~49, Hungary}
\author{Hosho Katsura}
\affiliation{Department of Physics, Graduate School of Science, The University of Tokyo, 7-3-1 Hongo, Tokyo 113-0033, Japan}
\affiliation{Institute for Physics of Intelligence, The University of Tokyo, 7-3-1 Hongo, Tokyo 113-0033, Japan}
\affiliation{Trans-scale Quantum Science Institute, The University of Tokyo, 7-3-1, Hongo, Tokyo 113-0033, Japan}
\begin{abstract}

We present a comprehensive theoretical study of the topological properties of triplon excitations in spin-$1/2$ dimer-singlet ground states defined on the maple leaf and star lattices. 
Our analysis is based on a model that includes Heisenberg interactions, Dzyaloshinskii-Moriya (DM) interactions, and an external magnetic field.
In the absence of an in-plane DM vector, we demonstrate that the triplon Hamiltonian maps onto the magnon Hamiltonian of the Kagome lattice, inheriting its nontrivial topological characteristics, including Berry curvature and topological invariants such as the $\mathbb{Z}_2$ invariant and Chern numbers.
This correspondence enables us to derive analytical 
expressions for the spin Nernst and thermal Hall conductivities at low temperatures.
Furthermore, we explore the effects of realistic finite in-plane DM interactions, uncovering multiple topological transitions and a complex thermal Hall conductivity behavior, including potential sign reversals as functions of magnetic field and temperature.
Using layer groups, we also provide a symmetry classification of the star and maple leaf lattices.

\end{abstract}

\maketitle

\section{Introduction}\label{sec:Intro}
Recently, there has been growing interest in exploring the topological band structures of bosonic excitations, analogous to those extensively studied in electronic systems \cite{kane2005$Z_2$, fu2007Topological, hasan2010Colloquium}. 
%
As in their electronic counterparts, the nontrivial band topology in bosonic systems originates from the Berry curvature, which gives rise to Hall-type transport phenomena.
%
These phenomena have been extensively investigated both theoretically and experimentally, providing crucial experimental signatures of nontrivial band topology in magnon 
\cite{katsura2010Theory, onose2010Observation, matsumoto2011Theoretical, ideue2012Effect, matsumoto2014Thermal, mook2014Magnon, cheng2016Spin, zyuzin2016Magnon, shiomi2017Experimental, murakami2017Thermal, han2017Spin, kondo2020NonHermiticity, mcclarty2022Topological, zhuoTopological}
, photon \cite{raghu2008Analogs, petrescu2012Anomalous, rechtsman2013Photonic, hafezi2013Imaging, ben-abdallah2016Photon}, phonon \cite{strohm2005Phenomenological, sheng2006Theory, kagan2008Anomalous, zhang2010Topological, zhang2011phonon, qin2012Berry}, triplon \cite{romhanyi2015Hall, malki2017Magnetic, mcclarty2017Topological, joshi2019$mathbbZ_2$, sun2021Negative, bhowmick2021Weyl, thomasen2021Fragility, esaki2024Electric}, and flavor-wave systems \cite{romhanyi2019Multipolar, furukawa2020Effects, ma2024Upperbranch, lu2024Spin}. 

In magnetic insulators, transverse thermal transport has emerged as a powerful experimental tool for probing topological spin excitations \cite{murakami2017Thermal, zhang2024Thermal}. 
When magnetic excitations exhibit nontrivial topology, a longitudinal temperature gradient induces a transverse spin current—a phenomenon known as the spin Nernst effect \cite{cheng2016Spin, zyuzin2016Magnon, shiomi2017Experimental}. 
In parallel, the same conditions also generate a transverse thermal current, referred to as the thermal Hall effect \cite{katsura2010Theory, onose2010Observation, matsumoto2011Theoretical}. 
Both phenomena have been observed in ordered magnets with Dzyaloshinskii–Moriya (DM) interactions, where the DM coupling effectively acts as a magnetic flux that can induce nontrivial topology in the magnon bands \cite{onose2010Observation, ideue2012Effect, mook2014Magnon}.

In contrast, much less is known about quantum dimer magnets in which neighboring $S=1/2$ spins form singlet valence bonds. Such systems exhibit a gapped singlet ground state with bosonic triplet excitations 
known as triplons. Several theoretical studies have predicted the thermal Hall effect of triplons in SrCu$_2$(BO$_3$)$_2$ \cite{romhanyi2015Hall, malki2017Magnetic, mcclarty2017Topological, sun2021Negative, bhowmick2021Weyl} and $X$CuCl$_3$ ($X$ =Tl, K) \cite{esaki2024Electric}. 
To date, however, no experimental confirmation of this effect has been reported \cite{suetsugu2022Intrinsic}, underscoring the need for deeper theoretical insight for identifying promising candidate materials.

In ordered ferromagnets, a no-go theorem 
rules out the magnon thermal Hall effect in edge-sharing geometries, such as square, honeycomb, and cubic lattices, even in the presence of Dzyaloshinskii–Moriya interactions \cite{katsura2010Theory, ideue2012Effect, kawano2019Thermal, esaki2024Electric, buzo2024Thermal}. 
%
%
However, this restriction does not straightforwardly extend to antiferromagnets or quantum dimer magnets \cite{kawano2019Thermal, esaki2024Electric, buzo2024Thermal}. Thus, it is crucial to clarify under what conditions triplon excitations in quantum dimer magnets can support robust topological thermal transport.

\begin{figure*}[bt]
    \includegraphics[width=0.95\textwidth]{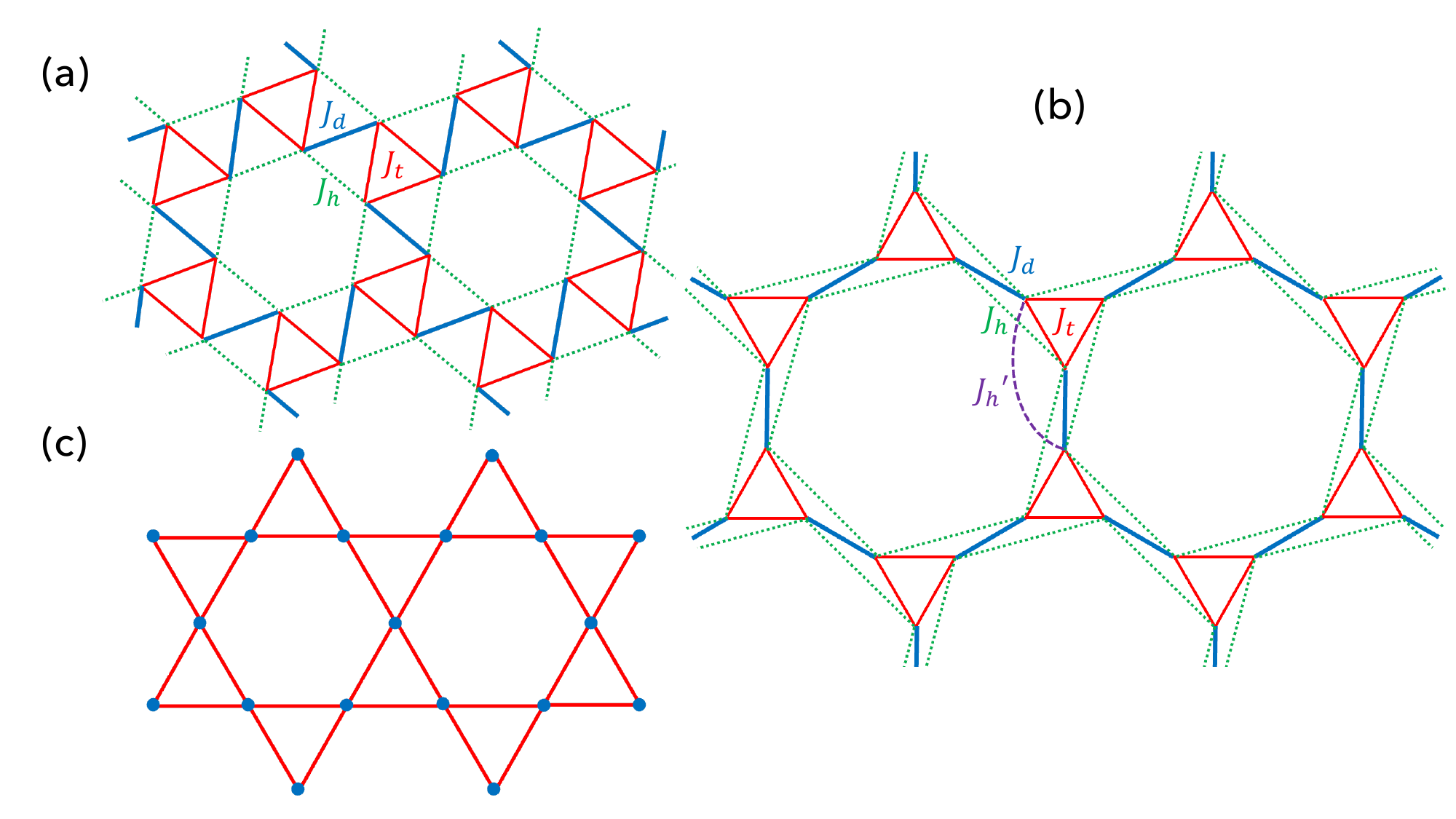}
    \caption{Schematic figures of (a) a maple-leaf and (b) a star lattice with three types of Heisenberg interactions $J_d$, $J_t$, and $J_h$ ($J^{\prime}_h$) \cite{commentfig1}. In this figure, the thick blue bonds represent $J_d$ (dimer bond), the solid red bonds denote $J_t$, and the dotted green (dashed purple) bonds indicate $J_h$ ($J^{\prime}_h$).
    Both lattices (a) and (b) are equivalent in terms of the connectivity when $J^{\prime}_{h} = 0$, and thus we treat them simultaneously in the following sections. When the midpoint of each $J_d$ (thick blue) bond is identified as a single effective site, the lattice structures reduce to that of a Kagome lattice (c).
    }
    \label{fig:Maple_Star_correspondence}
\end{figure*}

In this work, we provide a comprehensive theoretical framework to address this issue by focusing on quantum magnets on the maple-leaf \cite{penfield1890spangolite, miers1893Spangolite, frondel1949crystallography, hawthorne1993crystal, betts1995new, olmi1995crystal, schmalfuss2002spin$frac12$, fennell2011Spangolite, kampf2013Leadtellurium, mills2014Bluebellite, norman2018Copper, inosov2018Quantum, haraguchi2018Frustrated, venkatesh2020Magnetic, haraguchi2021Quantum, makuta2021Dimensional, ghosh2022Another, saha2023Twodimensional, gresista2023Candidate, ghosh2023Maple, schmoll2024Tensor, ghosh2024Triplon} and star lattices \cite{zheng2007Star, sorolla2020Synthesis, dornellas2024KitaevHeisenberg, ishikawa2024Geometric}, which have recently attracted considerable attention due to their unique magnetic properties. 
Specifically, we study a model incorporating Heisenberg interactions, symmetry-allowed DM interactions, and an out-of-plane magnetic field, 
focusing on a paramagnetic phase with a dimer singlet ground state. 
%
A salient feature of this model is that, when the midpoint of each $J_d$ (thick blue) bond is identified as a single effective site, both systems reduce to the Kagome lattice [see Fig.~\ref{fig:Maple_Star_correspondence}].
This correspondence considerably simplifies the analysis and provides an intuitive picture for the topological thermal transport of triplons.

In the absence of the in-plane DM vector, we show that the triplon Hamiltonian maps to that of the magnon Hamiltonian on the Kagome lattice, with a finite Berry curvature. 
Consequently, the system inherits nontrivial band topologies characterized by either a $\mathbb{Z}_2$ invariant (in zero magnetic field) or a Chern number (in finite magnetic fields) of the underlying Kagome magnon model, allowing us to derive analytical formulas for spin Nernst and thermal Hall conductivities at low temperatures.
Importantly, even in the presence of realistic finite in-plane DM interactions, the triplon–magnon correspondence provides valuable insight into the behavior of the thermal Hall conductivity, predicting multiple topological transitions and rich variety of thermal Hall conductivity behaviors as a function of the magnetic field, including a possible sign reversal with varying temperature.



The remainder of the paper is organized as follows. In Sec. \ref{sec:Model_analysis}, we introduce the model and detail the bond-operator formalism used to describe triplon excitations in the dimer singlet phase. In Sec. \ref{sec:Dxy_0}, we establish the correspondence between the triplon Hamiltonian and the magnon Hamiltonian on the Kagome lattice, elucidating the topological nature of the triplon bands through Chern numbers and providing analytical and numerical evidence for the spin Nernst and thermal Hall effects. Section \ref{sec:Dxy} explores numerically how finite in-plane DM interactions modify the topological invariants and thermal Hall conductivity, explicitly focusing on the role of the external magnetic field. 
Section \ref{sec:Conclusion} summarizes our findings, discusses their implications, and suggests promising directions for future research. Additional technical details, including symmetry analyses, explicit triplon Hamiltonian formulations, comparisons with exact diagonalization of small clusters, and supplementary derivations, are provided in the appendices.

\section{Model and analysis}\label{sec:Model_analysis}
\subsection{Lattice structure and the model Hamiltonian}\label{subsec:Lattice_Ham}
\begin{figure}[thb]
    \centering
    \includegraphics[width=1.0\linewidth]{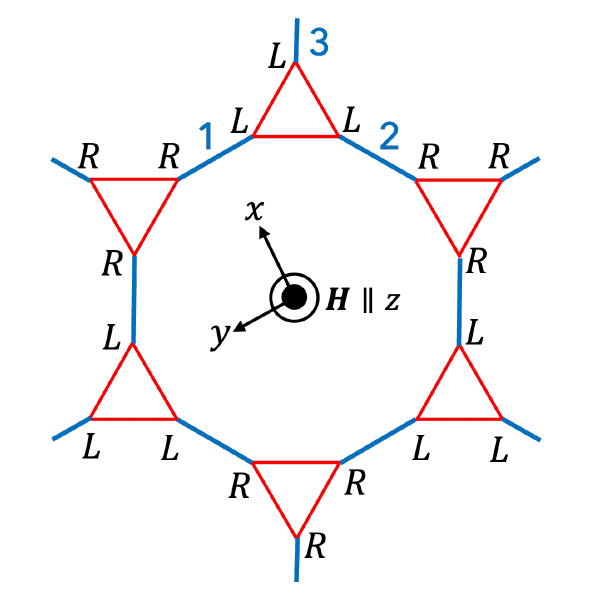}
    \caption{Schematic picture of dimers in our model. The symbols $L$ and $R$ denote the left and right spins of each dimer (thick blue bond), and the indices 1, 2, and 3 are the sublattice indices. 
    The coordinate system represented by $x$, $y$, and $z$ axes are also indicated in the figure. 
    }
    \label{fig:Model_schematic}
\end{figure}

Our model consists of a two-dimensional interacting dimer system on the (deformed) maple-leaf and star lattices [Fig. \ref{fig:Model_schematic}], where the $S = 1/2$ spins form dimers due to strong intradimer interactions. 
The unit cell contains three equivalent dimers, belonging to three different sublattices. 
Here we assume that our system is described by the space group $R\overline{3}$, which can be realized in realistic magnets with the maple-leaf and star lattice structures \cite{haraguchi2018Frustrated, haraguchi2021Quantum, ishikawa2024Geometric}. The Hamiltonian of the system in an out-of-plane magnetic field $\bm{H}\parallel z$ is given by

\begin{equation}\label{Eq:Spin_Hamiltonian_Model}
    \mathcal{H} = \mathcal{H}_{\rm Hei} + \mathcal{H}_{\rm DM} + \mathcal{H}_{\rm Zee}
\end{equation}
with
\begin{align}\label{Eq:HHei}
    \mathcal{H}_{\rm Hei} & = \sum_{\alpha = d,t,h} J_\alpha \sum_{\ev{i,j} \in B_\alpha} \bm{S}_i \cdot \bm{S}_j, \\  \label{Eq:HDM}
    \mathcal{H}_{\rm DM} & = \sum_{\ev{i,j} \in B_t} \bm{D}_{ij} \cdot (\bm{S}_i \times \bm{S}_j),\\ \label{Eq:HZee}
    \mathcal{H}_{\rm Zee} & = -g\mu_B \sum_i \bm{H}\cdot \bm{S}_i,
\end{align}
where $\ev{i,j}$ denotes the bond between sites $i$ and $j$. In 
Eqs.~(\ref{Eq:HHei}) and (\ref{Eq:HDM}), $B_{\alpha = d}$ 
represents the set of thick blue bonds, $B_{\alpha = t}$ 
the set of solid red bonds, and $B_{\alpha = h}$ 
the set of dashed green bonds, as depicted in Fig. \ref{fig:Maple_Star_correspondence}. In the Hamiltonian $H_{\rm Hei}$ (\ref{Eq:HHei}), $J_d$ describes the antiferromagnetic intradimer Heisenberg interaction, whereas $J_t$ and $J_h$ represent the antiferromagnetic interdimer Heisenberg interactions. The model also includes the symmetry-allowed interdimer DM interaction term $\mathcal{H}_{\rm DM}$ (\ref{Eq:HDM}), where $\bm{D}_{ij}$ are related by the symmetry operations of the space group $R\overline{3}$, i.e., \cite{ishikawa2024Geometric} 
\begin{equation}\label{Eq:Def_DM_vec}
    \bm{D}_{ij} = 
    \begin{cases}
        \bm{D} & i\in V_{1,\alpha},\ j\in V_{2,\alpha}, \\
        U\bm{D} & i\in V_{2,\alpha},\ j\in V_{3,\alpha}, \\
        U^{-1}\bm{D} & i\in V_{3,\alpha},\ j\in V_{1,\alpha}, \\
    \end{cases}
\end{equation}
where $\bm{D} = (D_x, D_y, D_z)^T$. Here we denote by $V_{l,\alpha}$ the set of spin sites on the sublattice  
$l=1,2,3$, where $\alpha = L, R$ indicates whether the spin lies on the left or right side of each dimer. 
The DM vectors 
satisfy $\bm{D}_{ij} = -\bm{D}_{ji}$. 
The orthogonal matrix $U$ is explicitly given by
\begin{equation}\label{Eq:Def_U_matrix}
   U = \begin{pmatrix}
        -\frac{1}{2} & -\frac{\sqrt{3}}{2} & 0 \\
        \frac{\sqrt{3}}{2} & -\frac{1}{2} & 0 \\
        0 & 0 & 1
    \end{pmatrix}.
\end{equation}
The Hamiltonian $\mathcal{H}_{\mathrm{Zee}}$ describes the Zeeman term, where $g$ is the g-factor, and $\mu_{B}$ is the Bohr magneton. In 
later sections, we simply refer to $g\mu_B \abs{\bm{H}}$ as $h$.

In Appendix~\ref{Appendix:Symmetry}, we classify the star and maple-leaf lattices embedded in three-dimensional crystals using layer groups. The symmetry classification reveals that a variety of anisotropic interaction terms are, in principle, allowed by symmetry. To simplify the treatment, we restrict our attention to terms that are linear in spin-orbit coupling—namely, the antisymmetric DM interaction—and neglect higher-order contributions such as symmetric anisotropic exchange terms, which arise at second order in spin-orbit coupling \cite{moriya1960anisotropic}. Including these higher-order terms would significantly complicate the analysis and is beyond the scope of the present work (see, for example, Ref.~\onlinecite{thomasen2021Fragility} for a more detailed treatment of a similar model).

Since we are interested in the paramagnetic phase, which is well represented by a product state of singlets, the magnetic field is assumed to be small enough such that the spin gap is finite. The analytical expression of the spin gap at $\Gamma$ without $D_{x,y}$ is given by
\begin{equation}\label{Eq:Spin_gap}
    E_{\mathrm{gap}} = \sqrt{J_d (J_d - J_t + J_h)} - \frac{\sqrt{3}}{2}\sqrt{\frac{J_d}{J_d - J_t + J_h}} \abs{D_z} - h.
\end{equation}
See Appendix \ref{Appendix:Analytical_expression} for the detailed derivation. Additionally, we assume that $J_d \gg J_t \gg J_h$, $\abs{D_z} \geq 0$. The condition that $J_d$ is much larger than other interactions is necessary to stabilize the dimer order \footnote{An exact dimer ground state is realized only when $J_t = J_h$ and without the DM interaction \cite{ghosh2022Another}. However, the ground state is well approximated by the dimer ground state even when $J_t \gg J_h$ and with the DM interaction. See Appendix \ref{Appendix:ED_vs_Triplon} for details.}. The condition $J_t \gg J_h, \abs{D_z}$ is relevant in realistic magnets with maple-leaf and star lattice structures \cite{haraguchi2018Frustrated, haraguchi2021Quantum, ishikawa2024Geometric}.

\subsection{Bond-operator formalism}\label{subsec:Bond_operator}
In the paramagnetic phase described by a product state of singlets, the low-lying excitations are spin-$1$ triplet excitations on each dimer, which we refer to as 
triplons. To describe their band structure, 
we introduce bond operators $s^{l\dag}_{\bm{R}}$ and $t^{l\dag}_{m,\bm{R}}$ ($m = \pm 1,0$) that create the singlet state $\ket{s}^{l}_{\bm{R}}$ and the three triplet states $\ket{t_{m}}^{l}_{\bm{R}}$ out of the vacuum $\ket{0}^{l}_{\bm{R}}$ on each dimer \cite{sachdev1990Bondoperator}: 
\begin{equation}
    \begin{split}
        &\ket{s}^{l}_{\bm{R}} = s^{l\dag}_{\bm{R}}\ket{0}^{l}_{\bm{R}}=\frac{1}{\sqrt{2}}(\ket{\uparrow\downarrow}^{l}_{\bm{R}}-\ket{\downarrow\uparrow}^{l}_{\bm{R}}),\\
        &\ket{t_{+1}}^{l}_{\bm{R}} = t^{l\dag}_{+1,\bm{R}}\ket{0}^{l}_{\bm{R}}=-\ket{\uparrow\uparrow}^{l}_{\bm{R}},\\ &\ket{t_{0}}^{l}_{\bm{R}} = t^{l\dag}_{0,\bm{R}}\ket{0}^{l}_{\bm{R}}=\frac{1}{\sqrt{2}}(\ket{\uparrow\downarrow}^{l}_{\bm{R}}+\ket{\downarrow\uparrow}^{l}_{\bm{R}}),\\ &\ket{t_{-1}}^{l}_{\bm{R}}= t^{l\dag}_{-1,\bm{R}}\ket{0}^{l}_{\bm{R}}=\ket{\downarrow\downarrow}^{l}_{\bm{R}},
    \end{split}\label{Eq:bond_operator}
\end{equation}
where $\bm{R}$ and $l = 1, 2, 3$ denote the position of the unit cell and the sublattice indices. The operators defined in Eq. (\ref{Eq:bond_operator}) obey Bose statistics and are subject to the particle number constraint $s^{l\dag}_{\bm{R}}s^{l}_{\bm{R}}+\sum_{m=\pm 1,0} t^{l\dag}_{m,\bm{R}}t^{l}_{m,\bm{R}}=1$ on each dimer. We follow the standard procedure and replace $s^{l\dag}_{\bm{R}}s^{l}_{\bm{R}}$ with $1-(1/N) \sum_{m,\bm{R}}t^{l\dag}_{m,\bm{R}}t^{l}_{m,\bm{R}}$, where $N$ is the number of dimers on the sublattice $l$ 
, and the total number of spin sites is $6N$. This procedure is justified at low temperatures. 

Let us rewrite the Hamiltonian (\ref{Eq:Spin_Hamiltonian_Model}) in terms of the bond operators. By introducing the Fourier transform: $t^{l\dag}_{m,\bm{R}}=(1/N) \sum_{m,\bm{k}}t^{l\dag}_{m,\bm{k}}e^{i\bm{k}\cdot\bm{R}^{l}}$
($\bm{R}^{1}=\bm{R}$, $\bm{R}^{2}=\bm{R} + \bm{a}_1$, and $\bm{R}^{3} = \bm{R} - \bm{a}_3$)
and retaining 
up to quadratic order in $t^{\dag l}_{m,\bm{k}}$ and $t^{l}_{m,\bm{k}}$, the Hamiltonian (\ref{Eq:Spin_Hamiltonian_Model}) takes the form $\mathcal{H}=\Hamzero + \Hamtwo$. Here, $\Hamzero$ is the constant term with respect to the bosonic operators \footnote{We note that the constant term $\Hamzero$ is not the ground state energy. The appropriate expression of the ground state energy is given in Eq. (\ref{Eq:Correction_GS}).}, 
and the quadratic term $\Hamtwo$ represents 
the bosonic Bogoliubov-de Gennes (BdG) Hamiltonian of the form:
\begin{equation}
\label{Eq:Hamtwo}
\Hamtwo = \frac{1}{2}\sum_{\bm{k}} \begin{pmatrix}
    \tilde{\bm{t}}^{\dag}_{\bm{k}} \\
    \tilde{\bm{t}}_{-\bm{k}}\\
\end{pmatrix}^{T} H^{\mathrm{tot}}_{\mathrm{BdG}}(\bm{k})\begin{pmatrix}
    \tilde{\bm{t}}_{\bm{k}} \\
    \tilde{\bm{t}}^{\dag}_{-\bm{k}}\\
\end{pmatrix}.
\end{equation}
Here, the vector $\tilde{\bm{t}}_{\bm{k}} = (\bm{t}_{+1,\bm{k}},\bm{t}_{0,\bm{k}},\bm{t}_{-1,\bm{k}})^T$ with $\bm{t}_{m,\bm{k}} = (t^{1}_{m,\bm{k}},t^{2}_{m,\bm{k}},t^{3}_{m,\bm{k}})$ is a set of nine operators, corresponding to the three spin states ($m = \pm 1, 0$) and three sublattices ($l = 1, 2, 3$). The $18\times 18$ matrix $H^{\mathrm{tot}}_{\mathrm{BdG}}(\bm{k})$ takes the following form:
\begin{equation}\label{Eq:BdG_Form}
   H^{\mathrm{tot}}_{\mathrm{BdG}}(\bm{k}) = \begin{pmatrix} \Xi(\bm{k}) & \Pi(\bm{k})\\
   \Pi^{\ast}(-\bm{k}) & \Xi^{\ast}(-\bm{k})\\
   \end{pmatrix}.
\end{equation}
The explicit expression of the above matrix (\ref{Eq:BdG_Form}) is given in Appendix \ref{Appendix:Explicit_form}.

The BdG Hamiltonian (\ref{Eq:Hamtwo}) has to be diagonalized using a paraunitary matrix $T^{\mathrm{tot}}(\bm{k})$ \cite{colpa1978Diagonalization}, which satisfies
\begin{align}
    &T^{\mathrm{tot} \dag}(\bm{k})\Sigma^{\mathrm{tot}}_z T^{\mathrm{tot}}(\bm{k}) = \Sigma^{\mathrm{tot}}_z, \notag \\
    &\Sigma^{\mathrm{tot}}_z = \sigma_z \otimes I_{3\times 3} \otimes I_{3\times 3}, \label{Eq:paraunitary_matrix}
\end{align}
so as to preserve the bosonic commutation relation for the transformed operators $(\tilde{\bm{t}}^{\dag}_{\bm{k}},\tilde{\bm{t}}_{-\bm{k}})(T^{{\mathrm{tot}}\dag}(\bm{k}))^{-1}$. Here, $\sigma_a (a = x, y, z)$ is the $a$ component of the Pauli matrix acting on the particle-hole space, and the two $3\times 3$ identity matrices $I_{3\times 3}$ act on the spin space spanned by $S_{z} = \pm 1, 0$ and the sublattice degrees of freedom $(l = 1, 2, 3)$, respcetively.
Using the paraunitary matrix $T^{\mathrm{tot}}(\bm{k})$, 
the matrix (\ref{Eq:BdG_Form}) is diagonalized as
\begin{equation}
\label{Eq:Eigenproblem_BdG_tot_T}
T^{\mathrm{tot}\dag} (\bm{k}) H^{\mathrm{tot}}_{\mathrm{BdG}}(\bm{k})T^{\mathrm{tot}}(\bm{k})= E^{\mathrm{tot}}(\bm{k}),
\end{equation}
where 
$E^{\mathrm{tot}}(\bm{k})$ 
is a diagonal matrix given by
\begin{align}
E^{\mathrm{tot}}(\bm{k})=\mathrm{diag}(E^{\mathrm{tot}}_{1}(\bm{k}), \cdots, E^{\mathrm{tot}}_{9}(\bm{k}), \nonumber \\
E^{\mathrm{tot}}_{1}(-\bm{k}), \cdots, E^{\mathrm{tot}}_{9}(-\bm{k})), 
\end{align}
with the ordering $0\leq E^{\mathrm{tot}}_{1}(\bm{k}) \leq \cdots  \leq E^{\mathrm{tot}}_{9}(\bm{k})$. Acting with $T^{\mathrm{tot}}(\bm{k})\Sigma^{\mathrm{tot}}_z$ on the left side of Eq. (\ref{Eq:Eigenproblem_BdG_tot_T}), we obtain
\begin{equation}\label{Eq:Eigenproblem_BdG_tot}
    \Sigma^{\mathrm{tot}}_z H^{\mathrm{tot}}_{\mathrm{BdG}}(\bm{k}) T^{\mathrm{tot}}(\bm{k}) = T^{\mathrm{tot}}(\bm{k}) \Sigma^{\mathrm{tot}}_z E^{\mathrm{tot}}(\bm{k}).
\end{equation}
Namely, we can construct the paraunitary matrix $T^{\mathrm{tot}}(\bm{k})$ from the eigenvactors of $\Sigma^{\mathrm{tot}}_z H^{\mathrm{tot}}_{\mathrm{BdG}}(\bm{k})$. Thus, we should solve the following eigenvalue problem for the BdG Hamiltonian:
\begin{equation}\label{Eq:Eigenproblem_BdG_tot_Psi}
    \Sigma^{\mathrm{tot}}_z H^{\mathrm{tot}}_{\mathrm{BdG}}(\bm{k})\bm{\Psi}^{\mathrm{tot}}_{n,\sigma}(\bm{k}) = E^{\mathrm{tot}}_{n,\sigma}(\bm{k})\bm{\Psi}^{\mathrm{tot}}_{n,\sigma}(\bm{k}),
\end{equation}
where $E^{\mathrm{tot}}_{n,\sigma}(\bm{k}) = \sigma E^{\mathrm{tot}}_n (\sigma \bm{k})$ ($n = 1,\cdots,9$ and $\sigma = \pm$) are the particle ($\sigma = +$) and the hole ($\sigma = -$) bands, respectively. In order to analyze the transport of topological triplons, it is sufficient to consider the particle bands. The validity of the triplon wave analysis presented in this section is 
discussed in Appendix \ref{Appendix:ED_vs_Triplon}.

Here we note that, in the absence of $D_{x,y}$, the spin Hamiltonian (\ref{Eq:Spin_Hamiltonian_Model}) 
has a U(1) symmetry, which makes the $t_0$ modes decoupled from the $t_{\pm}$ modes. This can be understood 
as follows: By a U(1) rotation with an angle $\phi$ in the spin space $e^{-i\phi S_{z}}$, the basis $(t^{\dag}_{m,\bm{k}},t_{-m,-\bm{k}})$, which spans the $S_{z} = m$ ($m = \pm 1, 0$) subspace is transformed as
\begin{equation}\label{Eq:U1_basis}
    (t^{\dag}_{m,\bm{k}}, t_{-m,-\bm{k}}) \rightarrow e^{-im\phi} (t^{\dag}_{m,\bm{k}},t_{-m,-\bm{k}}).
\end{equation}
Hence, if the Hamiltonian preserves the U(1) symmetry, i.e., $H^{\mathrm{tot}}_{\mathrm{BdG}} (\bm{k})$ is invariant with respect to the above transformation (\ref{Eq:U1_basis}), the matrix elements between $t_{0}$ and $t_{\pm}$ modes must vanish. For this reason, we first consider the $D_{x,y} = 0$ case (Sec. \ref{sec:Dxy_0}), where the analysis 
is simpler, and then move on to the $D_{x,y}\ne 0$ case (Sec. \ref{sec:Dxy}) in the remaining sections.

\section{The high-symmetry case without in-plane DM vector}
\label{sec:Dxy_0}

\subsection{Triplon Hamiltonian and its symmetry}\label{subsec:Triplon_Ham}
Here we focus on the subspace of $S_{z} = \pm 1$ in the absence of $D_{x,y}$. 
The BdG Hamiltonian in this subspace can be expressed in terms of $3\times 3$ matrices $H_{\mathrm{mag}} (\bm{k})$ and $H_{\pm}(\bm{k})= H_{\mathrm{mag}} (\bm{k}) + J_d \mp h$ as follows:
\begin{equation}\label{Eq:Kagome_decouple}
    H_{\mathrm{BdG}}(\bm{k}) = \begin{pmatrix}
        H_+ (\bm{k}) &0&0& H_{\mathrm{mag}}(\bm{k})\\
        0&H^{\ast}_{-} (-\bm{k})&H^{\ast}_{\mathrm{mag}}(-\bm{k})&0\\
        0&H^{\ast}_{\mathrm{mag}}(-\bm{k})&H^{\ast}_{+}(-\bm{k})&0\\
        H_{\mathrm{mag}} (\bm{k})&0&0&H_{-} (\bm{k})
    \end{pmatrix}.
\end{equation}
The matrix $H_{\mathrm{mag}} (\bm{k})$ is given by
\begin{equation}\label{Eq:single_Kagome}
    H_{\mathrm{mag}} (\bm{k}) = \frac{J^{\prime}}{2} \begin{pmatrix}
        0& e^{i\phi}\cos{k_1}&e^{-i\phi}\cos{k_3}\\
        e^{-i\phi}\cos{k_1}&0&e^{i\phi}\cos{k_2}\\
        e^{i\phi}\cos{k_3}&e^{-i\phi}\cos{k_2}&0
    \end{pmatrix},
\end{equation}
with $J^{\prime} = \sqrt{(J_t - J_h)^2 + D_{z}^2}$, $\phi = (D_z / (J_t - J_h))^{-1}$, and $k_i = \bm{k} \cdot \bm{a}_i$, where $\bm{a}_1 = (1/2, \sqrt{3}/2)$, $\bm{a}_2 = (-1,0)$, and $\bm{a}_3 = (1/2, - \sqrt{3}/2)$. 
The above $3\times 3$ matrix (\ref{Eq:single_Kagome}) corresponds to the magnon Hamiltonian on the Kagome lattice \cite{katsura2010Theory, mook2014Edge, mook2014Magnon, lee2015Thermal}.

To obtain the Berry curvature of the triplon bands, we need to solve the eigenvalue problem for $\Sigma_z H_{\mathrm{BdG}}(\bm{k})$, where $\Sigma_z = \sigma_z \otimes I_{2\times 2} \otimes I_{3\times 3}$.

We find that the eigenvalues of this matrix 
can be expressed in terms of those of $H_{\mathrm{mag}} (\bm{k})$. 
Let $\lambda_{n}(\bm{k})$ ($n=1,2,3$) be the $n$th eigenvalue of $H_{\mathrm{mag}}(\bm{k})$ and let $\bm{\psi}_{n}(\bm{k})$ be the corresponding eigenvector. The eigenvalues are ordered such that $\lambda_1 (\bm{k}) \leq \lambda_2 (\bm{k}) \leq \lambda_3 (\bm{k})$. The eigenvalues of $\Sigma_z H_{\mathrm{BdG}}(\bm{k})$ are then given by 
\begin{align}\label{Eq:Energy_relation}
    E_{n,s,\sigma} (\bm{k}) &= \sigma (\sqrt{(J_d + \lambda_n (s\sigma\bm{k}))^2 - (\lambda_n (s\sigma\bm{k}))^2} -sh), 
\end{align}
where $s = \pm$ 
denotes the spin degrees of freedom and $\sigma = +$ ($-$) represents the particle 
(hole) band.
One can also express the eigenvectors of $\Sigma_z H_{\mathrm{BdG}}(\bm{k})$ in terms of those of $H_{\mathrm{mag}} (\bm{k})$. This allows us to establish a relation between the Berry curvatures of the triplon and magnon systems. Let $\bm{\Psi}_{n,s,\sigma}(\bm{k})$ be the eigenvector of $\Sigma_z H_{\mathrm{BdG}}(\bm{k})$ corresponding to the band $E_{n,s,\sigma} (\bm{k})$. For this band, the Berry curvature is defined by 
\begin{equation}\label{Eq:Def_Berry}
    \Omega_{n,s,\sigma}(\bm{k}) = -2\sigma\mathrm{Im}\left[\frac{\partial \bm{\Psi}^{\dag}_{n,s,\sigma}(\bm{k})}{\partial k_x} \Sigma_z \frac{\partial \bm{\Psi}_{n,s,\sigma}(\bm{k})}{\partial k_y}\right]. 
\end{equation}
One can show that this Berry curvature can be written as
\begin{align}
    \label{Eq:Berry_relation}
    \Omega_{n,s,\sigma} (\bm{k}) &= s\sigma\Omega^{\mathrm{mag}}_{n} (s\sigma\bm{k}),
\end{align}
where $\Omega^{\mathrm{mag}}_{n}(\bm{k}) = -2\mathrm{Im}\left[\frac{\partial \bm{\psi}^{\dag}_{n}(\bm{k})}{\partial k_x} \frac{\partial \bm{\psi}_{n}(\bm{k})}{\partial k_y}\right]$ is the Berry curvature associated with $\bm{\psi}_{n}(\bm{k})$.

The detailed derivation of Eqs. (\ref{Eq:Energy_relation}) and (\ref{Eq:Berry_relation}) is given in Appendix \ref{Appendix:Eigen_problem}.

\begin{figure}[thb]
    \centering
    \includegraphics[width=0.8\linewidth]{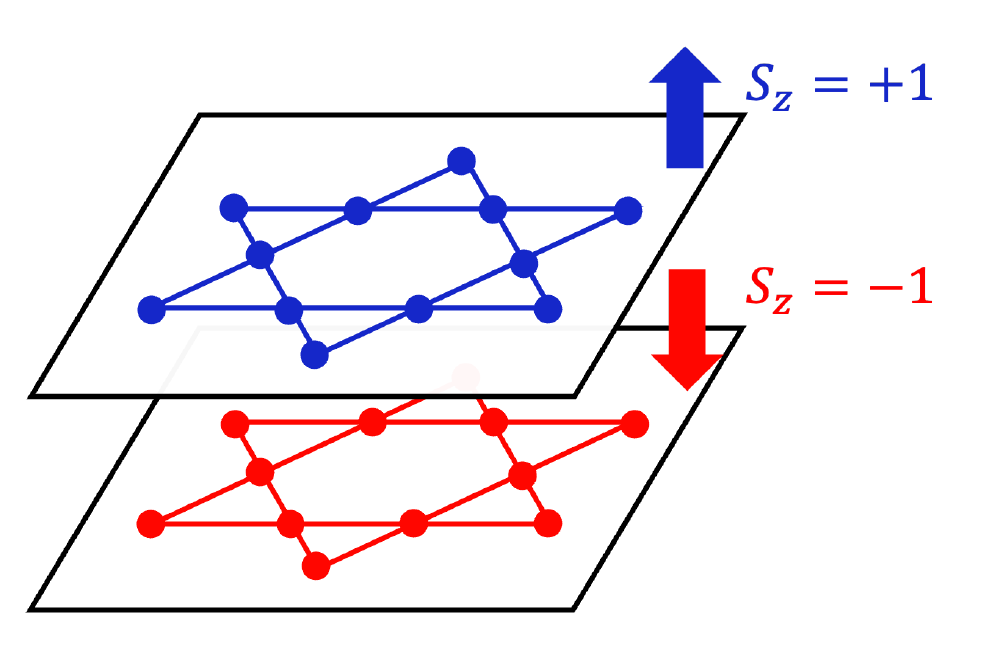}
    \caption{Schematic figure of the magnon system on a Kagome bilayer, which corresponds to the triplon system on maple-leaf and star lattices. 
    }
    \label{fig:Kagome_Bilayer}
\end{figure}

From the 
relation (\ref{Eq:Berry_relation}), we see that the band topology of the triplon Hamiltonian (\ref{Eq:Kagome_decouple}) can be understood from that of the simpler Hamiltonian (\ref{Eq:single_Kagome}), which has the same form as the magnon Hamiltonian on the Kagome lattice. More intuitively, we can regard the triplon system as the magnon system on a Kagome bilayer, as shown in Fig. \ref{fig:Kagome_Bilayer}.

It is important to note here that the argument in this section holds true regardless of the lattice structure as long as the spin Hamiltonian consists of Heisenberg and interdimer DM interactions that conserve the U(1) symmetry in the spin space. In such cases, $t_0$ modes, which are topologically trivial are decoupled from the $t_{\pm}$ modes, allowing us to understand the topology of the $t_{\pm}$ modes directly from the corresponding magnon system, which is obtained by treating each dimer as a single site. The triplon–magnon correspondence established in this section is one of our main findings. By exploiting this correspondence, we can analyze the topological thermal transport of triplons 
more clearly. 


As seen from Eq. (\ref{Eq:Energy_relation}), $\bm{\Psi}_{n,+,+}(\bm{k})$ and $\bm{\Psi}_{n,-,+}(-\bm{k})$ are degenerate without the magnetic field $h$, suggesting that these two modes form the $n$th Kramers pair. This leads to a $\mathbb{Z}_2$ topological classification of the triplon system, which is naturally expected from the physical picture discussed above [see Fig. \ref{fig:Kagome_Bilayer}]. To illustrate how the presence of such Kramers pairs is guaranteed, let us now examine the symmetry properties of the BdG Hamiltonian (\ref{Eq:Kagome_decouple}).
%
In the absence of a magnetic field, one finds the following three relations:
\begin{align}\label{Eq:Pseudo_TR_sym}
    \Sigma_z H_{\mathrm{BdG}}(-\bm{k})\Theta^{\prime} &= \Theta^{\prime} \Sigma_z H_{\mathrm{BdG}}(\bm{k}), \\ 
    \Sigma_z H_{\mathrm{BdG}}(-\bm{k})\Theta &= \Theta \Sigma_z H_{\mathrm{BdG}}(\bm{k}), \label{Eq:TR_sym}\\ 
    \Sigma_z H_{\mathrm{BdG}}(\bm{k})R &= R\Sigma_z H_{\mathrm{BdG}}(\bm{k}). \label{Eq:U1_sym}
\end{align}
Here, the operators $\Theta^{\prime}$ and $\Theta$ represent pseudo time-reversal and time-reversal operators, respectively. The paraunitary matrix $R$ corresponds to the U(1) symmetry in the spin space \footnote{In the $2\times 2$ spin subspace, the U(1) rotation is represented by $e^{-i\phi \sigma_z} \simeq \cos\phi~I_{2\times 2} - i\sin\phi~\sigma_z$, where $\phi$ is a rotation angle. Thus, the commutation relation (\ref{Eq:U1_sym}) ensures the U(1) symmetry in the spin space.}. 
The explicit forms of these operators are given by
\begin{align}\label{Eq:Expression_pseudo_TR}
    \Theta^{\prime} &= \sigma_z \otimes i\sigma_y \otimes I_{3\times 3} \cdot K,\\
    \Theta &= I_{2\times 2} \otimes \sigma_x \otimes I_{3\times 3} \cdot K, \label{Eq:Expression_TR}\\
    R &= \sigma_z \otimes \sigma_z \otimes I_{3\times 3},\label{Eq:Expression_U1}
\end{align}
where $K$ denotes the complex conjugation. Here, we note that the pseudo time-reversal operator $\Theta^{\prime}$ is given as 
the combination of time-reversal and U(1) symmetries, i.e., $\Theta^{\prime} = \Theta R$. The above pseudo time-reversal (\ref{Eq:Expression_pseudo_TR}) and time-reversal operators (\ref{Eq:Expression_TR}) satisfy the following relations:
\begin{equation}\label{Eq:Square_TR_operators}
    \Theta^{\prime 2} = -1, \quad \Theta^2 = 1,
\end{equation}
The first relation for $\Theta^{\prime}$ ensures the existence of Kramers pairs of bosons \cite{kondo2019$mathbbZ_2$, joshi2019$mathbbZ_2$, thomasen2021Fragility}, where each pair of bands is related by $\bm{\Psi}_{n,-1,+}(-\bm{k}) \propto \Theta^{\prime}\bm{\Psi}_{n,+1,+}(\bm{k})$. The finite magnetic field $h \neq 0$ breaks the pseudo time-reversal (\ref{Eq:Pseudo_TR_sym}) and time-reversal (\ref{Eq:TR_sym}) symmetries, opening the band gaps between each pair of bands.

\begin{figure}[thb]
    \centering
    \includegraphics[width=1.0\linewidth]{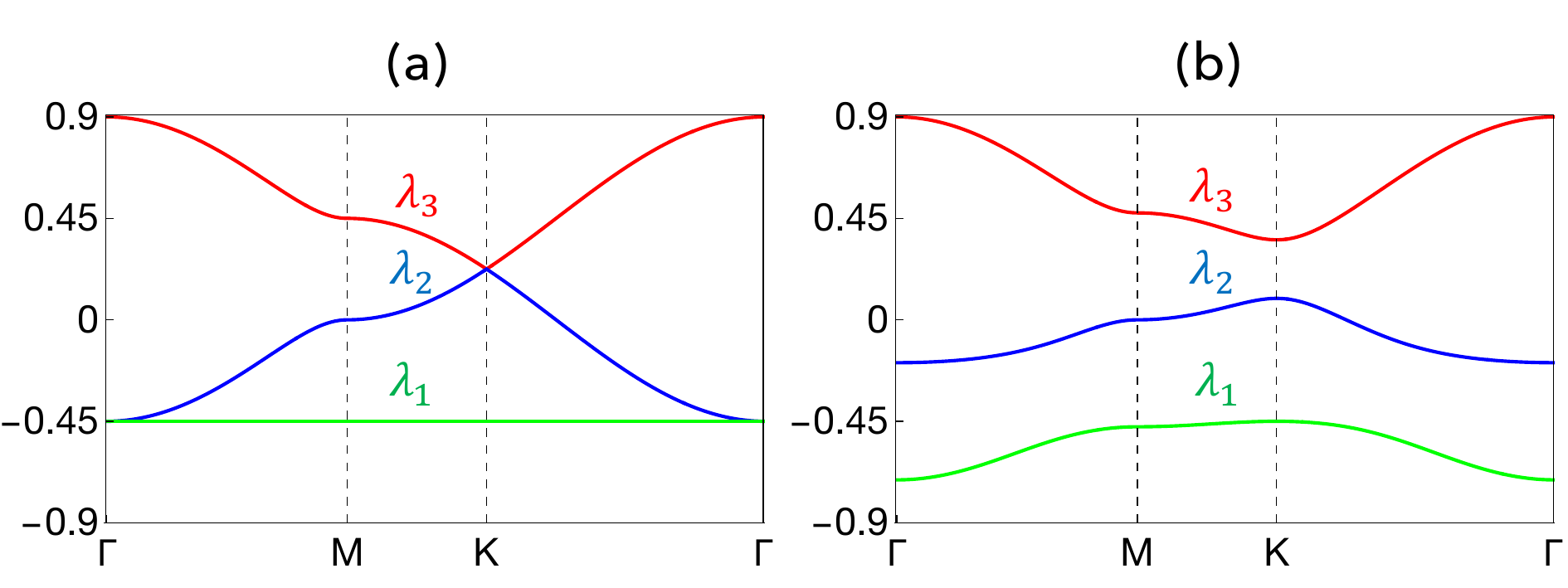}
    \caption{The three bands of $H_{\mathrm{mag}} (\bm{k})$ (\ref{Eq:single_Kagome}). The parameters used in the plot are $J_t = 1.0$, $J_h = 0.1$, and (a) $D_z = 0$, (b) $D_z = 0.3$. The high-symmetry points in reciprocal space are denoted by $\Gamma = (0,0)$, $M = (\frac{\pi}{2}, \frac{\pi}{2\sqrt{3}})$, and $K = (\frac{\pi}{3}, \frac{\pi}{\sqrt{3}})$.}
    \label{fig:Band_Kagome}
\end{figure}

\subsection{Phase diagram and the Chern numbers}\label{subsec:Phase_diagram}
\begin{figure}[thb]
    \centering
    \includegraphics[width=0.9\linewidth]{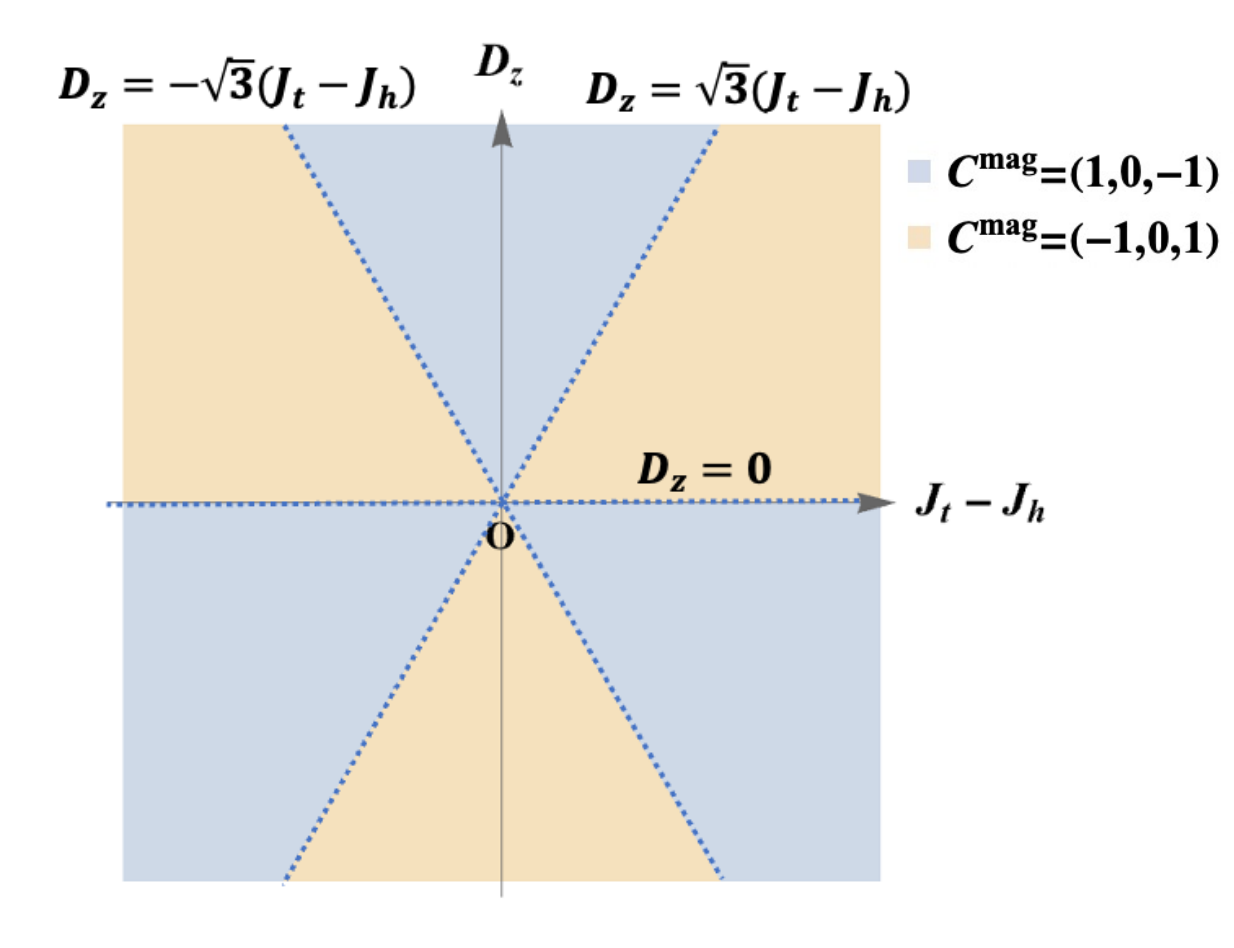}
    \caption{Phase diagram of $H_{\mathrm{mag}}(\bm{k})$ (\ref{Eq:single_Kagome}) 
    as a function of $J_t - J_h$ and $D_z$. 
    The components of $\bm{C}^{\mathrm{mag}} = (C^{\mathrm{mag}}_1, C^{\mathrm{mag}}_2, C^{\mathrm{mag}}_3)$ are the Chern numbers of the magnon bands $\lambda_1 (\bm{k})$, $\lambda_2(\bm{k})$, and $\lambda_3(\bm{k})$, respectively.
    }
    \label{fig:Phase_diagram_H0}
\end{figure}

In order to understand the band topology of the triplon system described by 
Eq. (\ref{Eq:Kagome_decouple}), it is useful to identify the band topology of the corresponding magnon system described by $H_{\mathrm{mag}}(\bm{k})$ in Eq. (\ref{Eq:single_Kagome}) as suggested in the previous section \ref{subsec:Triplon_Ham}. Hence, we here determine the phase diagram of $H_{\mathrm{mag}}(\bm{k})$ 
as a function of $J_t - J_h$ and $D_z$ based on the Chern number. 
The result is shown in Fig. \ref{fig:Phase_diagram_H0}.
To determine the phase boundaries, we used the fact that the band gaps of $H_{\mathrm{mag}} (\bm{k})$ 
close only at high symmetry points $\Gamma$ and $K$ [see Fig. \ref{fig:Band_Kagome}]. 
An explicit calculation shows that the eigenvalues of $H_{\mathrm{mag}} (\bm{k})$ at ${\bm k}=\Gamma$ and $K$ are given by
\begin{equation}\label{Eq:gap_closing_G}
    J_t - J_h, \frac{\sqrt{3}D_z - J_t + J_h}{2}, \frac{-\sqrt{3}D_z - J_t + J_h}{2},
\end{equation}
and
\begin{equation}\label{Eq:gap_closing_K}
    -\frac{J_t - J_h}{2}, -\frac{\sqrt{3}D_z - J_t + J_h}{4}, -\frac{-\sqrt{3}D_z - J_t + J_h}{4},
\end{equation}
respectively. Therefore, 
the gap closes along the lines
\begin{equation}\label{Eq:Phase_boundary}
\begin{cases}
    D_z = 0, \\
    D_z = \sqrt{3} (J_t - J_h) & (D_z > 0),\\
    D_z = -\sqrt{3} (J_t - J_h) & (D_z < 0),\\
\end{cases}
\end{equation}
which are shown as dotted lines in Fig. \ref{fig:Phase_diagram_H0}.

The set of Chern numbers of each phase is determined numerically \footnote{We used the method in Ref. \cite{fukui2005Chern}}.
Here, the Chern number of the $n$th band of $H_{\mathrm{mag}} (\bm{k})$ is defined by $C^{\mathrm{mag}}_n = \frac{1}{2\pi} \int_{\mathrm{BZ}} d^2 k \ \Omega^{\mathrm{mag}}_{n}(\bm{k})$. 
From Eq. (\ref{Eq:Berry_relation}), it follows that the Chern number of each triplon band 
defined by $C_{n,s,\sigma} = \frac{1}{2\pi} \int_{\mathrm{BZ}} d^2 k \ \Omega_{n,s,\sigma}(\bm{k})$ can be expressed in terms of $C^{\mathrm{mag}}_n$ as $C_{n,s,\sigma} = s\sigma C^{\mathrm{mag}}_n$.

\subsection{$\mathbb{Z}_2$ invariant and the spin Nernst conductivity}\label{subsec:Z_2}
\begin{figure}[b]
    \centering
    \includegraphics[width=0.9\linewidth]{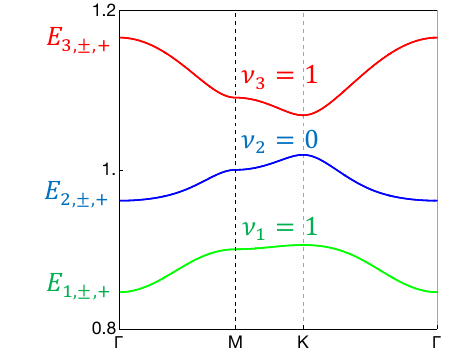}
    \caption{The band structure of $H_{\mathrm{BdG}} (\bm{k})$ (\ref{Eq:Kagome_decouple}) in the absence of the magnetic field. The parameters used in the plot are $J_d = 1.0$, $J_t = 0.2$, $J_h = 0.02$, and $D_z = 0.06$. The high-symmetry points in reciprocal space are denoted by $\Gamma = (0,0)$, $M = (\frac{\pi}{2}, \frac{\pi}{2\sqrt{3}})$, and $K = (\frac{\pi}{3}, \frac{\pi}{\sqrt{3}})$. Each Kramers pair of bands $E_{n,\pm, +} (\bm{k})$ ($n = 1, 2, 3$) expressed in Eq. (\ref{Eq:Energy_relation}) are degenerate in the entire Brillouin zone. The $\mathbb{Z}_2$ invariants are also indicated in the figure. 
    }
    \label{fig:Kramers_pair_band}
\end{figure}
In the absence of a magnetic field, two particle bands $E_{n,+,+}(\bm{k})$ and $E_{n,-,+}(\bm{k})$ form a Kramers pair because of the presence of the pseudo time-reversal symmetry $\Theta^{\prime}$, as explained in the previous section \ref{subsec:Triplon_Ham}. In addition, the triplon Hamiltonian (\ref{Eq:Kagome_decouple}) satisfies the inversion symmetry. As a result, each Kramers pair of bands $E_{n,+,+}(\bm{k})$ and $E_{n,-,+}(\bm{k})$ is doubly degenerate in the entire Brillouin zone, as shown in Fig. \ref{fig:Kramers_pair_band}. In this case, the total Chern number of each Kramers pair vanishes due to $\Omega_{n,+,+} + \Omega_{n,-,+} = 0$. (see Eq. (\ref{Eq:Berry_relation}).) In the presence of such (pseudo) time reversal symmetry, the topology of the triplon bands is characterized by a $\mathbb{Z}_2$ invariant. In an inversion symmetric system, the $\mathbb{Z}_2$ invariant $\nu_n$ of the $n$th Kramers pair of bands can be expressed in terms of the parity eigenvalues of the band \cite{fu2007Topological, hasan2010Colloquium, joshi2019$mathbbZ_2$, thomasen2021Fragility}, i.e.,
\begin{equation}\label{Eq:Def_Z2}
    \prod_{i = 1}^{4} \xi_n (\Gamma_{i}) = (-1)^{\nu_n},
\end{equation}
where $\xi_n (\Gamma_{i})$ represents the parity eigenvalues at the four time-reversal invariant momenta (TRIM). The TRIM in the present case are $\Gamma = (0,0)$, $M_1 = (\frac{\pi}{2},-\frac{\pi}{2\sqrt{3}})$, $M_2 = (\frac{\pi}{2}, \frac{\pi}{2\sqrt{3}})$, and $M_3 = (0, \frac{\pi}{\sqrt{3}})$. 
From the explicit calculation [see Appendix \ref{Appendix:Z_2}], we find that the $\mathbb{Z}_2$ invariants always satisfy $\nu_1 = \nu_3 = 1$ and $\nu_2 = 0$ 
for finite $D_z$. This implies that the nontrivial $\mathbb{Z}_2$ invariants come from the pair of the Chern numbers $\pm 1$ of each Kramers pair [see Fig. \ref{fig:Phase_diagram_H0}]. More specifically, the $\mathbb{Z}_2$ invariant (\ref{Eq:Def_Z2}) can be regarded as the spin Chern number $\nu_n = \frac{1}{2} (C_{n,+,+} - C_{n,-,+})\ (\mathrm{mod}\ 2)$, as in electronic systems with conservation of $S_z$ \cite{kane2005$Z_2$, hasan2010Colloquium, kondo2019$mathbbZ_2$, thomasen2021Fragility}.



\begin{figure}[thb]
    \centering
    \includegraphics[width=0.9\linewidth]{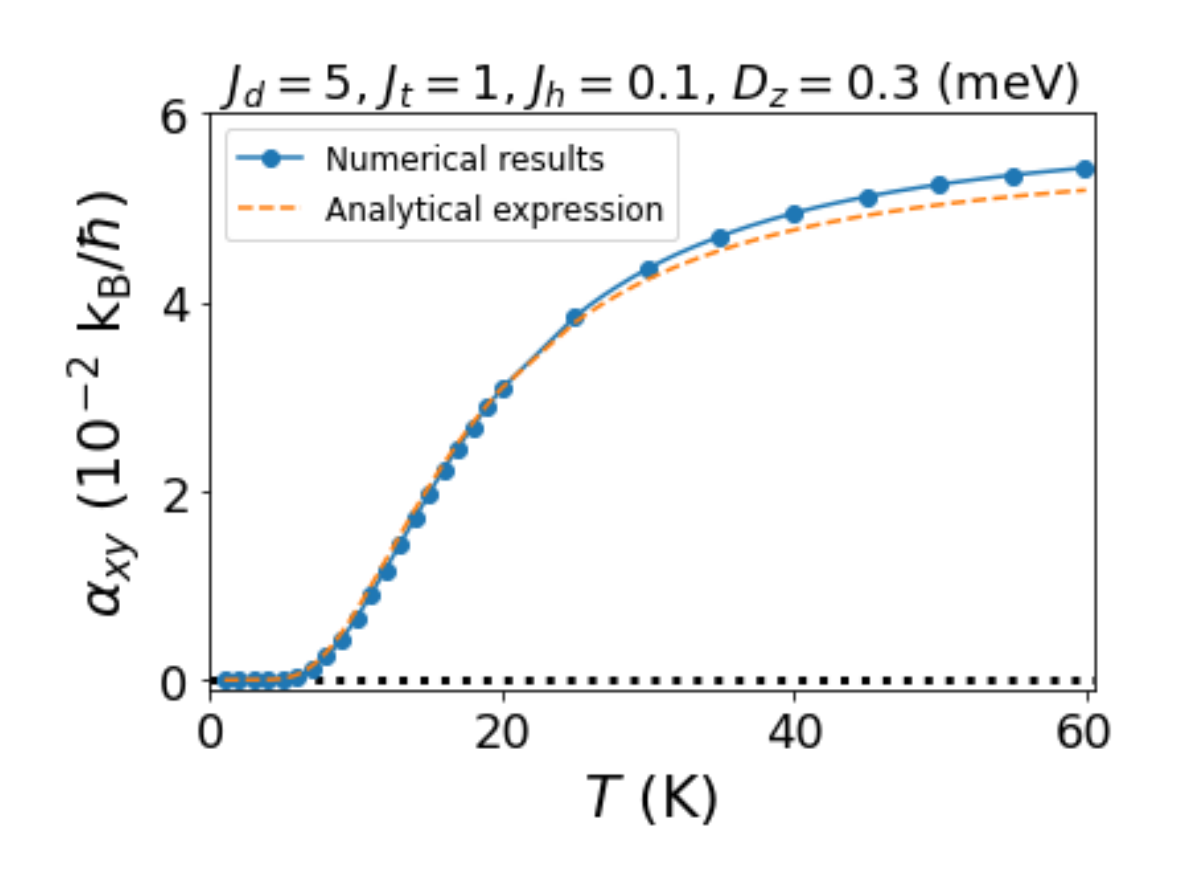}
    \caption{Comparison between the analytical expression (\ref{Eq:Analytical_Spin_Nernst}) and the numerical results of $\alpha_{xy}$ below $\SI{60}{K}$. The parameters used in the calculations are $J_d = \SI{5.0}{meV}$, $J_t = \SI{1.0}{meV}$, $J_h = \SI{0.1}{meV}$, $D_z = \SI{0.3}{meV}$, and $h = 0$.}
    \label{fig:Spin_Nernst}
\end{figure}

Reflecting the topology of the triplon bands, the system can exhibit Hall-type transport. 
According to Refs. \cite{matsumoto2014Thermal,cheng2016Spin, zyuzin2016Magnon, thomasen2021Fragility}, the spin Nernst and thermal Hall conductivities in the $x$-$y$ plane are, respectively, given by 
\begin{widetext}
\begin{equation}\label{Eq:Spin_Nernst_formula}
   \alpha_{xy} = -\frac{2k_B}{\hbar}\sum_{n} \int_{\mathrm{BZ}} \frac{d^2 k}{(2\pi)^2} c_1 (\rho(E_{n,+,+}(\bm{k}))) \Omega_{n,+,+} (\bm{k}),
\end{equation}
and
\begin{equation}\label{Eq:Thermal_Hall_formula}
    \kappa_{xy} = -\frac{k_{B}^2 T}{\hbar}\sum_{n,s} \int_{\mathrm{BZ}} \frac{d^2 k}{(2\pi)^2} \left[c_2 (\rho(E_{n,s,+}(\bm{k}))) - \frac{\pi^2}{3}\right] \Omega_{n,s,+} (\bm{k}),
\end{equation}
\end{widetext}
respectively. Here, $k_B$ is the Boltzmann constant, $T$ 
the temperature, $\hbar$ 
the Planck constant, and 
$\rho(E) = 1/(e^{\beta E} - 1)$ the Bose distribution function with $\beta$ being the inverse temperature. The explicit form of $c_n (\rho) (n = 1, 2)$ is given by $c_n (\rho) = \int_{0}^{\rho} \ln^{n}(1+t^{-1}) dt$. In the absence of 
magnetic field, the system exhibits a finite spin Nernst effect, reflecting the $\mathbb{Z}_2$ topological nature, whereas the thermal Hall effect vanishes 
due to the cancellation $\Omega_{n,+,+} + \Omega_{n,-,+} = 0$. The numerical results of the spin Nernst conductivity $\alpha_{xy}$ are shown in Fig. \ref{fig:Spin_Nernst}.

In order to gain a deeper understanding of the behavior of $\alpha_{xy}$, we obtain the following analytical expression: 
\begin{equation}\label{Eq:Analytical_Spin_Nernst}
    \alpha_{xy} \simeq \mathrm{sgn}(D_z)\frac{k_B}{\pi\hbar} \sum_{p = \Gamma, K} \sum_{\tau = \pm} \tau c_1 (\rho(E_{p,\tau,0})),
\end{equation}
where $\mathrm{sgn}(D_z)$ corresponds to the sign of $D_z$. We define $E_{\Gamma, \tau, \upsilon}$ and $E_{K, \tau, \upsilon}$ ($\upsilon = \pm, 0$) by
\begin{align}\nonumber
\label{Eq:E_G_symmetry}
    E_{\Gamma, \tau ,\upsilon}
    = &  \sqrt{J_d (J_d - J_t + J_h)} \\
    &- \tau\frac{\sqrt{3}}{2}\sqrt{\frac{J_d}{J_d - J_t + J_h}} \abs{D_z} -\upsilon h, 
\end{align} and
\begin{align}\nonumber
\label{Eq:E_K_symmetry}
    E_{K, \tau ,\upsilon}
    = & \sqrt{J_d \left(J_d + \frac{J_t-J_h}{2}\right)} \\
    & - \tau\frac{\sqrt{3}}{4}\sqrt{\frac{J_d}{J_d +\frac{J_t - J_h}{2}}} \abs{D_z} -\upsilon h, 
\end{align}
respectively. Here, $E_{\Gamma, +, +}$ corresponds to the spin gap (\ref{Eq:Spin_gap}). The detailed derivation of the above expression (\ref{Eq:Analytical_Spin_Nernst}) is given in Appendix \ref{Appendix:Analytical_expression}. Figure \ref{fig:Spin_Nernst} clearly shows that the above expression (\ref{Eq:Analytical_Spin_Nernst}) well explains the numerical results even at high temperatures. Here we note that from Eqs. (\ref{Eq:Analytical_Spin_Nernst})-(\ref{Eq:E_K_symmetry}), $\alpha_{xy}$ with a small $D_z$ linearly depends on $D_z$.

\subsection{Chern number and the thermal Hall conductivity}\label{subsec:Z_Dxy0}
\begin{figure}[thb]
    \centering
    \includegraphics[width=0.9\linewidth]{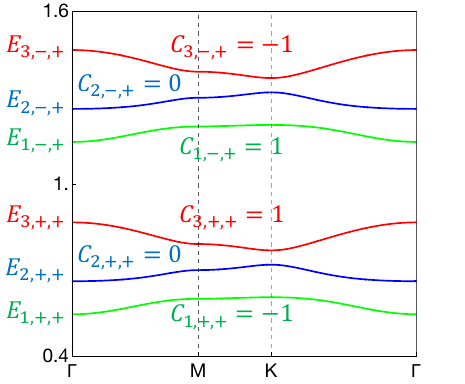}
    \caption{The band structure of $H_{\mathrm{BdG}} (\bm{k})$ (\ref{Eq:Kagome_decouple}) with a finite magnetic field. The parameters used in the plot are $J_d = 1.0$, $J_t = 0.2$, $J_h = 0.02$, $D_z = 0.06$, and $h = 0.3 $. The high-symmetry points in reciprocal space are denoted by $\Gamma = (0,0)$, $M = (\frac{\pi}{2}, \frac{\pi}{2\sqrt{3}})$, and $K = (\frac{\pi}{3}, \frac{\pi}{\sqrt{3}})$. Each pair of two bands $E_{n, +, +} (\bm{k})$ and $E_{n, -, +} (\bm{k})$ ($n = 1, 2, 3$) are no longer degenerate due to the absence of the pseudo time-reversal symmetry represented by $\Theta^{\prime}$. The Chern numbers are also indicated in the figure.}
    \label{fig:Band_h}
\end{figure}

\begin{figure}[thb]
    \centering
    \includegraphics[width=1.0\linewidth]{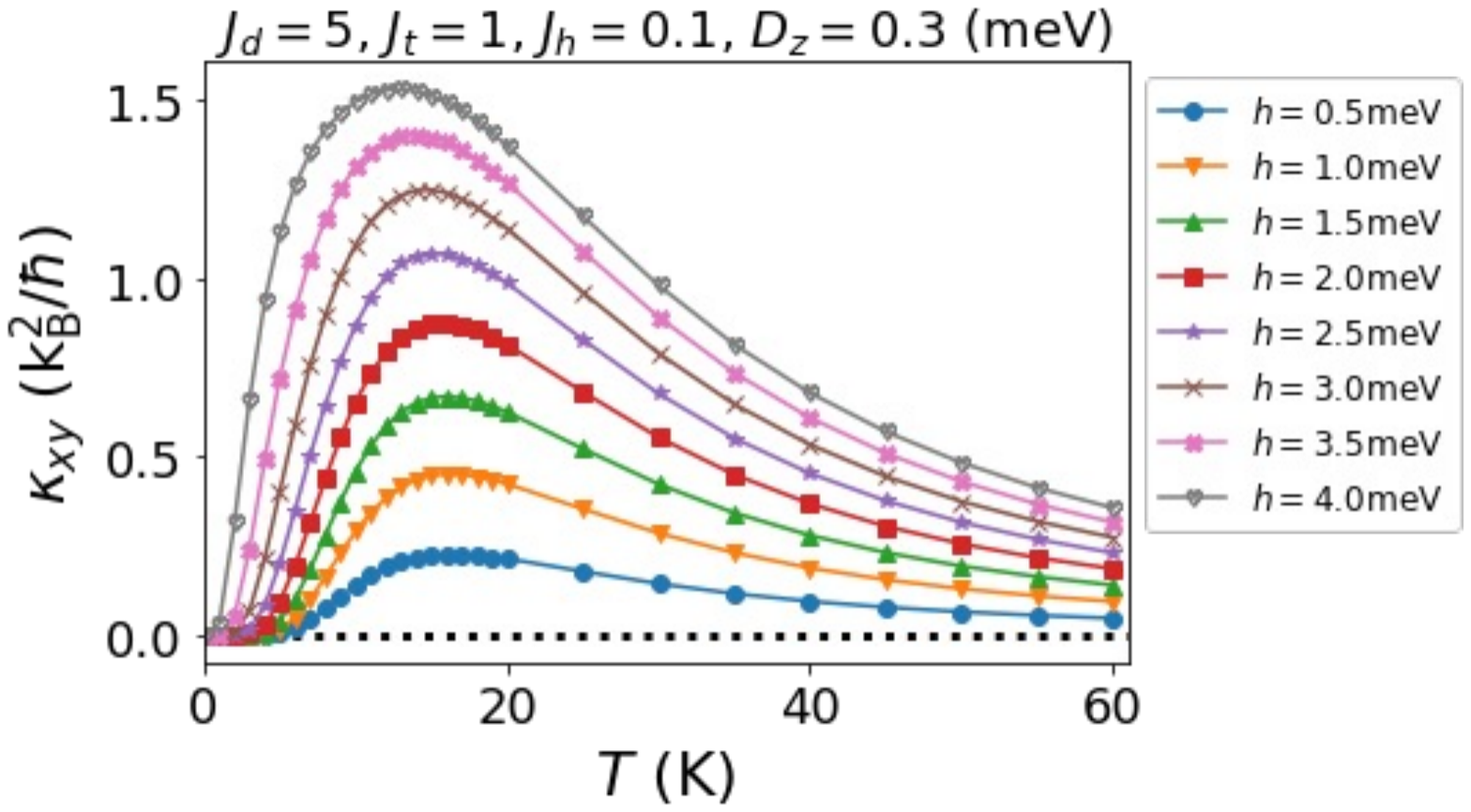}
    \caption{Numerical results for the thermal Hall conductivity $\kappa_{xy}$ (\ref{Eq:Thermal_Hall_formula}) below $\SI{60}{K}$ with $J_d = \SI{5.0}{meV}$, $J_t = \SI{1.0}{meV}$, $J_h = \SI{0.1}{meV}$, $D_z = \SI{0.3}{meV}$, and $h = 0.5 \times m~\si{meV}$ ($m = 1, 2, \cdots, 8$). The value of $1.0$ on the vertical axis of the graph corresponds to $\SI{1.2}{mW/K.m}$, 
    assuming a lattice constant $c = \SI{15}{\angstrom}$ in the $z$-direction.}
    \label{fig:Thermal_Hall_Dxy0_60K}
\end{figure}

In the presence of a finite magnetic field, the pseudo time-reversal symmetry represented by the operator $\Theta^{\prime}$ is broken, and thus the two bands $E_{n,+,+}$ and $E_{n,-,+}$ are no longer degenerate [see Fig. \ref{fig:Band_h}]. In this case, the topology of the triplon bands is characterized by the Chern number of each particle band $E_{n,s,+}$, i.e., $C_{n,s,+} = sC^{\mathrm{mag}}_n$ [see Sec. \ref{subsec:Phase_diagram}], resulting in a finite thermal Hall effect in the presence of $D_z$. The numerical results of the thermal Hall conductivity (\ref{Eq:Thermal_Hall_formula}) are shown in Fig. \ref{fig:Thermal_Hall_Dxy0_60K}. 

From the numerical results in Fig. \ref{fig:Thermal_Hall_Dxy0_60K}, we find that $\kappa_{xy}$ with realistic values of material parameters and magnetic fields are comparable to the values obtained in the previous experiments on the thermal Hall effect of magnons \cite{onose2010Observation, ideue2012Effect, hirschberger2015Thermal} and phonons \cite{strohm2005Phenomenological}. This indicates that quantum dimer magnets on the maple-leaf and star lattices could serve as promising candidates for the thermal Hall effect of triplons.

\begin{figure}[thb]
    \centering
    \includegraphics[width=1.0\linewidth]{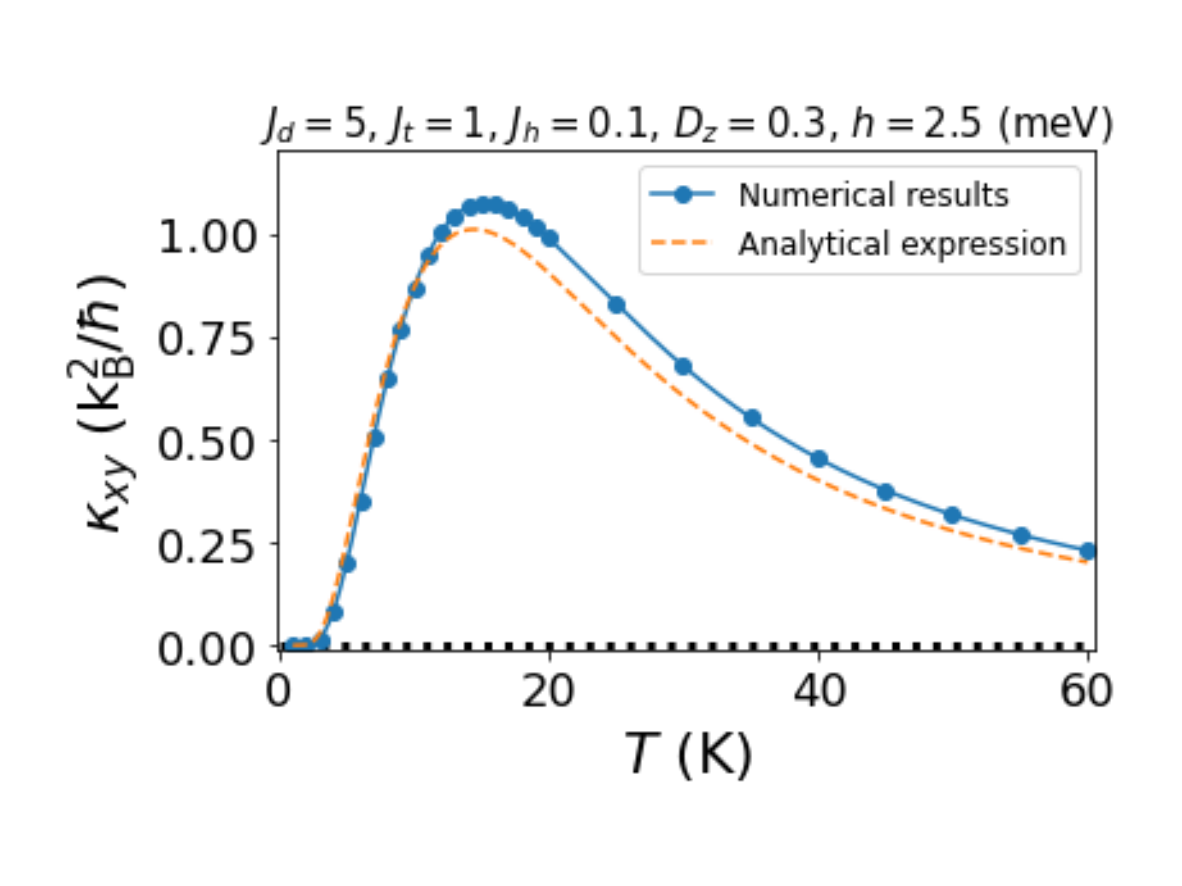}
    \caption{Comparison between the analytical expression (\ref{Eq:Analytical_Thermal_Hall}) and the numerical results of $\kappa_{xy}$ below $\SI{60}{K}$. The parameters used in the calculations are $J = \SI{5.0}{meV}$, $J_t = \SI{1.0}{meV}$, $J_h = \SI{0.1}{meV}$, $D_z = \SI{0.3}{meV}$, and $h = \SI{2.5}{meV}$.}
    \label{fig:Numerical_analytical_Dxy0}
\end{figure}

By a similar argument as in the previous section \ref{subsec:Z_2}, we derive the analytical expression for $\kappa_{xy}$. Using Eqs. (\ref{Eq:E_G_symmetry}) and (\ref{Eq:E_K_symmetry}), we have
\begin{equation}\label{Eq:Analytical_Thermal_Hall}
    \kappa_{xy} \simeq \mathrm{sgn}(D_z) \frac{k^2_B T}{2\pi\hbar} \sum_{p = \Gamma, K} \sum_{\tau = \pm} \sum_{\upsilon = \pm} \tau\upsilon c_2 (\rho(E_{p,\tau,\upsilon})).
\end{equation}
See Appendix \ref{Appendix:Analytical_expression} for the detailed derivation. From the above expression, 
we find that $\kappa_{xy}$ depends almost linearly on both $D_z$ and $h$. To verify the validity of 
the analytical formula, we compare it with numerical 
results in the low-temperature region, as shown in Fig. \ref{fig:Numerical_analytical_Dxy0}. From this figure, it is clear that the analytical expression in Eq. (\ref{Eq:Analytical_Thermal_Hall}) 
shows an excellent agreement with the numerical results.

\section{The low-symmetry case with a finite in-plane DM vector}
\label{sec:Dxy}

\begin{figure}[thb]
    \centering
    \includegraphics[width=1.0\linewidth]{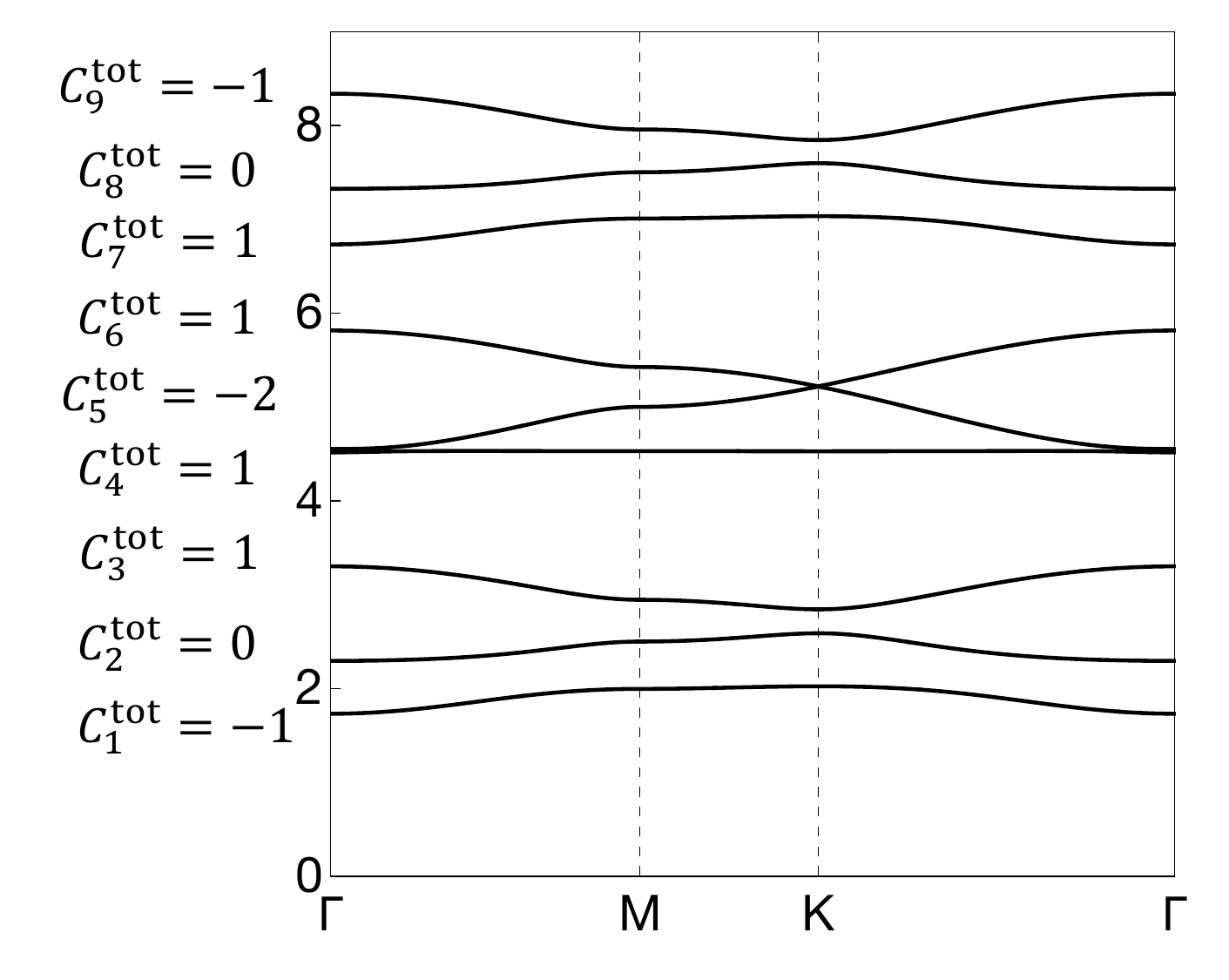}
    \caption{A representative band structure of $H^{\mathrm{tot}}_{\mathrm{BdG}} (\bm{k})$ (\ref{Eq:BdG_Form}) for a finite in-plane DM vector and magnetic field. The parameters used in the plot are $J_d = 5.0$, $J_t = 1.0$, $J_h = 0.1$, $D_x = D_y = 0.21$, and $D_z = 0.3$, and $h = 2.5$. The high-symmetry points in reciprocal space are denoted by $\Gamma = (0,0)$, $M = (\frac{\pi}{2}, \frac{\pi}{2\sqrt{3}})$, and $K = (\frac{\pi}{3}, \frac{\pi}{\sqrt{3}})$. The Chern numbers $(C^{\mathrm{tot}}_1, C^{\mathrm{tot}}_2, \cdots)$  are also indicated in the figure.}
    \label{fig:Band_Dxy}
\end{figure}

A finite in-plane DM vector $D_{x, y}$ breaks the U(1) symmetry in the spin space, introducing a coupling between the $t_0$ and $t_{\pm}$ modes.
%
As a result, the triplon–magnon correspondence established in Sec. \ref{sec:Dxy_0} no longer 
holds. Therefore, we perform a numerical analysis of the full BdG matrix (\ref{Eq:BdG_Form}).
%
%
Here, a finite magnetic field is essential to enable a topological classification of the system, as it typically results in nine fully gapped triplon bands, as illustrated in Fig. \ref{fig:Band_Dxy}. Note that although some band gaps may be too small to be 
visible at this scale, they generally remain open 
in a finite magnetic field. 
Therefore, in this section, we examine how 
the application of a magnetic field affects the triplon band topology and the associated thermal Hall effect. 
This analysis is particularly relevant since the magnetic field is typically the only experimentally tunable parameter.

\subsection{Phase diagram and the Chern numbers}\label{subsec:Phase_diagram_Dxy}

\begin{table}[thb]
    \centering
    \begin{ruledtabular}
    \begin{tabular}{cc}
    $h(\si{meV})$ & $\bm{C}^{\mathrm{tot}}$\\
    \hline
    $0.1$ & $(1,-3,2,-1,-6,9,0,-2,1)$ \\ 
    $0.2$ & $(1,0,-1,1,1,0,-3,0,1)$ \\
    $0.3$ & $(1,0,-3,3,-2,1,-1,0,1)$ \\
    $0.4$ & $(-1,3,-4,3,-2,1,-2,3,-1)$ \\
    $0.5, 0.6, 0.7$ & $(-1,3,-4,3,-1,-2,1,0,-1)$ \\
    $0.8 \leq h \leq 1.6$ & $(-1,0,1,1,-2,3,-1,0,-1)$ \\
    $1.6 \leq h \leq 4.0$ & $(-1,0,1,1,-2,1,1,0,-1)$ \\
    \end{tabular}
    \end{ruledtabular}
    \caption{Various phases with 
    varying magnetic fields, 
    characterized by the set of Chern numbers $\bm{C}^{\mathrm{tot}} = (C^{\mathrm{tot}}_1, C^{\mathrm{tot}}_2, 
    \dots, C^{\mathrm{tot}}_9)$ of the nine triplon bands. The parameters used in the calculations are $J_d = \SI{5.0}{meV}$, $J_t = \SI{1.0}{meV}$, $J_h = \SI{0.1}{meV}$, $D_x = D_y = \SI{0.21}{meV}$, $D_z = \SI{0.3}{meV}$; 
    the magnetic field $h$ is varied in steps of $0.1 \si{meV}$. Here, we do not determine the precise phase boundaries, i.e., the gap closing points. 
    }
    \label{Table:Phase_Diagram_Dxy}
\end{table}

Based on the set of Chern numbers of the nine triplon bands, we numerically identify multiple phases with distinct band topologies as the magnetic field varies, which is shown in Table \ref{Table:Phase_Diagram_Dxy}. Here, the Chern number of the $n$th band is defined by $C^{\mathrm{tot}}_n = \frac{1}{2\pi}\int_{\mathrm{BZ}} d^2 k \ \Omega^{\mathrm{tot}}_{n,+} (\bm{k})$, where the Berry curvature $\Omega^{\mathrm{tot}}_{n, \sigma} (\bm{k})$ is given using the eigenvector of $\Sigma^{\mathrm{tot}}_z H^{\mathrm{tot}}_{\mathrm{BdG}}(\bm{k})$ in Eq. (\ref{Eq:Eigenproblem_BdG_tot_Psi}) as $\Omega^{\mathrm{tot}}_{n, \sigma} (\bm{k}) = -2\sigma\mathrm{Im}\left[\frac{\partial \bm{\Psi}^{\mathrm{tot}\ \dag}_{n,\sigma}(\bm{k})}{\partial k_x} \Sigma^{\mathrm{tot}}_z \frac{\partial \bm{\Psi}_{n,\sigma}(\bm{k})}{\partial k_y}\right]$. It is noteworthy that finite $D_{x, y}$ can result in 
large Chern numbers, such as $C^{\mathrm{tot}}_5 = -6$ and $C^{\mathrm{tot}}_6 = 9$ with $h = \SI{0.1}{meV}$, which is in contrast 
to the $D_{x, y} = 0$ case, where the Chern number of each triplon band can only take values $\pm 1$ or $0$. 
We note in passing that a similarly rich phase diagram with a variety of large triplon Chern numbers has been found in the Kitaev-Heisenberg model on the star lattice \cite{dornellas2024KitaevHeisenberg}.


\subsection{Thermal Hall conductivity}\label{subsec:THE_Dxy}
\begin{figure*}[thb]
    \includegraphics[width=1.0\textwidth]{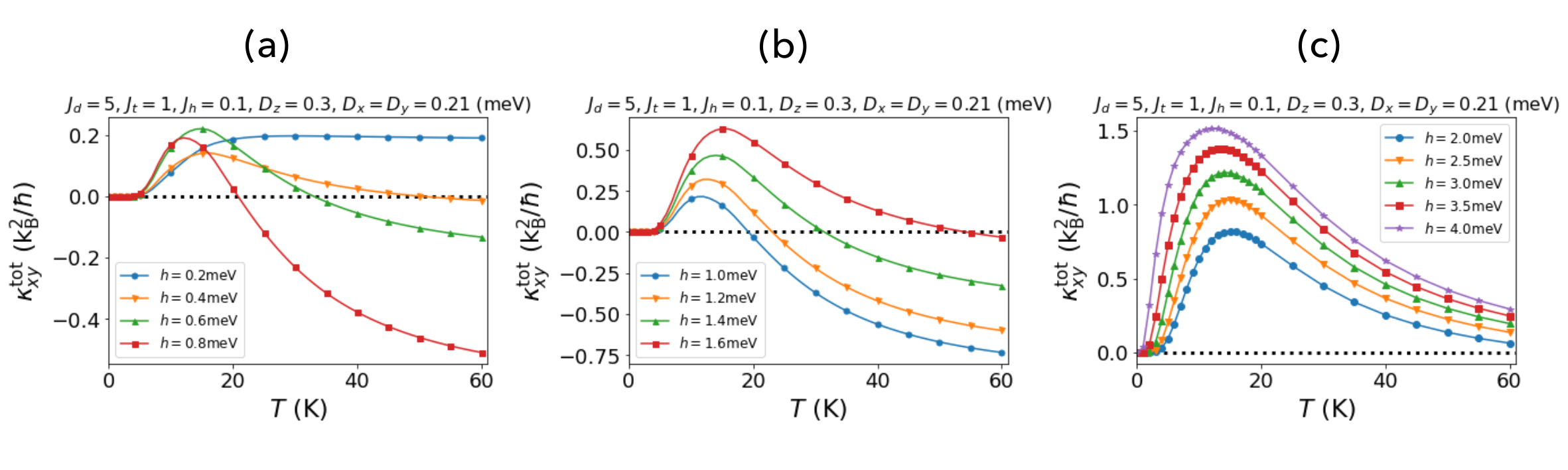}
    \caption{Numerical results of the thermal Hall conductivity $\kappa^{\mathrm{tot}}_{xy}$ below $\SI{60}{K}$ with $J_d = \SI{5.0}{meV}$, $J_t = \SI{1.0}{meV}$, $J_h = \SI{0.1}{meV}$, $D_x = D_y = \SI{0.21}{meV}$, $D_z = \SI{0.3}{meV}$, and (a) $h = 0.2 \times m~\si{meV}$ ($m = 1, 2, \cdots, 4$), (b) $h = 0.2 \times m~\si{meV}$ ($m = 5, 6, \cdots, 8$), and (c) $h = 0.5 \times m~\si{meV}$ ($m = 4, 5, \cdots, 8$).}
    \label{fig:THE_Dxy_60K}
\end{figure*}

\begin{figure}[thb]
    \centering
    \includegraphics[width=1.0\linewidth]{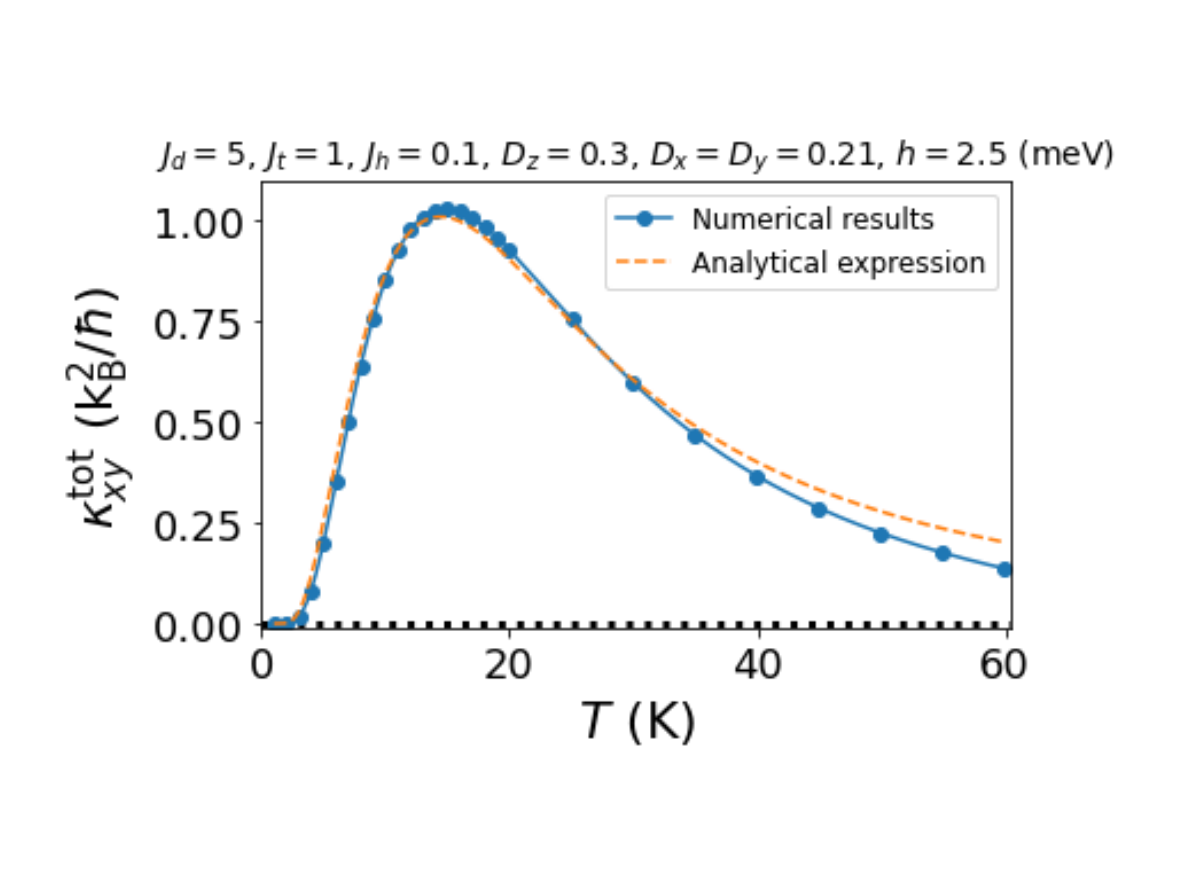}
    \caption{Comparison between the analytical expression [Eq. (\ref{Eq:Analytical_Thermal_Hall})] and the numerical results of $\kappa^{\mathrm{tot}}_{xy}$ below $\SI{60}{K}$. The parameters used in the calculations are $J = \SI{5.0}{meV}$, $J_t = \SI{1.0}{meV}$, $J_h = \SI{0.1}{meV}$, $D_x = D_y = \SI{0.21}{meV}$, $D_z = \SI{0.3}{meV}$, and $h = \SI{2.5}{meV}$.}
    \label{fig:Numerical_Analytical_Dxy}
\end{figure}

We numerically compute the thermal Hall conductivity 
\begin{align}
\kappa^{\mathrm{tot}}_{xy} \!=\! -\frac{k_{B}^{2} T}{\hbar}\sum_{n=1}^{9} \int_{\mathrm{BZ}} \!\frac{d^2 k}{(2\pi)^2}\! \left[c_2 (\rho(E^{\mathrm{tot}}_{n,+}(\bm{k}))) \!-\! \frac{\pi^{2}}{3}\right] \!\Omega^{\mathrm{tot}}_{n,+} (\bm{k}).
\end{align}
Figure \ref{fig:THE_Dxy_60K} shows the numerical results 
for several values of $h$.
Figure \ref{fig:THE_Dxy_60K} (a) and (b) indicate that $\kappa^{\mathrm{tot}}_{xy}$ in the low-magnetic field region where $\SI{1.6}{meV}\geq h \geq \SI{0.4}{meV}$ undergoes a sign reversal as the temperature increases. This implies that the contribution from the upper bands 
becomes dominant 
over that from the lowest band with increasing temperature. A similar sign reversal behavior has also been found in Kagome magnets with magnon excitations \cite{mook2014Magnon, lee2015Thermal, hirschberger2015Thermal}. 
In contrast, in the high-magnetic-field region where $h \geq \SI{2.0}{meV}$, $\kappa^{\mathrm{tot}}_{xy}$ does not exhibit such sign reversal behavior but shows a similar temperature and magnetic field dependence as in the absence of $D_{x, y}$ [see Fig. \ref{fig:THE_Dxy_60K} (c) and Fig. \ref{fig:Thermal_Hall_Dxy0_60K}]. More precisely, we find that $\kappa^{\mathrm{tot}}_{xy}$ in a high magnetic field is still well described by the same analytical expression Eq. (\ref{Eq:Analytical_Thermal_Hall}) as in the absence of $D_{x,y}$, which is shown in Fig. \ref{fig:Numerical_Analytical_Dxy}.

To clarify this point, let us consider the high magnetic field regime, where the gaps between the lowest three bands and the higher bands remain open even in the presence of $D_{x,y}$. In this case, the effective description discussed in Appendix \ref{Appendix:Analytical_expression} remains the same when the BdG matrix (\ref{Eq:BdG_Form}) is projected onto the subspace spanned by the lowest three modes obtained for $\bm{D} = 0$. This is because this subspace coincides with the $S_z= +1$ sector, and $D_{x}$ and $D_{y}$ do not introduce any matrix elements within that sector [see Appendix \ref{Appendix:Explicit_form}].

Consequently, in the low-temperature regime, where contributions from the lowest modes are dominant, the analytical expression Eq. (\ref{Eq:Analytical_Thermal_Hall}) still provides a good description of $\kappa^{\mathrm{tot}}_{xy}$ even with $D_{x,y}$. Indeed, the set of Chern numbers for the lowest modes $(C^{\mathrm{tot}}_1, C^{\mathrm{tot}}_2, C^{\mathrm{tot}}_3)$ given in the seventh row of Table. \ref{Table:Phase_Diagram_Dxy} is identical to that without $D_{x,y}$ [see also Fig. \ref{fig:Phase_diagram_H0}]. 

\section{Conclusion and outlook}\label{sec:Conclusion}

In this work, we have investigated the spin Nernst and thermal Hall effects of topological triplon excitations on the maple-leaf and star lattices. Our analysis is based on the insight that these lattices become equivalent to the Kagome lattice when each dimer bond is treated as a single effective site.

We considered two scenarios: one with higher symmetry, where the DM interaction lies perpendicular to the plane ($D_{x,y} = 0$), and one with reduced symmetry, where the DM vector has arbitrary orientation. In the higher symmetry case, the spin Hamiltonian preserves U(1) spin rotational symmetry, allowing us to directly relate the triplon band topology and thermal transport properties to those of a magnon Hamiltonian on the Kagome lattice. This correspondence provides a simple and powerful guiding principle. If a quantum dimer magnet can be effectively mapped to a magnon system with corner-sharing geometry and out-of-plane DM interactions, it is highly likely to host spin Nernst and thermal Hall effects driven by topological triplon bands.

Given the broad range of quantum dimer magnets identified experimentally, this result opens up promising directions for discovering materials exhibiting topological thermal transport. In particular, dimer magnets that preserve U(1) symmetry naturally support triplon bands with $\mathbb{Z}_2$ topology—making them strong candidates for realizing triplon analogs of $\mathbb{Z}_2$ topological insulators and the associated spin Nernst effect, which remain largely unexplored in experiments.

When the in-plane DM components are finite ($D_{x,y} \neq 0$), the U(1) symmetry is broken, and the simple mapping to the Kagome magnon model no longer applies. Nevertheless, the system exhibits even richer topological behavior, including multiple topological phase transitions and sign reversals of the thermal Hall conductivity as a function of magnetic field and temperature, even without reversing the direction of the field itself. This highlights the importance of studying realistic lattice geometries where symmetry is reduced, as such conditions can give rise to potentially useful transport phenomena.

On the material side, quantum magnets on the maple-leaf and star lattices have been actively explored in recent years. 
Examples include bluebellite Cu$_6$IO$_3$(OH)$_{10}$Cl \cite{mills2014Bluebellite, haraguchi2021Quantum}, mojaveite Cu$_6$TeO$_4$(OH)Cl \cite{mills2014Bluebellite}, fuettererite Pb$_3$Cu$_6$TeO$_6$(OH)$_7$Cl$_5$ \cite{kampf2013Leadtellurium}, sabelliite (Cu,Zn)$_2$Zn[(As,Sb)O$_4$](OH)$_3$ \cite{olmi1995crystal}, spangolite Cu$_6$Al(SO$_4$)(OH)$_{12}$Cl$\cdot3$H$_2$O \cite{penfield1890spangolite, miers1893Spangolite, frondel1949crystallography, hawthorne1993crystal, schmoll2024Tensor}, MgMn$_3$O$_7$$\cdot3$H$_2$O \cite{haraguchi2018Frustrated}, Na$_2$Mn$_3$O$_7$ \cite{venkatesh2020Magnetic, saha2023Twodimensional}, and dimethylammonium copper sulfate [(CH$_3$)$_2$(NH$_2$)]$_3$Cu$_3$(OH)(SO$_4$)$_4$$\cdot$0.24H$_2$O \cite{sorolla2020Synthesis, ishikawa2024Geometric}.
%
Among these, recent theoretical studies indicate that spangolite may host a dimer singlet ground state at low temperatures \cite{schmoll2024Tensor}, identifying it as a particularly promising candidate for realizing topological triplon transport. 
Moreover, even in materials that do not exhibit a dimer singlet ground state under ambient conditions, it may be possible to induce such a phase through external pressure. Similar physics is also expected in Kagome bilayer systems with dimer singlet phases \cite{thomasen2021Fragility}.
Theoretical investigations of these candidate materials are a natural next step. In developing realistic models, it will be essential to consider the role of symmetric exchange anisotropies, which can undermine the robustness of triplon band topology \cite{thomasen2021Fragility}. In addition, many-body interactions between triplons may have a significant impact on topological thermal transport \cite{suetsugu2022Intrinsic, koyama2023Flavorwave, koyama2023Nonlinear, choi2023Sizable, koyama2024Thermal, koyama2024Impact, Chatzichrysafis2025Thermal}.

Another exciting direction is to explore topological thermal transport in systems with flavor-wave excitations \cite{romhanyi2019Multipolar, furukawa2020Effects, koyama2023Flavorwave, ma2024Upperbranch, lu2024Spin, zhang2024Thermal}. Multi-flavor Mott insulators, where the local Hilbert space hosts multiple degrees of freedom, have recently garnered considerable interest \cite{chen2024Multiflavor, mila2024Mott}. In such systems, elementary excitations can be viewed as multiplet bosons—or generalized ``triplons"—emerging from singlet-like ground states \cite{zhang2024Thermal}. Analogous guiding principles to those developed here may apply in these contexts as well, pointing toward flavor-wave bosons as candidates for topological thermal transport. Since multi-flavor Mott insulators arise in both strongly correlated electron systems and ultracold atomic gases \cite{chen2024Multiflavor, mila2024Mott}, studying their flavor-wave excitations could significantly broaden the landscape of materials exhibiting topological thermal transport.

Finally, we emphasize that the analytical expressions derived from our $2\times 2$ effective Hamiltonian near the gap-closing points [see Appendix \ref{Appendix:Analytical_expression}] accurately capture the behavior of both spin Nernst and thermal Hall conductivities, even at relatively high temperatures. As such, our theoretical framework offers a versatile and robust tool for developing analytical formulas in a wider class of topological bosonic systems.

\begin{acknowledgments}
We thank Jung Hoon Han, Yasir Iqbal, Hana Schiff, and Judit Romh\'{a}nyi for valuable discussions. 
%
%
This work was supported by JSPS KAKENHI Grants No. JP23K25790 and No. JP24K00546; 
by MEXT KAKENHI Grants-in-Aid 
for Transformative Research Areas A “Extreme Universe” (Grant No. JP21H05191); by the Hungarian NKFIH OTKA Grant No. K 142652; and in part by the National Science Foundation under Grant No. PHY-2309135 to the Kavli Institute for Theoretical Physics (KITP). Part of this work was also carried out at the Aspen Center for Physics, supported by a grant from the Simons Foundation (1161654, Troyer).
N.E. was supported by Forefront Physics and Mathematics Program to Drive Transformation (FoPM), a World-leading Innovative Graduate Study (WINGS) Program, the University of Tokyo and JSR Fellowship, the University of Tokyo. Y.A. was supported by JST PRESTO Grant No. JPMJPR2251. 
 
\end{acknowledgments}

\medskip

\appendix

\begin{figure*}[bt]
\includegraphics[width=.95\textwidth]{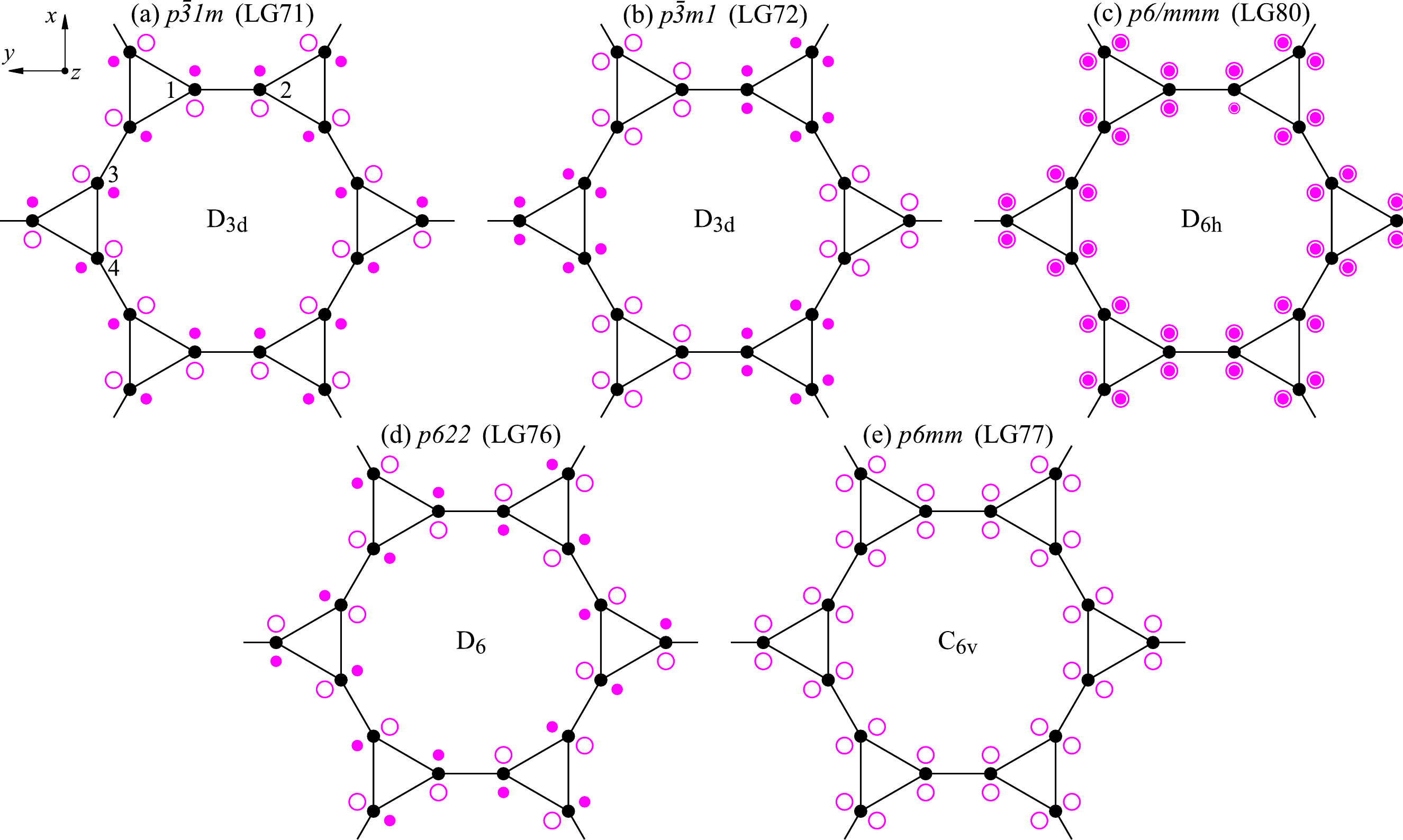}
\caption{\label{fig:star_layer_groups_hex}
The five layer groups of the star lattice, compatible with the wallpaper group $p6mm$ (WG17). 
In each panel, magenta open and solid circles mark the general point positions above ($z>0$) and below ($z<0$) the $xy$ plane, respectively.
The notation follows the International Tables for Crystallography \cite{ITCVolE}.
(a)–(c) show the achiral layer groups LG71, LG72, and LG80, each possessing inversion centers at bond centers and the centers of the hexagons, while (d) and (e) depict the chiral layer groups LG76 and LG77. 
LG80 (panel c) is the highest-symmetry layer group of the star lattice, appearing, for instance, in the space group $P6/mmm$ (No.~191). 
Its point group at the center of the hexagon is $D_{6h}$ in Schoenflies notation, with 24 symmetry elements. 
The other four layer groups—LG71, LG72, LG76, and LG77—are maximal subgroups of LG80. Their point groups ($D_{3d}$, $D_6$, and $C_{6v}$) each have 12 symmetry elements. Finally, (a) is also the layer group of the honeycomb lattice composed of edge-sharing octahedra, which realizes the Kitaev model.
}
\end{figure*}

\begin{figure*}[bt]
    \centering
\includegraphics[width=.95\textwidth]{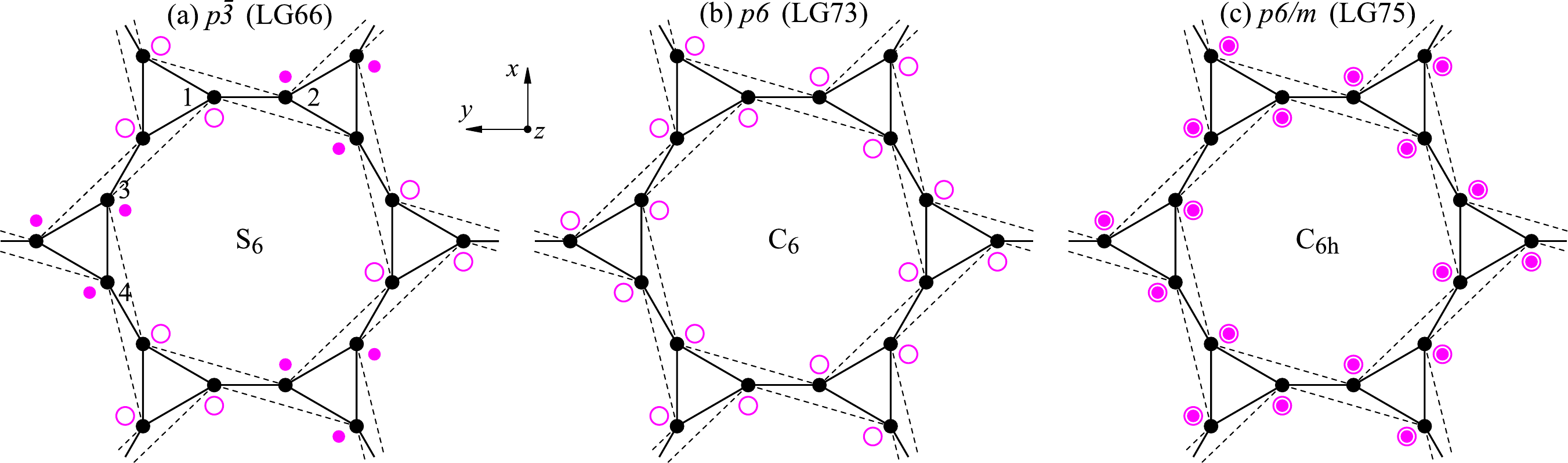}
\caption{\label{fig:mapleleaf_layer_groups_hex}
The three layer groups of the maple leaf lattice, compatible with the wallpaper group $p6$ (WG16). 
In each panel, magenta open and solid circles mark the general point positions above ($z>0$) and below ($z<0$) the $xy$ plane, respectively.
The notation follows the International Tables for Crystallography \cite{ITCVolE}.
(a) and (c) show the achiral layer groups LG66 and LG75, each possessing inversion centers at bond centers and the centers of the hexagons, while the layer group LG73 (panel b) is chiral. 
The highest symmetry layer group of the maple leaf lattice is LG75. Its point group at the center of the hexagon is $C_{6h}$ (Schoenflies notation) with 12 symmetry elements, and it appears in the space group $P6/mcc$ (No.~192). The remaining two, LG66 and LG73, are maximal subgroups of LG75, with point groups $S_{6}$ and  $C_{6}$, each having six symmetry elements.
The LG66 in panel (a) is the symmetry group of the layers normal to vector $(0~0~1)$ in the space group $R{\bar 3}$ (No.~148), for example.}
\end{figure*}

\section{Symmetry classification of the star and maple leaf lattices using layer groups}
\label{Appendix:Symmetry}

 In the mathematical sense, the two-dimensional star and maple leaf lattices are each associated with a particular wallpaper group. However, since these lattices form part of a three-dimensional crystal, a wallpaper group alone does not fully describe their symmetry properties. Instead, layer groups provide a more comprehensive classification, accounting for the three-dimensional environment of the two-dimensional lattices \cite{ITCVolE}.

 Layer groups provide complete information on the symmetries for the bonds in these lattices, enabling the identification of any anisotropic interactions between neighboring spins. 
 To determine the layer groups of the star lattice, we begin with the highest-symmetry layer group, $p6/mmm$ (LG80), which is straightforward to construct by extending the general point positions of the 
$p6mm$ wallpaper group symmetrically above and below the $xy$-plane. 
We obtain the lower-symmetry layer groups by systematically removing symmetry elements of the $D_{6h}$ point group.
In Fig.~\ref{fig:star_layer_groups_hex}, we illustrate all five layer groups that keep the wallpaper group of the star lattice (e.g., we discard the breathing star lattice):
$p\bar{3}1m$ (LG71), $p\bar{3}m1$ (LG72), $p622$ (LG76), $p6mm$ (LG77), and $p6/mmm$ (LG80), following the nomenclature and numbering from the International Tables for Crystallography and the Bilbao Crystallographic Server. 
The procedure replicates the surjective mapping from layer groups to wallpaper groups presented in Ref.~\onlinecite{SZE2025e01009}.

For the maple leaf lattice, we follow the same approach. Starting from its highest-symmetry layer group, $p6/m$ (LG75) of the maple leaf lattice, we obtain $p\bar 3$ (LG66) and $p6$ (LG73), as shown in Fig.~\ref{fig:mapleleaf_layer_groups_hex}. Let us note that the LG75 is a subgroup of the LG80, the layer group of the star lattice.
   
  The general positions in each layer group specify the point group symmetry of every bond, which in turn determines all possible anisotropic exchange interactions, including the antisymmetric Dzyaloshinskii–Moriya interaction and the symmetric $\Gamma$ terms.
   For any bond within a layer, the products $S^x_1 S^x_2$, $S^y_1 S^y_2$, and $S^z_1 S^z_2$ are invariants. They can thus appear in the Hamiltonian with distinct coefficients—reducing to the usual isotropic Heisenberg form when these coefficients are equal.
  In addition to these, Table~\ref{tab:syms12} shows how the different--symmetric and antisymmetric--combinations of spin operators of sites 1 and 2 transform under the different symmetry operations of a bond connecting two triangles in Figs.~\ref{fig:star_layer_groups_hex}(a) and \ref{fig:mapleleaf_layer_groups_hex}(a). 
  A similar analysis applies to the spins at sites 3 and 4 within a triangle, where Table \ref{tab:syms34} lists the additional invariants.

For example, for the star lattice at its highest-symmetry layer group $p6/mmm$ (LG80), the spin Hamiltonians for two representative bonds are:
\begin{subequations}
\begin{align}
  \mathcal{H}^{(1,2)}_{d} &= J^{xx}_{d} S_1^x S_2^x + J^{yy}_{d} S_1^y S_2^y + J^{zz}_{d} S_1^z S_2^z \,, 
  \\
  \mathcal{H}^{(3,4)}_{t} &= J^{xx}_{t} S_3^x S_4^x + J^{yy}_{t} S_3^y S_4^y + J^{zz}_{t} S_3^z S_4^z  \nonumber \\
      & \phantom{=} + D^z (S_3^x S_4^y - S_3^y S_4^x)   \,.
\end{align}
\end{subequations}
One gets the corresponding Hamiltonians for the other bonds in the lattice by applying the point group operations similarly to Eqs.~(\ref{Eq:Def_DM_vec}) and (\ref{Eq:Def_U_matrix}).

In the maple leaf lattice with highest-symmetry layer group $p6/m$ (LG75), the analogous Hamiltonians are:
\begin{subequations}
\begin{align}
  \mathcal{H}^{(1,2)}_{d} &= J^{xx}_{d} S_1^x S_2^x + J^{yy}_{d} S_1^y S_2^y + J^{zz}_{d} S_1^z S_2^z \nonumber \\
        & \phantom{=} + \Gamma_d (S_1^x S_2^y + S_1^y S_2^x)\,,\\
  \mathcal{H}^{(3,4)}_{t} &= J^{xx}_{t} S_3^x S_4^x + J^{yy}_{t} S_3^y S_4^y + J^{zz}_{t} S_3^z S_4^z \nonumber \\
      & \phantom{=} + \Gamma_{t} (S_3^x S_4^y + S_3^y S_4^x) + D^z (S_3^x S_4^y - S_3^y S_4^x) \,.  
\end{align}
\end{subequations}
Once again, the remaining bonds can be treated by symmetry operations, ensuring a complete description of the exchange interactions throughout the lattice. The lowering of the symmetry group of the bonds increases the number of anisotropic terms in the Hamiltonian.

\begin{table*}[bt]
\caption{Anisotropic exchanges on the bonds between the triangles connecting sites 1 and 2, as shown in Figs.~\ref{fig:star_layer_groups_hex}(a) and \ref{fig:mapleleaf_layer_groups_hex}(a). The first column lists the symmetry elements, followed by five columns describing how each operation transforms the sites and spin operators. The last six columns indicate by a $\checkmark$ whether a symmetric or antisymmetric combination is invariant under each symmetry operation. The final two rows specify the layer groups of the star and maple leaf lattices in which these invariants occur.}
\label{tab:syms12}
\begin{ruledtabular}
\begin{tabular}{ccccccccccccrrrrr}
$g$ & 1 & 2 & $S^x$ & $S^y$ & $S^z$ & $S^y_1 S^z_2 + S^z_1 S^y_2 $ & $S^x_1 S^z_2 + S^z_1 S^x_2 $  & $S^x_1 S^y_2 + S^y_1 S^x_2$ & $S^y_1 S^z_2 - S^z_1 S^y_2 $ & $S^z_1 S^x_2 - S^x_1 S^z_2 $  & $S^x_1 S^y_2 - S^y_1 S^x_2 $\\
\hline
Id.	& 1 & 2 & $S^x$ & $S^y$ & $S^z$ & $\checkmark$ & $\checkmark$ & $\checkmark$ & $\checkmark$ & $\checkmark$ & $\checkmark$\\
$C_2(x)$ & 2 & 1 & $S^x$ & $-S^y$ & $-S^z$ & $\checkmark$ & - & - & - & $\checkmark$ & $\checkmark$\\
$C_2(y)$ & 1 & 2 & $-S^x$ & $S^y$ & $-S^z$ & - & $\checkmark$ & - & - & $\checkmark$ & - \\
$C_2(z)$ & 2 & 1 & $-S^x$ & $-S^y$ & $S^z$ & - & - & $\checkmark$ & $\checkmark$ & $\checkmark$ & -\\
$\sigma_{yz}$ & 1 & 2 & $S^x$ & $-S^y$ & $-S^z$ & $\checkmark$ & - & - & $\checkmark$ & - & -\\
$\sigma_{xz}$ & 2 & 1 & $-S^x$ & $S^y$ & $-S^z$ & - & $\checkmark$ & - & $\checkmark$ & - & $\checkmark$\\
$\sigma_{xy}$ & 1 & 2 & $-S^x$ & $-S^y$ & $S^z$ & - & - & $\checkmark$ & - & - & $\checkmark$\\
I & 2 & 1 & $S^x$ & $S^y$ & $S^z$ & $\checkmark$ & $\checkmark$ & $\checkmark$ & - & - & - \\ 
\hline
star LG & & & & & & 72 & 71 & - & 77 & 76 & - \\ 
maple leaf LG & & & & & & 66 & 66 & 66,73,75 & 73 & 73 & - \\ 
\end{tabular}
\end{ruledtabular}
\end{table*}%

\begin{table*}[bt]
\caption{The same as Table~\ref{tab:syms12}, but for the triangular bond connecting sites 3 and 4, as shown in Figs.~\ref{fig:star_layer_groups_hex}(a) and \ref{fig:mapleleaf_layer_groups_hex}(a).}
\label{tab:syms34}
\begin{ruledtabular}
\begin{tabular}{ccccccccccccrrrrr}
$g$ & 3 & 4 & $S^x$ & $S^y$ & $S^z$ & $S^y_3 S^z_4 + S^z_3 S^y_4 $ & $S^x_3 S^z_4 + S^z_3 S^x_4 $  & $S^x_3 S^y_4 + S^y_3 S^x_4$ & $S^y_3 S^z_4 - S^z_3 S^y_4 $ & $S^z_3 S^x_4 - S^x_3 S^z_4 $  & $S^x_3 S^y_4 - S^y_3 S^x_4 $\\
\hline
Id.	& 3 & 4 & $S^x$ & $S^y$ & $S^z$ & $\checkmark$ & $\checkmark$ & $\checkmark$ & $\checkmark$ & $\checkmark$ & $\checkmark$\\
$C_2(y)$ & 4 & 3 & $-S^x$ & $S^y$ & $-S^z$ & - & $\checkmark$ & - & $\checkmark$ & - & $\checkmark$\\
$\sigma_{yz}$ & 4 & 3 & $S^x$ & $-S^y$ & $-S^z$ & $\checkmark$ & - & - & - & $\checkmark$ & $\checkmark$\\
$\sigma_{xy}$ & 3 & 4 & $-S^x$ & $-S^y$ & $S^z$ & - & - & $\checkmark$ & - & - & $\checkmark$\\
\hline
star LG & & & & & & 72,77 & 71,76 & - & 71,76 & 72,77  & 71,72,76,77,80 \\ 
maple leaf LG & & & & & & 66,73 & 66,73 & 66,73,75 & 66,73 & 66,73 & 66,73,75 \\ 
\end{tabular}
\end{ruledtabular}
\end{table*}%

\begin{widetext}

\section{Explicit expression of the triplon Hamiltonian (\ref{Eq:BdG_Form})} \label{Appendix:Explicit_form}
Here we give an explicit expression for the triplon Hamiltonian. The submatrices $\Xi(\bm{k})$ and $\Pi(\bm{k})$ defined in Eq. (\ref{Eq:BdG_Form}) are given by
\begin{align}
    \Xi (\bm{k}) &= I_{3\times 3} \otimes [J_d I_{3\times 3} + H^{\mathrm{Hei}}_{\mathrm{mag}} (\bm{k})] + \left(\frac{1}{2}\lambda_3 + \frac{\sqrt{3}}{2}\lambda_8 \right)\otimes [H^{\mathrm{DM}}_{\mathrm{mag}} (\bm{k}) - hI_{3\times 3}] \notag \\ &+ i[(\lambda_2 - \lambda_7) \otimes \mathrm{Re}(H^{\mathrm{in}}(\bm{k})) +  (\lambda_1 - \lambda_6) \otimes \mathrm{Im}(H^{\mathrm{in}}(\bm{k}))], \notag \\
    \Pi (\bm{k}) &= \left(\frac{1}{3} I_{3\times 3} - \frac{1}{2} \lambda_3 + \lambda_4 + \frac{\sqrt{3}}{6}\lambda_8 \right) \otimes H^{\mathrm{Hei}}_{\mathrm{mag}}(\bm{k}) - i\lambda_5 \otimes H^{\mathrm{DM}}_{\mathrm{mag}}(\bm{k})\notag \\ &+ i[(\lambda_2 - \lambda_7) \otimes \mathrm{Re}(H^{\mathrm{in}}(\bm{k})) + i (\lambda_2 + \lambda_7) \otimes \mathrm{Im}(H^{\mathrm{in}}(\bm{k}))], \label{Eq:HBdG_Explicit}
\end{align}
where $H^{\mathrm{Hei}}_{\mathrm{mag}}(\bm{k})$ and $H^{\mathrm{DM}}_{\mathrm{mag}}(\bm{k})$ correspond to $H_{\mathrm{mag}}(\bm{k})$ defined in Eq. (\ref{Eq:single_Kagome}) without the DM interaction and without the Heisenberg interaction, respectively. Here, we use the Gell-Mann matrices $\lambda_{i}$ $(i = 1,\cdots, 8)$, which are defined by
\begin{align}
    \lambda_1 &= \begin{pmatrix}
        0 & 1 & 0 \\
        1 & 0 & 0 \\
        0 & 0 & 0 
    \end{pmatrix}, \quad
    \lambda_2 = \begin{pmatrix}
        0 & -i & 0 \\
        i & 0 & 0 \\
        0 & 0 & 0 
    \end{pmatrix}, \quad
    \lambda_3 = \begin{pmatrix}
        1 & 0 & 0 \\
        0 & -1 & 0 \\
        0 & 0 & 0 
    \end{pmatrix}, \quad
    \lambda_4 = \begin{pmatrix}
        0 & 0 & 1 \\
        0 & 0 & 0 \\
        1 & 0 & 0 
    \end{pmatrix}, \notag \\
    \lambda_5 &= \begin{pmatrix}
        0 & 0 & i \\
        0 & 0 & 0 \\
        -i & 0 & 0 
    \end{pmatrix}, \quad
    \lambda_6 = \begin{pmatrix}
        0 & 0 & 0 \\
        0 & 0 & 1 \\
        0 & 1 & 0 
    \end{pmatrix}, \quad
    \lambda_7 = \begin{pmatrix}
        0 & 0 & 0 \\
        0 & 0 & -i \\
        0 & i & 0 
    \end{pmatrix}, \quad 
    \lambda_8 = \frac{1}{\sqrt{3}} \begin{pmatrix}
        1 & 0 & 0 \\
        0 & 1 & 0 \\
        0 & 0 & -2 
    \end{pmatrix}. \label{Eq:Gell_mann}
\end{align}
The matrix $H^{\mathrm{in}}(\bm{k})$ represents the in-plane component of the DM interaction, which is expressed by

\begin{equation}\label{Eq:H_in_explicit}
    H^{\mathrm{in}}(\bm{k})=
\frac{D_{\mathrm{in}}}{2\sqrt2} e^{-i\phi_{\mathrm{in}}}
\begin{pmatrix}
0 &
-e^{i\frac{\pi}{2}}\cos k_{1} &
-e^{i\frac{\pi}{6}}\cos k_{3}\\
e^{i\frac{\pi}{2}}\cos k_{1} &
0 &
-e^{-i\frac{\pi}{6}}\cos k_{2}\\
e^{i\frac{\pi}{6}}\cos k_{3} &
e^{-i\frac{\pi}{6}}\cos k_{2} &
0
\end{pmatrix},
\end{equation}
where $D_{\mathrm{in}} = \sqrt{D^2_x + D^2_y}$ and $\phi_{\mathrm{in}} = \arctan(D_y/D_x)$.

\section{Validity of the triplon-wave analysis}\label{Appendix:ED_vs_Triplon}

\begin{figure}[H]
    \centering
    \includegraphics[width=0.8\linewidth]{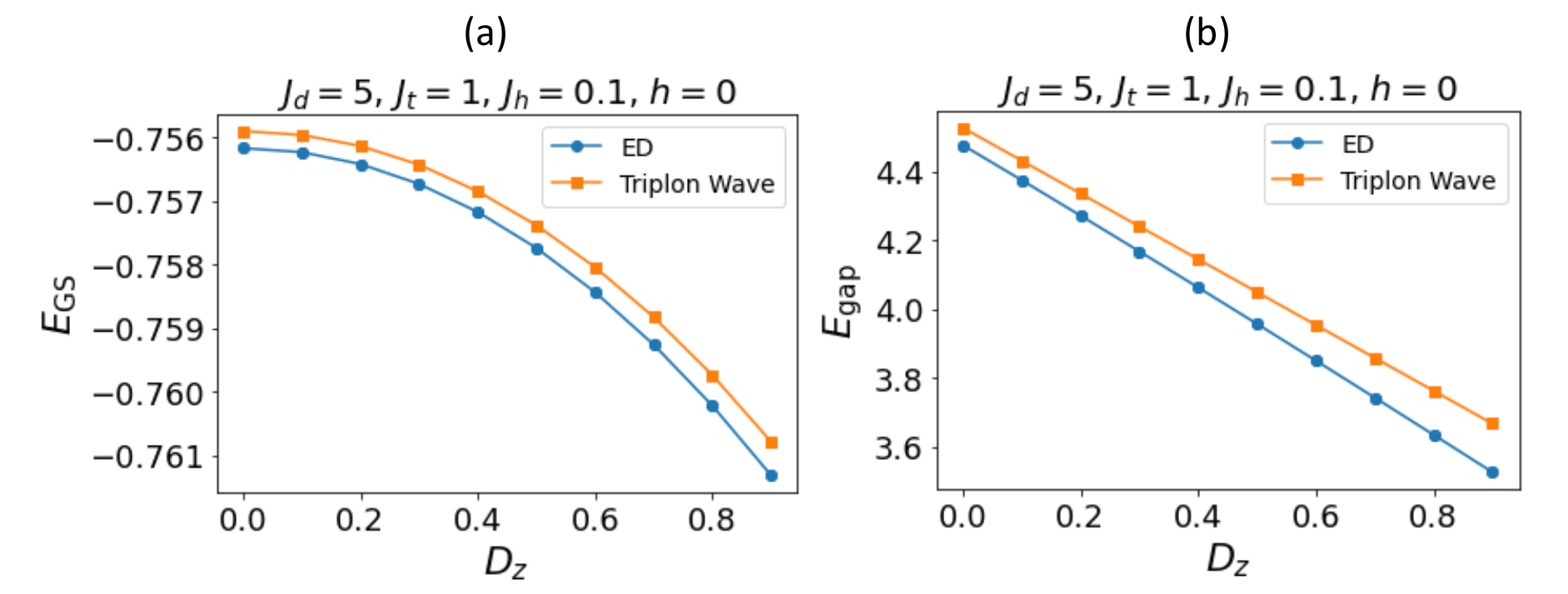}
    \caption{(a) The numerical results of the ground-state energy 
    per dimer 
    as a function of $D_z$ 
    obtained from exact diagonalization (ED) and from Eq. (\ref{Eq:Correction_GS}). (b) The spin gap 
    as a function of $D_z$ 
    obtained from ED and the analytical expression (\ref{Eq:Spin_gap}). The parameters used in the calculations are $J_d = 5.0$, $J_t = 1.0$, $J_h = 0.1$, and $h = 0$.}
    \label{fig:ED_VS_Triplon}
\end{figure}

\begin{figure}[H]
    \centering
    \includegraphics[width=0.8\linewidth]{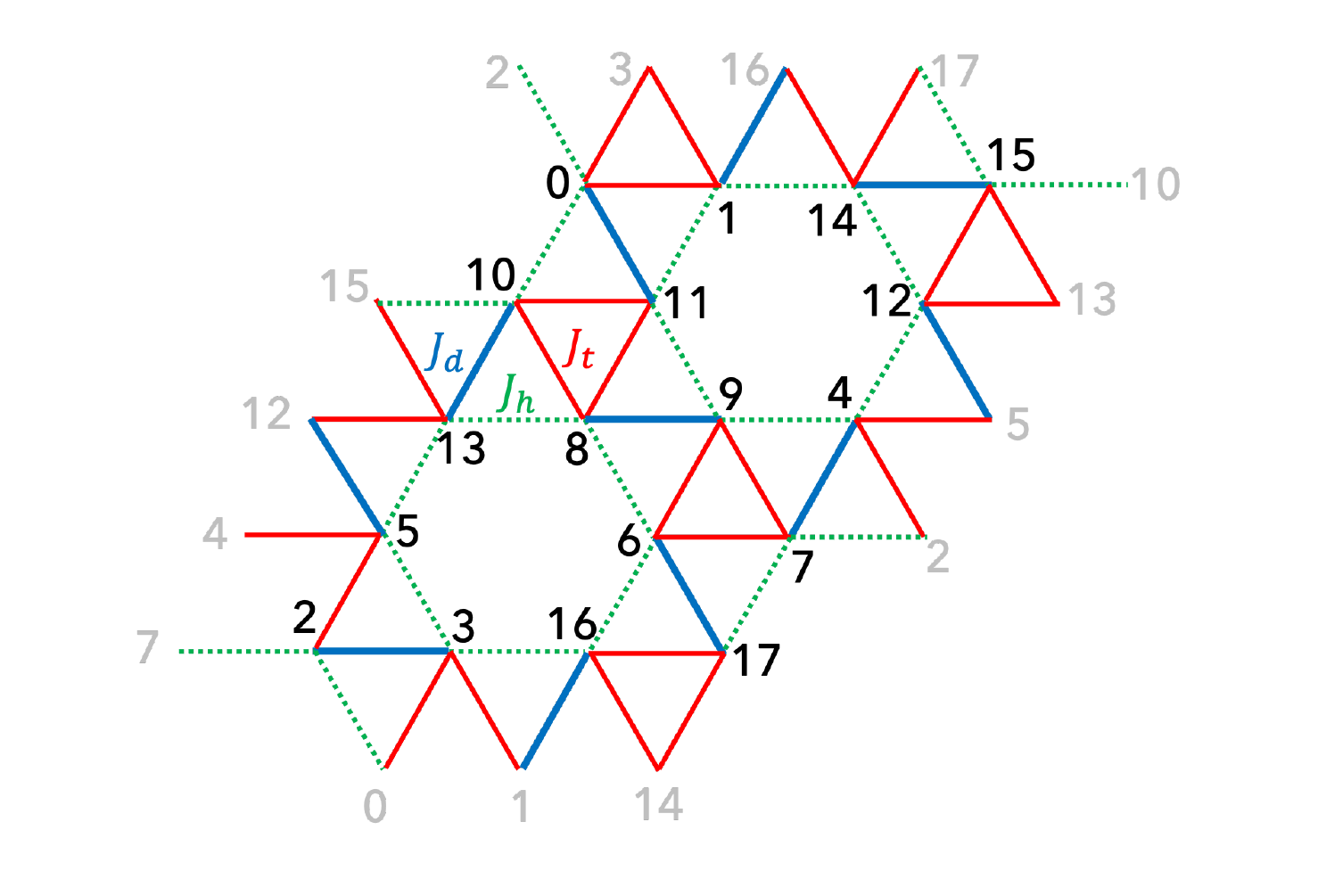}
    \caption{The 18-site (with three unit cells) cluster with periodic boundary conditions used in the exact diagonalization calculation. 
    }
    \label{fig:Maple_cluster}
\end{figure}

In order to verify the validity of the triplon-wave analysis presented in 
the main text, we perform exact diagonalization of a cluster with $18$ sites (Fig. \ref{fig:Maple_cluster}) and compare the resulting ground-state energy and spin gap with those obtained from the triplon-wave analysis. 
The results are shown in Fig. \ref{fig:ED_VS_Triplon}. 

Here, we do not consider the in-plane component $D_x$ and $D_y$ of the DM vector, for simplicity. The ground-state energy per 
dimer and the spin gap at $\Gamma$ obtained from the triplon-wave analysis are given by \cite{neumann2020Orbital}
\begin{equation}\label{Eq:Correction_GS}
    E_{\mathrm{GS}} = -\frac{3}{4} J_d -\frac{1}{4N_{\mathrm{Dimer}}}\sum_{\bm{k}} \mathrm{Tr} H^{\mathrm{tot}}_{\mathrm{BdG}}(\bm{k}) + \frac{1}{2N_{\mathrm{Dimer}}}\sum_{n=1}^{9} \sum_{\bm{k}} E^{\mathrm{tot}}_{n,+} (\bm{k}),
\end{equation}
and Eq. (\ref{Eq:Spin_gap}), respectively. Here, $N_{\mathrm{Dimer}}$ denotes the number of dimers, which is chosen 
large enough to ensure the convergence of the results.

Clearly, 
Fig. \ref{fig:ED_VS_Triplon} 
shows that the triplon-wave analysis 
captures the ground-state energy and spin gap even 
for moderate values of $D_z \sim J_t - J_h$, 
demonstrating its ability to describe the low-energy physics. It should be noted that with the parameters used in our numerical analysis, i.e., $J_h / J_t = 10$, the ground state is not an exact dimer singlet even without $D_z$; 
the exact dimer ground state is realized only when $J_t = J_h$ and $D_z = 0$. Nevertheless, Fig. \ref{fig:ED_VS_Triplon} suggests that the ground state is still well 
approximated by the dimer singlet state even when $J_t \ne J_h$ and in the presence of a large $D_z$.

\section{Derivation of Eqs. (\ref{Eq:Energy_relation}) and (\ref{Eq:Berry_relation})}\label{Appendix:Eigen_problem}
Here we derive Eqs. (\ref{Eq:Energy_relation}) and (\ref{Eq:Berry_relation}) by solving the following eigenvalue problem for the BdG Hamiltonian (\ref{Eq:Kagome_decouple}):
\begin{equation}\label{Eq:Eigenvalue_problem_BdG_pm}
    \Sigma_z H_{\mathrm{BdG}} (\bm{k}) \bm{\Psi}_{n,s,\sigma} (\bm{k}) = E_{n,s,\sigma}(\bm{k}) \bm{\Psi}_{n,s,\sigma}(\bm{k}).
\end{equation}
To this end, we make the following ansatz for the eigenvectors of $\Sigma_z H_{\mathrm{BdG}} (\bm{k})$:
\begin{eqnarray}
{\bm \Psi}^{(1)}_n ({\bm k}) = \begin{pmatrix}
        \cosh (\vartheta_n (\bm{k}))\bm{\psi}_n (\bm{k})\\
        0\\
        0\\
        \sinh(\vartheta_n (\bm{k}))\bm{\psi}_n (\bm{k})
    \end{pmatrix}, \quad 
{\bm \Psi}^{(2)}_n ({\bm k}) = \begin{pmatrix}
        0\\
        \cosh(\vartheta_n (-\bm{k}))\bm{\psi}^{\ast}_n (-\bm{k})\\
        \sinh(\vartheta_n (-\bm{k}))\bm{\psi}^{\ast}_n (-\bm{k})\\
        0
    \end{pmatrix},
\end{eqnarray}
where $\bm{\psi}_n (\bm{k})$ is the eigenvector of $H_{\rm mag} (\bm{k})$ [Eq. (\ref{Eq:single_Kagome})] with eigenvalue $\lambda_n (\bm{k})$. A straightforward calculation shows that for ${\bm \Psi}^{(1)}_n ({\bm k})$ to be an eigenvector of $\Sigma_z H_{\mathrm{BdG}} (\bm{k})$, $\tanh (\vartheta_n (\bm{k}))$ must be a solution of the quadratic equation in $x$
\begin{eqnarray}
    \lambda_n (\bm{k}) \,x^2 + 2 (J_d + \lambda_n (\bm{k})) \,x + \lambda_n (\bm{k}) = 0.
\end{eqnarray}
The two solutions are given by
\begin{eqnarray}
    x_{\pm} = \frac{-(J_d + \lambda_n (\bm{k})) \pm \sqrt{(J_d + \lambda_n (\bm{k}))^2 - (\lambda_n (\bm{k}))^2}}{\lambda_n (\bm{k})}.
    \label{eq:quadratic_eq}
\end{eqnarray}
We denote by $\bm{\Psi}_{n,+,+} (\bm{k})$ [resp. $\bm{\Psi}_{n,-,-} (\bm{k})$] the eigenvector ${\bm \Psi}^{(1)}_n ({\bm k})$ corresponding to the solution $x=x_+$ [resp. $x_-$]. Similarly, one finds that ${\bm \Psi}^{(2)}_n ({\bm k})$ is an eigenvector of $\Sigma_z H_{\mathrm{BdG}} (\bm{k})$ when $\tanh (\vartheta_n (-\bm{k}))$ is a solution of
\begin{eqnarray}
    \lambda_n (-{\bm k})\, x^2 + 2 (J_x + \lambda_n (-{\bm k})) \, x + \lambda_n (-{\bm k}) = 0.
\end{eqnarray}
Comparing with Eq. \eqref{eq:quadratic_eq}, we see that $x=\tanh \theta_n (-{\bm k})$ and $\coth \theta_n (-{\bm k})$ solve the above equation. We denote by $\bm{\Psi}_{n,-,+} (\bm{k})$ and $\bm{\Psi}_{n,+,-} (\bm{k})$ the eigenvectors ${\bm \Psi}^{(2)}_n ({\bm k})$ corresponding to these solutions.

To summarize, the eigenvalues of $\Sigma_z H_{\mathrm{BdG}} (\bm{k})$ and the corresponding eigenvectors are given as follows:
\begin{align}\label{Eq:Eigenvalue_+1}
    E_{n,+,\sigma} (\bm{k}) &= \sigma (\sqrt{(J_d + \lambda_n (\sigma\bm{k}))^2 - (\lambda_n (\sigma\bm{k}))^2} -h), \\
    E_{n,-,\sigma} (\bm{k}) &= \sigma (\sqrt{(J_d + \lambda_n (-\sigma\bm{k}))^2 - (\lambda_n (-\sigma\bm{k}))^2} +h), \label{Eq:Eigenvalue_-1}\\
    \bm{\Psi}_{n,+,+} (\bm{k}) &= \begin{pmatrix}
        \cosh(\theta_n (\bm{k}))\bm{\psi}_n (\bm{k})\\
        0\\
        0\\
        \sinh(\theta_n (\bm{k}))\bm{\psi}_n (\bm{k})
    \end{pmatrix}, \label{Eq:Eigenvector_+1_+}\\
    \bm{\Psi}_{n,-,+} (\bm{k}) &= \begin{pmatrix}
        0\\
        \cosh(\theta_n (-\bm{k}))\bm{\psi}^{\ast}_n (-\bm{k})\\
        \sinh(\theta_n (-\bm{k}))\bm{\psi}^{\ast}_n (-\bm{k})\\
        0
    \end{pmatrix}, \label{Eq:Eigenvector_-1_+}\\
    \bm{\Psi}_{n,+,-} (\bm{k}) &= \begin{pmatrix}
        0\\
        \sinh(\theta_n (-\bm{k}))\bm{\psi}^{\ast}_n (-\bm{k})\\
        \cosh(\theta_n (-\bm{k}))\bm{\psi}^{\ast}_n (-\bm{k})\\
        0
    \end{pmatrix}, \label{Eq:Eigenvector_+1_-}\\
    \bm{\Psi}_{n,-,-} (\bm{k}) &= \begin{pmatrix}
        \sinh(\theta_n (\bm{k}))\bm{\psi}_n (\bm{k})\\
        0\\
        0\\
        \cosh(\theta_n (\bm{k}))\bm{\psi}_n (\bm{k})
    \end{pmatrix}. \label{Eq:Eigenvector_-1_-}
\end{align}
Here, $\theta_n (\bm{k})$ is defined by
\begin{equation}\label{Eq:theta_eigen_decouple}
    \tanh(\theta_n (\bm{k})) = \frac{-(J_d + \lambda_n (\bm{k}))+\sqrt{(J_d + \lambda_n (\bm{k}))^2 - (\lambda_n (\bm{k}))^2}}{\lambda_n (\bm{k})}.
\end{equation}
We now compute the Berry connection and curvature 
associated with the eigenvectors $\bm{\Psi}_{n,s,\sigma}(\bm{k})$. They are defined by 
\begin{align}\label{Eq:Def_Berry_connection}
    \bm{A}_{n,s,\sigma} &= i\sigma\ev{\ev{\bm{\Psi}_{n,s,\sigma}(\bm{k}),\nabla_{\bm{k}}\bm{\Psi}_{n,s,\sigma}(\bm{k})}} \\
    \Omega_{n,s,\sigma} &= (\nabla_{\bm{k}}\times \bm{A}_{n,s,\sigma}(\bm{k}))_{z}, \label{Eq:Def_Berry_curvature}
\end{align}
where the inner product is defined by $\ev{\ev{\bm{\Phi},\bm{\Psi}}} = \bm{\Phi}^{\dag}\Sigma_z \bm{\Psi}$. By substituting Eqs. (\ref{Eq:Eigenvector_+1_+})-(\ref{Eq:Eigenvector_-1_-}) into Eq. (\ref{Eq:Def_Berry_connection}), we obtain the following relations \cite{kondo2019$mathbbZ_2$}:

\begin{align}
    &\bm{A}_{n,+,+}(\bm{k}) = \bm{A}_{n,-,-}(\bm{k}) \notag \\
    &=i\ev{\cosh(\theta_n (\bm{k}))\bm{\psi}_n (\bm{k}),\nabla_{\bm{k}}\cosh(\theta_n (\bm{k}))\bm{\psi}_n (\bm{k})}-i\ev{\sinh(\theta_n (\bm{k}))\bm{\psi}_n (\bm{k}),\nabla_{\bm{k}}\sinh(\theta_n (\bm{k}))\bm{\psi}_n (\bm{k})} \notag \\
    &= i\ev{\bm{\psi}_n (\bm{k}), \nabla_{\bm{k}} \bm{\psi}_n (\bm{k})} \notag \\ 
    &= \bm{A}^{\mathrm{mag}}_{n} (\bm{k}), \label{Eq:Berry_connection_relation_+}\\
    &\bm{A}_{n,-,+}(\bm{k}) = \bm{A}_{n,+,-}(\bm{k}) \notag \\
    &=i\ev{\cosh(\theta_n (-\bm{k}))\bm{\psi}_n (-\bm{k}),\nabla_{\bm{k}}\cosh(\theta_n (-\bm{k}))\bm{\psi}_n (-\bm{k})}^{\ast}-i\ev{\sinh(\theta_n (-\bm{k}))\bm{\psi}_n (-\bm{k}),\nabla_{\bm{k}}\sinh(\theta_n (-\bm{k}))\bm{\psi}_n (-\bm{k})}^{\ast} \notag \\
    &= i\ev{\bm{\psi}_n (-\bm{k}), \nabla_{\bm{k}} \bm{\psi}_n (-\bm{k})}^{\ast} \notag \\
    &=i\ev{\bm{\psi}_n (-\bm{k}), \nabla_{-\bm{k}} \bm{\psi}_n (-\bm{k})} \notag \\
    &= \bm{A}^{\mathrm{mag}}_{n} (-\bm{k}), \label{Eq:Berry_connection_relation_-}
\end{align}
where $\ev{\bm{\phi},\bm{\psi}} = \bm{\phi}^{\dag} \bm{\psi}$ is the standard inner product, and $\bm{A}^{\mathrm{mag}}_{n} (\bm{k}) = i\ev{\bm{\psi}_n (\bm{k}), \nabla_{\bm{k}} \bm{\psi}_n (\bm{k})}$ is the Berry connection of $H_{\mathrm{mag}} (\bm{k})$ (\ref{Eq:single_Kagome}). Using the Berry curvature of $H_{\mathrm{mag}} (\bm{k})$ defined by $\Omega^{\mathrm{mag}}(\bm{k}) = (\nabla_{\bm{k}} \times \bm{A}^{\mathrm{mag}}_n (\bm{k}))_z$, we can derive Eq. (\ref{Eq:Berry_relation}) from Eqs. (\ref{Eq:Def_Berry_curvature})-(\ref{Eq:Berry_connection_relation_-}).


\section{Detailed calculation of $\mathbb{Z}_2$ invariant (\ref{Eq:Def_Z2})}\label{Appendix:Z_2}
\begin{figure}[H]
    \centering
    \includegraphics[width=0.45\linewidth]{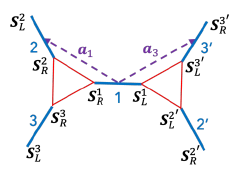}
    \caption{The schematic figure of dimers considered here. Thick blue bonds represent dimers, denoted by $1$, $2$, $2^{\prime}$, $3$, and $3^{\prime}$. Each dimer $l$ ($l = 1, 2, 2^{\prime}, 3, 3^{\prime}$) consists of two spins $S^{l}_{L}$ and $S^{l}_{R}$. The Dimers $l$ and $l^{\prime}$ ($l = 2, 3$) belong to the same sublattice. The parity operator $P$ acts on each spin as $P : S^{1}_{L/R}\rightarrow S^{1}_{R/L}$, $P : S^{2}_{L/R}\rightarrow S^{2^{\prime}}_{R/L}$, and $P : S^{3}_{L/R} \rightarrow S^{3^{\prime}}_{R/L}$, resulting in Eq. (\ref{Eq:P_rep_k}).}
    \label{fig:Def_Parity_operator}
\end{figure}

Here we calculate the $\mathbb{Z}_2$ invariant defined in Eq. (\ref{Eq:Def_Z2}). To compute the parity eigenvalues at each TRIM $\xi_n (\Gamma_i)$, we first define the parity operator $P$ as the inversion through the center of the dimer 1 \cite{thomasen2021Fragility} [see Fig. \ref{fig:Def_Parity_operator}]. 
The operator $P$ swaps the two spins $S^{1}_L$ and $S^{1}_R$ on dimer 1, while shifting dimers 2 and 3 by $-2\bm{a}_1 = -(1,\sqrt{3})$ and $2\bm{a}_3 = (1,-\sqrt{3})$, respectively.
Since the exchange of the two spins constituting a dimer does not affect the triplets, which are even with respect to the inversion through the center of the dimer, the momentum space representation of $P$ 
reads
\begin{equation}\label{Eq:P_rep_k}
    P_{\bm{k}} = I_{2\times 2}\otimes I_{2\times 2} \otimes p_{\bm{k}},
\end{equation}
where $p_{\bm{k}} = \mathrm{diag} (1, e^{-i 2k_1}, e^{i 2k_3})$. From Eqs. (\ref{Eq:Eigenvector_+1_+})-(\ref{Eq:Eigenvector_-1_-}) and (\ref{Eq:P_rep_k}), we find that $\xi_n (\Gamma_i)$ can be determined by the parity eigenvalues of $p_{\bm{k}}$. For this reason, we calculate the 
explicit expressions of $\bm{\psi}_n (\bm{k})$ at $M_1$, $M_2$, and $M_3$ as
\begin{align}
    M_1 : \begin{cases}
         \lambda_1 = -\varepsilon, \quad \bm{\psi}_1 = \left(-\frac{\zeta}{2}, \frac{1}{2}, 0\right)\\
        \lambda_2 = 0, \quad \bm{\psi}_2 = (0,0,1) \\
        \lambda_3 = \varepsilon, \quad \bm{\psi}_3 = \left(\frac{\zeta}{2}, \frac{1}{2}, 0\right),
        \end{cases} \label{Eq:Eigen_M1}
\end{align}

\begin{align}
    M_2 : \begin{cases}
        \lambda_1 = -\varepsilon, \quad \bm{\psi}_1 = \left(-\frac{\zeta^{\ast}}{2}, 0, \frac{1}{2} \right)\\
        \lambda_2 = 0, \quad \bm{\psi}_2 = (0,1,0) \\
        \lambda_3 = \varepsilon, \quad \bm{\psi}_3 = \left(\frac{\zeta^{\ast}}{2}, 0, \frac{1}{2}\right),
        \end{cases} \label{Eq:Eigen_M2}
\end{align}


\begin{align}
    M_3 : \begin{cases}
        \lambda_1 = -\varepsilon, \quad \bm{\psi}_3 = \left(0, -\frac{\zeta}{2}, \frac{1}{2}\right) \\
        \lambda_2 = 0, \quad \bm{\psi}_2 = (1,0,0) \\
        \lambda_3 = \varepsilon, \quad \bm{\psi}_3 = \left(0, \frac{\zeta}{2}, \frac{1}{2}\right),
        \end{cases} \label{Eq:Eigen_M3}
\end{align}
where $\varepsilon$ and $\zeta$ are given by
\begin{equation}\label{Eq:explicit_epsilon}
   \varepsilon = \frac{\sqrt{(J_t - J_h)^2 + D^{2}_{z}}}{2}, 
\end{equation}
and
\begin{equation}\label{Eq:explicit_zeta}
   \zeta = \frac{(J_t - J_h)+iD_z}{\sqrt{(J_t - J_h)^2 + D^{2}_{z}}},
\end{equation}
respectively. Here we note that $p_{\bm{k}}$ at $\Gamma$ is just an identity matrix, and thus the parity eigenvalue at $\Gamma$ for all Kramers pairs is $1$. Using the above expressions (\ref{Eq:Eigen_M1})-(\ref{Eq:explicit_zeta}), we obtain the parity eigenvalues $\xi_n (\Gamma_i)$ and $\nu_n$ in Eq. (\ref{Eq:Def_Z2}), as shown in Table \ref{Table:Parity_Z2}.
\begin{table}[t]
    \centering
    \begin{tabular}{ccccccc}\hline\hline
    Kramers pair & $n=1$ & $n=2$ & $n=3$ \\
    \hline
    $\xi_n (\Gamma)$ & $1$ & $1$ & $1$ \\ 
    $\xi_n (M_1)$ & $1$ & $-1$ & $1$ \\
    $\xi_n (M_2)$ & $1$ & $-1$ & $1$ \\
    $\xi_n (M_3)$ & $-1$ & $1$ & $-1$ \\
    \hline
    $\nu_n$ & 1 & 0 &1 \\ \hline\hline
    \end{tabular}
    \caption{The parity eigenvalues of each Kramers pair at TRIM $\xi_n (\Gamma_i)$ ($n = 1, 2, 3$) and the resulting $\mathbb{Z}_2$ invariant $\nu_n$.}
    \label{Table:Parity_Z2}
\end{table}

\section{Derivation of the spin gap (\ref{Eq:Spin_gap}) and the analytical expressions of the conductivities (\ref{Eq:Analytical_Spin_Nernst}) and (\ref{Eq:Analytical_Thermal_Hall})}\label{Appendix:Analytical_expression}
Here, we derive the analytical expressions for the spin Nernst (\ref{Eq:Analytical_Spin_Nernst}) and thermal Hall conductivities (\ref{Eq:Analytical_Thermal_Hall}) by using the fact that the Berry curvature is 
concentrated in the vicinity of the gap-closing points [see Fig. \ref{fig:Band_h}]. Consequently, from Eqs. (\ref{Eq:Eigenvalue_+1}) and (\ref{Eq:Eigenvector_+1_+}), the problem reduces to 
evaluating $\Omega^{\mathrm{mag}}_n (\bm{k})$ and $\lambda_n (\bm{k})$ around these points.

First, we construct an effective description of the lowest two bands near $\Gamma$. Specifically, we follow the approach outlined in Ref. \cite{thomasen2021Fragility}, i.e., reducing $H_{\mathrm{mag}}(\bm{k})$ (\ref{Eq:single_Kagome}) around $\Gamma$ to the following $2\times 2$ matrix involving only the two bands $\lambda_1 (\bm{k})$ and $\lambda_2 (\bm{k})$: 
\begin{equation}\label{Eq:Effective_Ham_Gamma}
    M^{\mathrm{eff}}_{\Gamma}(\bm{k}) = d_0 (\bm{k}) I_{2\times 2} + \bm{d} (\bm{k}) \cdot \bm{\sigma},
\end{equation}
where $\bm{\sigma} = (\sigma_x, \sigma_y, \sigma_z)$. To derive the above effective matrix (\ref{Eq:Effective_Ham_Gamma}), we introduce the following unitary matrix $U_{\Gamma}$ that diagonalizes $H_{\mathrm{mag}} (\bm{k})$ at $\Gamma$ without $D_z$:
\begin{equation}\label{Eq:Transformation_matrix_Gamma}
    U_{\Gamma} = \begin{pmatrix}
        \frac{1}{\sqrt{3}}&-\frac{1}{\sqrt{6}}&-\frac{1}{\sqrt{2}}\\
        \frac{1}{\sqrt{3}}&-\frac{1}{\sqrt{6}}&\frac{1}{\sqrt{2}}\\
        \frac{1}{\sqrt{3}}&\frac{2}{\sqrt{6}}&0 
    \end{pmatrix}.
\end{equation}
Using the above matrix ($\ref{Eq:Transformation_matrix_Gamma}$), we transform $H_{\mathrm{mag}} (\bm{k})$ and focus only on the subspace formed by the two modes corresponding to $\lambda_1 (\bm{k})$ and $\lambda_2 (\bm{k})$ near $\Gamma$. Let $\hat{P}_{12}$ be the projection operator to the subspace spanned by two modes $\bm{\psi}_1(\bm{k})$ and $\bm{\psi}_2(\bm{k})$ at $\Gamma$. By keeping only the leading term with respect to $k_x$ and $k_y$, 
the effective matrix (\ref{Eq:Effective_Ham_Gamma}) is found to be
\begin{align}
    M^{\mathrm{eff}}_{\Gamma}(\bm{k}) = & \hat{P}^{\dag}_{12}U^{\dag}_{\Gamma}\cdot H_{\mathrm{mag}} (\bm{k})\cdot U_{\Gamma}\hat{P}_{12} \notag \\
    \simeq & -\frac{J_t - J_h}{128}[64-16(k_{x}^2 + k_{y}^2)]\cdot I_{2\times 2} -\frac{J_t - J_h}{128}[16k_x k_y + 8\sqrt{3}(k_{x}^2 - k_{y}^2)]\cdot \sigma_x \notag \\
    & + \frac{\sqrt{3}}{2}D_z \cdot \sigma_y -\frac{J_t - J_h}{128} [16\sqrt{3}k_x k_y - 8(k_{x}^2 -k_{y}^2)]\cdot \sigma_z, 
\end{align}
which yields
\begin{align}
d_0 (\bm{k}) &=-\frac{J_t - J_h}{128}[64-16(k_{x}^2 + k_{y}^2)], \quad 
d_1 (\bm{k})  =-\frac{J_t - J_h}{128}[16k_x k_y + 8\sqrt{3}(k_{x}^2 - k_{y}^2)], \\
d_2 (\bm{k}) &= \frac{\sqrt{3}}{2}D_z, \quad\quad\quad\quad\quad\quad\quad\quad\quad\quad
d_3 (\bm{k})  =-\frac{J_t - J_h}{128} [16\sqrt{3}k_x k_y - 8(k_{x}^2 -k_{y}^2)].
\end{align}
The eigenvalues and the Berry curvature of the two-level system represented by $M^{\mathrm{eff}}_{\Gamma}(\bm{k})$ are then given by 
\begin{align}\notag
    \lambda_1 (\bm{k}) = d_0 (\bm{k}) - \abs{\bm{d}(\bm{k})} 
    \simeq \frac{J_h - J_t}{2} - \frac{(J_h - J_t)k^2 + \sqrt{48D_{z}^2 + (J_h - J_t)^2 k^4}}{8},\\
    \lambda_2 (\bm{k}) = d_0 (\bm{k}) + \abs{\bm{d}(\bm{k})} 
    \simeq \frac{J_h - J_t}{2} - \frac{(J_h - J_t)k^2 - \sqrt{48D_{z}^2 + (J_h - J_t)^2 k^4}}{8}, \label{Eq:lambda_Gamma_analytical}
\end{align}
and
\begin{align}\notag
    \Omega^{\mathrm{mag}}_1 (k_x, k_y) &= \frac{1}{2} \frac{\bm{\mathrm{d}}(\bm{k}) \cdot (\frac{\partial \bm{\mathrm{d}}(\bm{k})}{\partial k_x} \times \frac{\partial \bm{\mathrm{d}}(\bm{k})}{\partial k_y})}{{(\bm{\mathrm{d}}(\bm{k})}\cdot \bm{\mathrm{d}}(\bm{k}))^{3/2}} = \frac{-8\sqrt{3}D_z (J_t - J_h)^2 (k_{x}^{2}+k_{y}^2)}{(48D_{z}^2 + (J_t - J_h)^2 (k_{x}^2 + k_{y}^2)^2)^{3/2}}, \\
    \Omega^{\mathrm{mag}}_2 (k_x, k_y) &= 
    -\Omega^{\mathrm{mag}}_1 (k_x, k_y),
    \label{Eq:Berry_curvature_mag_Gamma_analytical}
\end{align}
respectively. By substituting Eqs. (\ref{Eq:lambda_Gamma_analytical}) and (\ref{Eq:Berry_curvature_mag_Gamma_analytical}) into Eqs. (\ref{Eq:Energy_relation}) and (\ref{Eq:Berry_relation}), we obtain 
\begin{align}\notag
    E_{1,\pm,+}(\bm{k}) &= \sqrt{(J_d + \lambda_1 (\bm{k}))^2 - (\lambda_1 (\bm{k}))^2} \mp h \notag \\
    &\simeq \sqrt{J_d (J_d - J_t + J_h)} \mp h - \frac{J_d}{8\sqrt{J_d(J_d - J_t + J_h)}}[(J_h - J_t)k^2 + \sqrt{48D_{z}^2 + (J_t - J_h)^2 k^4}], \notag \\
    E_{2,\pm,+}(\bm{k}) &= \sqrt{(J_d + \lambda_2 (\bm{k}))^2 - (\lambda_2 (\bm{k}))^2} \mp h \notag \\
    &\simeq \sqrt{J_d (J_d - J_t + J_h)} \mp h - \frac{J_d}{8\sqrt{J_d(J_d - J_t + J_h)}}[(J_h - J_t)k^2 - \sqrt{48D_{z}^2 + (J_t - J_h)^2 k^4}], \label{Eq:Energy_Gamma_analytical}
\end{align}
and
\begin{align}\notag
    \Omega_{1,\pm,+} (k_x, k_y) &= \pm \Omega^{\mathrm{mag}}_1 (k_x, k_y),\\
    \Omega_{2,\pm,+} (k_x, k_y) &= \pm \Omega^{\mathrm{mag}}_2(k_x, k_y), \label{Eq:Berry_curvature_Gamma_analytical}
\end{align}
which are valid in the vicinity of ${\bm k}=\Gamma$. Here we note that $E_{1,+,+}(\bm{k})$ at $\Gamma$ gives the spin gap (\ref{Eq:Spin_gap}).

In the same way, we obtain the following $2\times 2$ matrix involving only the two bands $\lambda_2 (\bm{k})$ and $\lambda_3 (\bm{k})$ near $K$
\begin{align}
    M^{\mathrm{eff}}_{K}(\bm{k}) &= \hat{P}^{\dag}_{23}U^{\dag}_{K}\cdot H_{\mathrm{mag}} (\bm{k})\cdot U_{K}\hat{P}_{23} \notag \\
    &\simeq -\frac{J_t - J_h}{4}\cdot I_{2\times 2} +\frac{J_t - J_h}{8}(-3q_x + \sqrt{3}q_y)\cdot \sigma_x - \frac{\sqrt{3}}{4}D_z \cdot \sigma_y 
    +\frac{J_t - J_h}{8} (\sqrt{3}q_x + 3q_y)\cdot \sigma_z, \label{Eq:Explicit_Effective_Ham_K}
\end{align}
where $\bm{q} = (q_x, q_y)$ is the crystal momentum measured from each $K$ point, and $\hat{P}_{23}$ is the projection operator onto the subspace spanned by the two modes $\bm{\psi}_2(\bm{k})$ and $\bm{\psi}_3(\bm{k})$ at $K$. Here, $U_{K}$ is the following unitary matrix that diagonalizes $H_{\mathrm{mag}} (\bm{k})$ at $K$ without $D_z$:
\begin{equation}\label{Eq:Transformation_matrix_K}
    U_{K} = \begin{pmatrix}
        -\frac{1}{\sqrt{3}}&\frac{1}{\sqrt{6}}&\frac{1}{\sqrt{2}}\\
        \frac{1}{\sqrt{3}}&-\frac{1}{\sqrt{6}}&\frac{1}{\sqrt{2}}\\
        \frac{1}{\sqrt{3}}&\frac{2}{\sqrt{6}}&0 
    \end{pmatrix}.
\end{equation}
Therefore, the eigenvalues and the Berry curvature of the two level system represented by $M^{\mathrm{eff}}_{K}(\bm{k})$ are given by 
\begin{align}\notag
    \lambda_2 (\bm{q}) &= \frac{J_t - J_h}{4} - \sqrt{3}\sqrt{(J_t - J_h)^2 (q^2_x + q^2_y) + D^2_z},\\
    \lambda_3 (\bm{q}) &= \frac{J_t - J_h}{4} + \sqrt{3}\sqrt{(J_t - J_h)^2 (q^2_x + q^2_y) + D^2_z}, \label{Eq:lambda_K_analytical}
\end{align}
and
\begin{align}\notag
    \Omega^{\mathrm{mag}}_2 (q_x, q_y) &= \frac{-D_z (J_t - J_h)^2}{2((J_t - J_h)^2 (q^2_x + q^2_y) + D^2_z)^{3/2}}, \\
    \Omega^{\mathrm{mag}}_3 (q_x, q_y) &= - \Omega^{\mathrm{mag}}_2 (q_x, q_y), \label{Eq:Berry_curvature_mag_K_analytical}
\end{align}
respectively. Substituting Eqs. (\ref{Eq:lambda_K_analytical}) and (\ref{Eq:Berry_curvature_mag_K_analytical})  into Eqs. (\ref{Eq:Energy_relation}) and (\ref{Eq:Berry_relation}) yields 
\begin{align}\notag
    E_{2,\pm,+}(\bm{q}) &= \sqrt{(J_d + \lambda_2 (\bm{q}))^2 - (\lambda_2 (\bm{q}))^2}\mp h \notag \\
    &\simeq \sqrt{J_d \left(J_d + \frac{J_t - J_h}{2}\right)} \mp h - \frac{\sqrt{3}}{4}\frac{\sqrt{(J_t - J_h)^2 (q^2_x + q^2_y) + D^2_z}}{\sqrt{1 + \frac{J_t - J_h}{2J_d}}}, \notag \\
    E_{3,\pm,+}(\bm{q}) &= \sqrt{(J_d + \lambda_3 (\bm{q}))^2 - (\lambda_3 (\bm{q}))^2} \mp h \notag \\
    &\simeq \sqrt{J_d \left(J_d + \frac{J_t - J_h}{2}\right)} \mp h + \frac{\sqrt{3}}{4}\frac{\sqrt{(J_t - J_h)^2 (q^2_x + q^2_y) + D^2_z}}{\sqrt{1 + \frac{J_t - J_h}{2J_d}}}. \label{Eq:Energy_K_analytical}
\end{align}
and 
\begin{align}\notag
    \Omega_{2, \pm, +} (q_x, q_y) &= \pm \Omega^{\mathrm{mag}}_2 (q_x, q_y), \\
    \Omega_{3, \pm, +} (q_x, q_y) &= \pm \Omega^{\mathrm{mag}}_3 (q_x, q_y). \label{Eq:Berry_curvature_K_analytical}
\end{align}
Note that they are valid only in the vicinity of $\bm{q}=0$.

With the above expressions (\ref{Eq:Energy_Gamma_analytical}), (\ref{Eq:Berry_curvature_Gamma_analytical}), (\ref{Eq:Energy_K_analytical}), and (\ref{Eq:Berry_curvature_K_analytical}), let us derive the analytical expression of the conductivities $\alpha_{xy}$ (\ref{Eq:Spin_Nernst_formula}) and $\kappa_{xy}$ (\ref{Eq:Thermal_Hall_formula}). Based on the fact that the Berry curvature is localized around $\Gamma$ and $K$, we can transform $\alpha_{xy}$ (\ref{Eq:Spin_Nernst_formula}) and $\kappa_{xy}$ (\ref{Eq:Thermal_Hall_formula}) as follows:
\begin{align}
    \alpha_{xy} &= -\frac{2k_B}{\hbar}\sum_{n} \int_{\mathrm{BZ}} \frac{d^2 k}{(2\pi)^2} c_1 (\rho) \Omega_{n,+,+} (\bm{k}) \notag \\
    &\simeq -\frac{k_B}{\pi\hbar} \sum_{p = \Gamma, K} \sum_{\tau = \pm} c_1(\rho(E_{p, \tau, 0}))\tau C_{p}, \label{Eq:alpha_integral}
\end{align}
\begin{align}
    \kappa_{xy} &= -\frac{k_{B}^2 T}{\hbar}\sum_{n,\sigma} \int_{\mathrm{BZ}} \frac{d^2 k}{(2\pi)^2} \left[c_2 (\rho) - \frac{\pi^2}{3}\right] \Omega_{n,\sigma,+} (\bm{k}) \notag \\
    &\simeq -\frac{k_{B}^2 T}{2\pi \hbar} \sum_{p = \Gamma, K} \sum_{\tau = \pm} \sum_{\sigma} c_2(\rho(E_{p, \tau, \sigma})) \tau \sigma C_{p}, \label{Eq:kappa_integral}
\end{align}
where, $E_{p, \tau, 0}$ [see Eqs. (\ref{Eq:E_G_symmetry}) and (\ref{Eq:E_K_symmetry})] are obtained by substituting Eqs. (\ref{Eq:Energy_Gamma_analytical}) and (\ref{Eq:Energy_K_analytical}) into $\bm{k} = 0$ and $\bm{q} = 0$, respectively, with $h = 0$. Meanwhile, $E_{p, \tau, \sigma}$ ($\sigma = \pm$) [see Eqs. (\ref{Eq:E_G_symmetry}) and (\ref{Eq:E_K_symmetry})] are obtained by making the same substitutions into Eqs. (\ref{Eq:Energy_Gamma_analytical}) and (\ref{Eq:Energy_K_analytical}), but without setting $h = 0$. In the above expressions (\ref{Eq:alpha_integral}) and (\ref{Eq:kappa_integral}), $C_{\Gamma}$ is given by the following integral of $\Omega_{1, +, +} (\bm{k})$ around $\Gamma$ (\ref{Eq:Berry_curvature_Gamma_analytical}): 

\begin{align}
    C_{\Gamma} &= \frac{1}{2\pi} \int_{\mathrm{BZ}} d^2 k \frac{-8\sqrt{3}D_z (J_t - J_h)^2 (k_{x}^{2}+k_{y}^2)}{(48D_{z}^2 + (J_t - J_h)^2 (k_{x}^2 + k_{y}^2)^2)^{3/2}} \notag \\ 
    &\simeq \int_{0}^{\infty} dk \frac{-8\sqrt{3}D_z (J_t - J_h)^2 k^3}{(48D_{z}^2 + (J_t - J_h)^2 k^4)^{3/2}} \notag \\
    &= -8\sqrt{3}D_z (J_t - J_h)^2 \frac{1}{2\sqrt{48D^2_z}(J_t - J_h)^2} = -\mathrm{sgn}(D_z). \label{Eq:C_Gamma}
\end{align}
In going from the first line to the second, we replace the integration over the Brillouin zone with that over all $\bm{k}$. From the second to the third line, we use the identity $\int_{0}^{\infty} dk \frac{k^3}{(a+ck^4)^{3/2}} = \frac{1}{2\sqrt{a}c}$. Similarly, $C_K$ is given by the following integral of $\Omega_{2, +, +} (\bm{k})$ around $K$ (\ref{Eq:Berry_curvature_K_analytical}):
\begin{align}
    C_{K} &= \frac{2}{2\pi} \int_{\mathrm{BZ}} d^2 q \frac{-D_z (J_t - J_h)^2}{2((J_t - J_h)^2 (q^2_x + q^2_y) + D^2_z)^{3/2}} \notag \\
    &\simeq \int_{0}^{\infty} dq \frac{-D_z (J_t - J_h)^2 q}{((J_t - J_h)^2 q^2 + D^2_z)^{3/2}} \notag \\
    &= -D_z(J_t - J_h)^2 \frac{1}{(J_t - J_h)^2 |D_z|} = -\mathrm{sgn}(D_z). \label{Eq:C_K}
\end{align}
To derive the second line from the first, we extend the integration from the Brillouin zone to the entire momentum space. From the second to the third line, we use the identity $\int_{0}^{\infty} dq \frac{q}{(aq^2 + b)^{3/2}} = \frac{1}{a\sqrt{b}}$. Here, it should be noted that $C_{\Gamma}$ and $C_K$ give the Chern numbers. More specifically, $C^{\mathrm{mag}}_n$ are expressed using $C_\Gamma$ and $C_K$ as
\begin{align}
    C^{\mathrm{mag}}_1 &= C_{\Gamma} = -\mathrm{sgn}(D_z), \notag \\
    C^{\mathrm{mag}}_2 &= - C_{\Gamma} + C_K = 0,  \notag \\
    C^{\mathrm{mag}}_3 &= -C_K = \mathrm{sgn}(D_z), \label{Eq:Chern_Analytical}
\end{align}
which are consistent with the phase diagram in Fig. \ref{fig:Phase_diagram_H0}. Therefore, by substituting Eq. (\ref{Eq:Chern_Analytical}) into Eqs. (\ref{Eq:alpha_integral}) and (\ref{Eq:kappa_integral}), we obtain the analytical formulas for the conductivities given in Eqs. (\ref{Eq:Analytical_Spin_Nernst}) and (\ref{Eq:Analytical_Thermal_Hall}).

\end{widetext}

\bibliography{Reference_thermal_Hall_triplon.bib,karlo}

\begin{thebibliography}{94}%
\makeatletter
\providecommand \@ifxundefined [1]{%
 \@ifx{#1\undefined}
}%
\providecommand \@ifnum [1]{%
 \ifnum #1\expandafter \@firstoftwo
 \else \expandafter \@secondoftwo
 \fi
}%
\providecommand \@ifx [1]{%
 \ifx #1\expandafter \@firstoftwo
 \else \expandafter \@secondoftwo
 \fi
}%
\providecommand \natexlab [1]{#1}%
\providecommand \enquote  [1]{``#1''}%
\providecommand \bibnamefont  [1]{#1}%
\providecommand \bibfnamefont [1]{#1}%
\providecommand \citenamefont [1]{#1}%
\providecommand \href@noop [0]{\@secondoftwo}%
\providecommand \href [0]{\begingroup \@sanitize@url \@href}%
\providecommand \@href[1]{\@@startlink{#1}\@@href}%
\providecommand \@@href[1]{\endgroup#1\@@endlink}%
\providecommand \@sanitize@url [0]{\catcode `\\12\catcode `\$12\catcode `\&12\catcode `\#12\catcode `\^12\catcode `\_12\catcode `\%12\relax}%
\providecommand \@@startlink[1]{}%
\providecommand \@@endlink[0]{}%
\providecommand \url  [0]{\begingroup\@sanitize@url \@url }%
\providecommand \@url [1]{\endgroup\@href {#1}{\urlprefix }}%
\providecommand \urlprefix  [0]{URL }%
\providecommand \Eprint [0]{\href }%
\providecommand \doibase [0]{https://doi.org/}%
\providecommand \selectlanguage [0]{\@gobble}%
\providecommand \bibinfo  [0]{\@secondoftwo}%
\providecommand \bibfield  [0]{\@secondoftwo}%
\providecommand \translation [1]{[#1]}%
\providecommand \BibitemOpen [0]{}%
\providecommand \bibitemStop [0]{}%
\providecommand \bibitemNoStop [0]{.\EOS\space}%
\providecommand \EOS [0]{\spacefactor3000\relax}%
\providecommand \BibitemShut  [1]{\csname bibitem#1\endcsname}%
\let\auto@bib@innerbib\@empty
\bibitem [{\citenamefont {Kane}\ and\ \citenamefont {Mele}(2005)}]{kane2005$Z_2$}%
  \BibitemOpen
  \bibfield  {author} {\bibinfo {author} {\bibfnamefont {C.~L.}\ \bibnamefont {Kane}}\ and\ \bibinfo {author} {\bibfnamefont {E.~J.}\ \bibnamefont {Mele}},\ }\bibfield  {title} {\bibinfo {title} {${Z}_{2}$ topological {{Order}} and the {{Quantum Spin Hall Effect}}},\ }\href {https://doi.org/10.1103/PhysRevLett.95.146802} {\bibfield  {journal} {\bibinfo  {journal} {Physical Review Letters}\ }\textbf {\bibinfo {volume} {95}},\ \bibinfo {pages} {146802} (\bibinfo {year} {2005})}\BibitemShut {NoStop}%
\bibitem [{\citenamefont {Fu}\ and\ \citenamefont {Kane}(2007)}]{fu2007Topological}%
  \BibitemOpen
  \bibfield  {author} {\bibinfo {author} {\bibfnamefont {L.}~\bibnamefont {Fu}}\ and\ \bibinfo {author} {\bibfnamefont {C.~L.}\ \bibnamefont {Kane}},\ }\bibfield  {title} {\bibinfo {title} {Topological insulators with inversion symmetry},\ }\href {https://doi.org/10.1103/PhysRevB.76.045302} {\bibfield  {journal} {\bibinfo  {journal} {Physical Review B}\ }\textbf {\bibinfo {volume} {76}},\ \bibinfo {pages} {045302} (\bibinfo {year} {2007})}\BibitemShut {NoStop}%
\bibitem [{\citenamefont {Hasan}\ and\ \citenamefont {Kane}(2010)}]{hasan2010Colloquium}%
  \BibitemOpen
  \bibfield  {author} {\bibinfo {author} {\bibfnamefont {M.~Z.}\ \bibnamefont {Hasan}}\ and\ \bibinfo {author} {\bibfnamefont {C.~L.}\ \bibnamefont {Kane}},\ }\bibfield  {title} {\bibinfo {title} {Colloquium: {{Topological}} insulators},\ }\href {https://doi.org/10.1103/RevModPhys.82.3045} {\bibfield  {journal} {\bibinfo  {journal} {Reviews of Modern Physics}\ }\textbf {\bibinfo {volume} {82}},\ \bibinfo {pages} {3045} (\bibinfo {year} {2010})}\BibitemShut {NoStop}%
\bibitem [{\citenamefont {Katsura}\ \emph {et~al.}(2010)\citenamefont {Katsura}, \citenamefont {Nagaosa},\ and\ \citenamefont {Lee}}]{katsura2010Theory}%
  \BibitemOpen
  \bibfield  {author} {\bibinfo {author} {\bibfnamefont {H.}~\bibnamefont {Katsura}}, \bibinfo {author} {\bibfnamefont {N.}~\bibnamefont {Nagaosa}},\ and\ \bibinfo {author} {\bibfnamefont {P.~A.}\ \bibnamefont {Lee}},\ }\bibfield  {title} {\bibinfo {title} {Theory of the {{Thermal Hall Effect}} in {{Quantum Magnets}}},\ }\href {https://doi.org/10.1103/PhysRevLett.104.066403} {\bibfield  {journal} {\bibinfo  {journal} {Physical Review Letters}\ }\textbf {\bibinfo {volume} {104}},\ \bibinfo {pages} {066403} (\bibinfo {year} {2010})}\BibitemShut {NoStop}%
\bibitem [{\citenamefont {Onose}\ \emph {et~al.}(2010)\citenamefont {Onose}, \citenamefont {Ideue}, \citenamefont {Katsura}, \citenamefont {Shiomi}, \citenamefont {Nagaosa},\ and\ \citenamefont {Tokura}}]{onose2010Observation}%
  \BibitemOpen
  \bibfield  {author} {\bibinfo {author} {\bibfnamefont {Y.}~\bibnamefont {Onose}}, \bibinfo {author} {\bibfnamefont {T.}~\bibnamefont {Ideue}}, \bibinfo {author} {\bibfnamefont {H.}~\bibnamefont {Katsura}}, \bibinfo {author} {\bibfnamefont {Y.}~\bibnamefont {Shiomi}}, \bibinfo {author} {\bibfnamefont {N.}~\bibnamefont {Nagaosa}},\ and\ \bibinfo {author} {\bibfnamefont {Y.}~\bibnamefont {Tokura}},\ }\bibfield  {title} {\bibinfo {title} {Observation of the {{Magnon Hall Effect}}},\ }\href {https://doi.org/10.1126/science.1188260} {\bibfield  {journal} {\bibinfo  {journal} {Science}\ }\textbf {\bibinfo {volume} {329}},\ \bibinfo {pages} {297} (\bibinfo {year} {2010})}\BibitemShut {NoStop}%
\bibitem [{\citenamefont {Matsumoto}\ and\ \citenamefont {Murakami}(2011)}]{matsumoto2011Theoretical}%
  \BibitemOpen
  \bibfield  {author} {\bibinfo {author} {\bibfnamefont {R.}~\bibnamefont {Matsumoto}}\ and\ \bibinfo {author} {\bibfnamefont {S.}~\bibnamefont {Murakami}},\ }\bibfield  {title} {\bibinfo {title} {Theoretical {{Prediction}} of a {{Rotating Magnon Wave Packet}} in {{Ferromagnets}}},\ }\href {https://doi.org/10.1103/PhysRevLett.106.197202} {\bibfield  {journal} {\bibinfo  {journal} {Physical Review Letters}\ }\textbf {\bibinfo {volume} {106}},\ \bibinfo {pages} {197202} (\bibinfo {year} {2011})}\BibitemShut {NoStop}%
\bibitem [{\citenamefont {Ideue}\ \emph {et~al.}(2012)\citenamefont {Ideue}, \citenamefont {Onose}, \citenamefont {Katsura}, \citenamefont {Shiomi}, \citenamefont {Ishiwata}, \citenamefont {Nagaosa},\ and\ \citenamefont {Tokura}}]{ideue2012Effect}%
  \BibitemOpen
  \bibfield  {author} {\bibinfo {author} {\bibfnamefont {T.}~\bibnamefont {Ideue}}, \bibinfo {author} {\bibfnamefont {Y.}~\bibnamefont {Onose}}, \bibinfo {author} {\bibfnamefont {H.}~\bibnamefont {Katsura}}, \bibinfo {author} {\bibfnamefont {Y.}~\bibnamefont {Shiomi}}, \bibinfo {author} {\bibfnamefont {S.}~\bibnamefont {Ishiwata}}, \bibinfo {author} {\bibfnamefont {N.}~\bibnamefont {Nagaosa}},\ and\ \bibinfo {author} {\bibfnamefont {Y.}~\bibnamefont {Tokura}},\ }\bibfield  {title} {\bibinfo {title} {Effect of lattice geometry on magnon {{Hall}} effect in ferromagnetic insulators},\ }\href {https://doi.org/10.1103/PhysRevB.85.134411} {\bibfield  {journal} {\bibinfo  {journal} {Physical Review B}\ }\textbf {\bibinfo {volume} {85}},\ \bibinfo {pages} {134411} (\bibinfo {year} {2012})}\BibitemShut {NoStop}%
\bibitem [{\citenamefont {Matsumoto}\ \emph {et~al.}(2014)\citenamefont {Matsumoto}, \citenamefont {Shindou},\ and\ \citenamefont {Murakami}}]{matsumoto2014Thermal}%
  \BibitemOpen
  \bibfield  {author} {\bibinfo {author} {\bibfnamefont {R.}~\bibnamefont {Matsumoto}}, \bibinfo {author} {\bibfnamefont {R.}~\bibnamefont {Shindou}},\ and\ \bibinfo {author} {\bibfnamefont {S.}~\bibnamefont {Murakami}},\ }\bibfield  {title} {\bibinfo {title} {Thermal {{Hall}} effect of magnons in magnets with dipolar interaction},\ }\href {https://doi.org/10.1103/PhysRevB.89.054420} {\bibfield  {journal} {\bibinfo  {journal} {Physical Review B}\ }\textbf {\bibinfo {volume} {89}},\ \bibinfo {pages} {054420} (\bibinfo {year} {2014})}\BibitemShut {NoStop}%
\bibitem [{\citenamefont {Mook}\ \emph {et~al.}(2014{\natexlab{a}})\citenamefont {Mook}, \citenamefont {Henk},\ and\ \citenamefont {Mertig}}]{mook2014Magnon}%
  \BibitemOpen
  \bibfield  {author} {\bibinfo {author} {\bibfnamefont {A.}~\bibnamefont {Mook}}, \bibinfo {author} {\bibfnamefont {J.}~\bibnamefont {Henk}},\ and\ \bibinfo {author} {\bibfnamefont {I.}~\bibnamefont {Mertig}},\ }\bibfield  {title} {\bibinfo {title} {Magnon {{Hall}} effect and topology in kagome lattices: {{A}} theoretical investigation},\ }\href {https://doi.org/10.1103/PhysRevB.89.134409} {\bibfield  {journal} {\bibinfo  {journal} {Physical Review B}\ }\textbf {\bibinfo {volume} {89}},\ \bibinfo {pages} {134409} (\bibinfo {year} {2014}{\natexlab{a}})}\BibitemShut {NoStop}%
\bibitem [{\citenamefont {Cheng}\ \emph {et~al.}(2016)\citenamefont {Cheng}, \citenamefont {Okamoto},\ and\ \citenamefont {Xiao}}]{cheng2016Spin}%
  \BibitemOpen
  \bibfield  {author} {\bibinfo {author} {\bibfnamefont {R.}~\bibnamefont {Cheng}}, \bibinfo {author} {\bibfnamefont {S.}~\bibnamefont {Okamoto}},\ and\ \bibinfo {author} {\bibfnamefont {D.}~\bibnamefont {Xiao}},\ }\bibfield  {title} {\bibinfo {title} {Spin {{Nernst Effect}} of {{Magnons}} in {{Collinear Antiferromagnets}}},\ }\href {https://doi.org/10.1103/PhysRevLett.117.217202} {\bibfield  {journal} {\bibinfo  {journal} {Physical Review Letters}\ }\textbf {\bibinfo {volume} {117}},\ \bibinfo {pages} {217202} (\bibinfo {year} {2016})}\BibitemShut {NoStop}%
\bibitem [{\citenamefont {Zyuzin}\ and\ \citenamefont {Kovalev}(2016)}]{zyuzin2016Magnon}%
  \BibitemOpen
  \bibfield  {author} {\bibinfo {author} {\bibfnamefont {V.~A.}\ \bibnamefont {Zyuzin}}\ and\ \bibinfo {author} {\bibfnamefont {A.~A.}\ \bibnamefont {Kovalev}},\ }\bibfield  {title} {\bibinfo {title} {Magnon {{Spin Nernst Effect}} in {{Antiferromagnets}}},\ }\href {https://doi.org/10.1103/PhysRevLett.117.217203} {\bibfield  {journal} {\bibinfo  {journal} {Physical Review Letters}\ }\textbf {\bibinfo {volume} {117}},\ \bibinfo {pages} {217203} (\bibinfo {year} {2016})}\BibitemShut {NoStop}%
\bibitem [{\citenamefont {Shiomi}\ \emph {et~al.}(2017)\citenamefont {Shiomi}, \citenamefont {Takashima},\ and\ \citenamefont {Saitoh}}]{shiomi2017Experimental}%
  \BibitemOpen
  \bibfield  {author} {\bibinfo {author} {\bibfnamefont {Y.}~\bibnamefont {Shiomi}}, \bibinfo {author} {\bibfnamefont {R.}~\bibnamefont {Takashima}},\ and\ \bibinfo {author} {\bibfnamefont {E.}~\bibnamefont {Saitoh}},\ }\bibfield  {title} {\bibinfo {title} {Experimental evidence consistent with a magnon {{Nernst}} effect in the antiferromagnetic insulator {{MnPS}}$_3$},\ }\href {https://doi.org/10.1103/PhysRevB.96.134425} {\bibfield  {journal} {\bibinfo  {journal} {Physical Review B}\ }\textbf {\bibinfo {volume} {96}},\ \bibinfo {pages} {134425} (\bibinfo {year} {2017})}\BibitemShut {NoStop}%
\bibitem [{\citenamefont {Murakami}\ and\ \citenamefont {Okamoto}(2017)}]{murakami2017Thermal}%
  \BibitemOpen
  \bibfield  {author} {\bibinfo {author} {\bibfnamefont {S.}~\bibnamefont {Murakami}}\ and\ \bibinfo {author} {\bibfnamefont {A.}~\bibnamefont {Okamoto}},\ }\bibfield  {title} {\bibinfo {title} {Thermal {{Hall Effect}} of {{Magnons}}},\ }\href {https://doi.org/10.7566/JPSJ.86.011010} {\bibfield  {journal} {\bibinfo  {journal} {Journal of the Physical Society of Japan}\ }\textbf {\bibinfo {volume} {86}},\ \bibinfo {pages} {011010} (\bibinfo {year} {2017})}\BibitemShut {NoStop}%
\bibitem [{\citenamefont {Han}\ and\ \citenamefont {Lee}(2017)}]{han2017Spin}%
  \BibitemOpen
  \bibfield  {author} {\bibinfo {author} {\bibfnamefont {J.~H.}\ \bibnamefont {Han}}\ and\ \bibinfo {author} {\bibfnamefont {H.}~\bibnamefont {Lee}},\ }\bibfield  {title} {\bibinfo {title} {Spin {{Chirality}} and {{Hall-Like Transport Phenomena}} of {{Spin Excitations}}},\ }\href {https://doi.org/10.7566/JPSJ.86.011007} {\bibfield  {journal} {\bibinfo  {journal} {Journal of the Physical Society of Japan}\ }\textbf {\bibinfo {volume} {86}},\ \bibinfo {pages} {011007} (\bibinfo {year} {2017})}\BibitemShut {NoStop}%
\bibitem [{\citenamefont {Kondo}\ \emph {et~al.}(2020)\citenamefont {Kondo}, \citenamefont {Akagi},\ and\ \citenamefont {Katsura}}]{kondo2020NonHermiticity}%
  \BibitemOpen
  \bibfield  {author} {\bibinfo {author} {\bibfnamefont {H.}~\bibnamefont {Kondo}}, \bibinfo {author} {\bibfnamefont {Y.}~\bibnamefont {Akagi}},\ and\ \bibinfo {author} {\bibfnamefont {H.}~\bibnamefont {Katsura}},\ }\bibfield  {title} {\bibinfo {title} {Non-{{Hermiticity}} and topological invariants of magnon {{Bogoliubov}}--de {{Gennes}} systems},\ }\href {https://doi.org/10.1093/ptep/ptaa151} {\bibfield  {journal} {\bibinfo  {journal} {Progress of Theoretical and Experimental Physics}\ }\textbf {\bibinfo {volume} {2020}},\ \bibinfo {pages} {12A104} (\bibinfo {year} {2020})}\BibitemShut {NoStop}%
\bibitem [{\citenamefont {McClarty}(2022)}]{mcclarty2022Topological}%
  \BibitemOpen
  \bibfield  {author} {\bibinfo {author} {\bibfnamefont {P.~A.}\ \bibnamefont {McClarty}},\ }\bibfield  {title} {\bibinfo {title} {Topological {{Magnons}}: {{A Review}}},\ }\href {https://doi.org/10.1146/annurev-conmatphys-031620-104715} {\bibfield  {journal} {\bibinfo  {journal} {Annual Review of Condensed Matter Physics}\ }\textbf {\bibinfo {volume} {13}},\ \bibinfo {pages} {171} (\bibinfo {year} {2022})}\BibitemShut {NoStop}%
\bibitem [{\citenamefont {Zhuo}\ \emph {et~al.}()\citenamefont {Zhuo}, \citenamefont {Kang}, \citenamefont {Manchon},\ and\ \citenamefont {Cheng}}]{zhuoTopological}%
  \BibitemOpen
  \bibfield  {author} {\bibinfo {author} {\bibfnamefont {F.}~\bibnamefont {Zhuo}}, \bibinfo {author} {\bibfnamefont {J.}~\bibnamefont {Kang}}, \bibinfo {author} {\bibfnamefont {A.}~\bibnamefont {Manchon}},\ and\ \bibinfo {author} {\bibfnamefont {Z.}~\bibnamefont {Cheng}},\ }\bibfield  {title} {\bibinfo {title} {Topological {{Phases}} in {{Magnonics}}},\ }\href {https://doi.org/10.1002/apxr.202300054} {\bibfield  {journal} {\bibinfo  {journal} {Advanced Physics Research}\ }\textbf {\bibinfo {volume} {n/a}},\ \bibinfo {pages} {2300054}}\BibitemShut {NoStop}%
\bibitem [{\citenamefont {Raghu}\ and\ \citenamefont {Haldane}(2008)}]{raghu2008Analogs}%
  \BibitemOpen
  \bibfield  {author} {\bibinfo {author} {\bibfnamefont {S.}~\bibnamefont {Raghu}}\ and\ \bibinfo {author} {\bibfnamefont {F.~D.~M.}\ \bibnamefont {Haldane}},\ }\bibfield  {title} {\bibinfo {title} {Analogs of quantum-{{Hall-effect}} edge states in photonic crystals},\ }\href {https://doi.org/10.1103/PhysRevA.78.033834} {\bibfield  {journal} {\bibinfo  {journal} {Physical Review A}\ }\textbf {\bibinfo {volume} {78}},\ \bibinfo {pages} {033834} (\bibinfo {year} {2008})}\BibitemShut {NoStop}%
\bibitem [{\citenamefont {Petrescu}\ \emph {et~al.}(2012)\citenamefont {Petrescu}, \citenamefont {Houck},\ and\ \citenamefont {Le~Hur}}]{petrescu2012Anomalous}%
  \BibitemOpen
  \bibfield  {author} {\bibinfo {author} {\bibfnamefont {A.}~\bibnamefont {Petrescu}}, \bibinfo {author} {\bibfnamefont {A.~A.}\ \bibnamefont {Houck}},\ and\ \bibinfo {author} {\bibfnamefont {K.}~\bibnamefont {Le~Hur}},\ }\bibfield  {title} {\bibinfo {title} {Anomalous {{Hall}} effects of light and chiral edge modes on the {{Kagom}}{\'e} lattice},\ }\href {https://doi.org/10.1103/PhysRevA.86.053804} {\bibfield  {journal} {\bibinfo  {journal} {Physical Review A}\ }\textbf {\bibinfo {volume} {86}},\ \bibinfo {pages} {053804} (\bibinfo {year} {2012})}\BibitemShut {NoStop}%
\bibitem [{\citenamefont {Rechtsman}\ \emph {et~al.}(2013)\citenamefont {Rechtsman}, \citenamefont {Zeuner}, \citenamefont {Plotnik}, \citenamefont {Lumer}, \citenamefont {Podolsky}, \citenamefont {Dreisow}, \citenamefont {Nolte}, \citenamefont {Segev},\ and\ \citenamefont {Szameit}}]{rechtsman2013Photonic}%
  \BibitemOpen
  \bibfield  {author} {\bibinfo {author} {\bibfnamefont {M.~C.}\ \bibnamefont {Rechtsman}}, \bibinfo {author} {\bibfnamefont {J.~M.}\ \bibnamefont {Zeuner}}, \bibinfo {author} {\bibfnamefont {Y.}~\bibnamefont {Plotnik}}, \bibinfo {author} {\bibfnamefont {Y.}~\bibnamefont {Lumer}}, \bibinfo {author} {\bibfnamefont {D.}~\bibnamefont {Podolsky}}, \bibinfo {author} {\bibfnamefont {F.}~\bibnamefont {Dreisow}}, \bibinfo {author} {\bibfnamefont {S.}~\bibnamefont {Nolte}}, \bibinfo {author} {\bibfnamefont {M.}~\bibnamefont {Segev}},\ and\ \bibinfo {author} {\bibfnamefont {A.}~\bibnamefont {Szameit}},\ }\bibfield  {title} {\bibinfo {title} {Photonic {{Floquet}} topological insulators},\ }\href {https://doi.org/10.1038/nature12066} {\bibfield  {journal} {\bibinfo  {journal} {Nature}\ }\textbf {\bibinfo {volume} {496}},\ \bibinfo {pages} {196} (\bibinfo {year} {2013})}\BibitemShut {NoStop}%
\bibitem [{\citenamefont {Hafezi}\ \emph {et~al.}(2013)\citenamefont {Hafezi}, \citenamefont {Mittal}, \citenamefont {Fan}, \citenamefont {Migdall},\ and\ \citenamefont {Taylor}}]{hafezi2013Imaging}%
  \BibitemOpen
  \bibfield  {author} {\bibinfo {author} {\bibfnamefont {M.}~\bibnamefont {Hafezi}}, \bibinfo {author} {\bibfnamefont {S.}~\bibnamefont {Mittal}}, \bibinfo {author} {\bibfnamefont {J.}~\bibnamefont {Fan}}, \bibinfo {author} {\bibfnamefont {A.}~\bibnamefont {Migdall}},\ and\ \bibinfo {author} {\bibfnamefont {J.~M.}\ \bibnamefont {Taylor}},\ }\bibfield  {title} {\bibinfo {title} {Imaging topological edge states in silicon photonics},\ }\href {https://doi.org/10.1038/nphoton.2013.274} {\bibfield  {journal} {\bibinfo  {journal} {Nature Photonics}\ }\textbf {\bibinfo {volume} {7}},\ \bibinfo {pages} {1001} (\bibinfo {year} {2013})}\BibitemShut {NoStop}%
\bibitem [{\citenamefont {{Ben-Abdallah}}(2016)}]{ben-abdallah2016Photon}%
  \BibitemOpen
  \bibfield  {author} {\bibinfo {author} {\bibfnamefont {P.}~\bibnamefont {{Ben-Abdallah}}},\ }\bibfield  {title} {\bibinfo {title} {Photon {{Thermal Hall Effect}}},\ }\href {https://doi.org/10.1103/PhysRevLett.116.084301} {\bibfield  {journal} {\bibinfo  {journal} {Physical Review Letters}\ }\textbf {\bibinfo {volume} {116}},\ \bibinfo {pages} {084301} (\bibinfo {year} {2016})}\BibitemShut {NoStop}%
\bibitem [{\citenamefont {Strohm}\ \emph {et~al.}(2005)\citenamefont {Strohm}, \citenamefont {Rikken},\ and\ \citenamefont {Wyder}}]{strohm2005Phenomenological}%
  \BibitemOpen
  \bibfield  {author} {\bibinfo {author} {\bibfnamefont {C.}~\bibnamefont {Strohm}}, \bibinfo {author} {\bibfnamefont {G.~L. J.~A.}\ \bibnamefont {Rikken}},\ and\ \bibinfo {author} {\bibfnamefont {P.}~\bibnamefont {Wyder}},\ }\bibfield  {title} {\bibinfo {title} {Phenomenological {{Evidence}} for the {{Phonon Hall Effect}}},\ }\href {https://doi.org/10.1103/PhysRevLett.95.155901} {\bibfield  {journal} {\bibinfo  {journal} {Physical Review Letters}\ }\textbf {\bibinfo {volume} {95}},\ \bibinfo {pages} {155901} (\bibinfo {year} {2005})}\BibitemShut {NoStop}%
\bibitem [{\citenamefont {Sheng}\ \emph {et~al.}(2006)\citenamefont {Sheng}, \citenamefont {Sheng},\ and\ \citenamefont {Ting}}]{sheng2006Theory}%
  \BibitemOpen
  \bibfield  {author} {\bibinfo {author} {\bibfnamefont {L.}~\bibnamefont {Sheng}}, \bibinfo {author} {\bibfnamefont {D.~N.}\ \bibnamefont {Sheng}},\ and\ \bibinfo {author} {\bibfnamefont {C.~S.}\ \bibnamefont {Ting}},\ }\bibfield  {title} {\bibinfo {title} {Theory of the {{Phonon Hall Effect}} in {{Paramagnetic Dielectrics}}},\ }\href {https://doi.org/10.1103/PhysRevLett.96.155901} {\bibfield  {journal} {\bibinfo  {journal} {Physical Review Letters}\ }\textbf {\bibinfo {volume} {96}},\ \bibinfo {pages} {155901} (\bibinfo {year} {2006})}\BibitemShut {NoStop}%
\bibitem [{\citenamefont {Kagan}\ and\ \citenamefont {Maksimov}(2008)}]{kagan2008Anomalous}%
  \BibitemOpen
  \bibfield  {author} {\bibinfo {author} {\bibfnamefont {{\relax Yu}.}~\bibnamefont {Kagan}}\ and\ \bibinfo {author} {\bibfnamefont {L.~A.}\ \bibnamefont {Maksimov}},\ }\bibfield  {title} {\bibinfo {title} {Anomalous {{Hall Effect}} for the {{Phonon Heat Conductivity}} in {{Paramagnetic Dielectrics}}},\ }\href {https://doi.org/10.1103/PhysRevLett.100.145902} {\bibfield  {journal} {\bibinfo  {journal} {Physical Review Letters}\ }\textbf {\bibinfo {volume} {100}},\ \bibinfo {pages} {145902} (\bibinfo {year} {2008})}\BibitemShut {NoStop}%
\bibitem [{\citenamefont {Zhang}\ \emph {et~al.}(2010)\citenamefont {Zhang}, \citenamefont {Ren}, \citenamefont {Wang},\ and\ \citenamefont {Li}}]{zhang2010Topological}%
  \BibitemOpen
  \bibfield  {author} {\bibinfo {author} {\bibfnamefont {L.}~\bibnamefont {Zhang}}, \bibinfo {author} {\bibfnamefont {J.}~\bibnamefont {Ren}}, \bibinfo {author} {\bibfnamefont {J.-S.}\ \bibnamefont {Wang}},\ and\ \bibinfo {author} {\bibfnamefont {B.}~\bibnamefont {Li}},\ }\bibfield  {title} {\bibinfo {title} {Topological {{Nature}} of the {{Phonon Hall Effect}}},\ }\href {https://doi.org/10.1103/PhysRevLett.105.225901} {\bibfield  {journal} {\bibinfo  {journal} {Physical Review Letters}\ }\textbf {\bibinfo {volume} {105}},\ \bibinfo {pages} {225901} (\bibinfo {year} {2010})}\BibitemShut {NoStop}%
\bibitem [{\citenamefont {Zhang}\ \emph {et~al.}(2011)\citenamefont {Zhang}, \citenamefont {Ren}, \citenamefont {Wang},\ and\ \citenamefont {Li}}]{zhang2011phonon}%
  \BibitemOpen
  \bibfield  {author} {\bibinfo {author} {\bibfnamefont {L.}~\bibnamefont {Zhang}}, \bibinfo {author} {\bibfnamefont {J.}~\bibnamefont {Ren}}, \bibinfo {author} {\bibfnamefont {J.-S.}\ \bibnamefont {Wang}},\ and\ \bibinfo {author} {\bibfnamefont {B.}~\bibnamefont {Li}},\ }\bibfield  {title} {\bibinfo {title} {The phonon {{Hall}} effect: Theory and application},\ }\href {https://doi.org/10.1088/0953-8984/23/30/305402} {\bibfield  {journal} {\bibinfo  {journal} {Journal of Physics: Condensed Matter}\ }\textbf {\bibinfo {volume} {23}},\ \bibinfo {pages} {305402} (\bibinfo {year} {2011})}\BibitemShut {NoStop}%
\bibitem [{\citenamefont {Qin}\ \emph {et~al.}(2012)\citenamefont {Qin}, \citenamefont {Zhou},\ and\ \citenamefont {Shi}}]{qin2012Berry}%
  \BibitemOpen
  \bibfield  {author} {\bibinfo {author} {\bibfnamefont {T.}~\bibnamefont {Qin}}, \bibinfo {author} {\bibfnamefont {J.}~\bibnamefont {Zhou}},\ and\ \bibinfo {author} {\bibfnamefont {J.}~\bibnamefont {Shi}},\ }\bibfield  {title} {\bibinfo {title} {Berry curvature and the phonon {{Hall}} effect},\ }\href {https://doi.org/10.1103/PhysRevB.86.104305} {\bibfield  {journal} {\bibinfo  {journal} {Physical Review B}\ }\textbf {\bibinfo {volume} {86}},\ \bibinfo {pages} {104305} (\bibinfo {year} {2012})}\BibitemShut {NoStop}%
\bibitem [{\citenamefont {Romh{\'a}nyi}\ \emph {et~al.}(2015)\citenamefont {Romh{\'a}nyi}, \citenamefont {Penc},\ and\ \citenamefont {Ganesh}}]{romhanyi2015Hall}%
  \BibitemOpen
  \bibfield  {author} {\bibinfo {author} {\bibfnamefont {J.}~\bibnamefont {Romh{\'a}nyi}}, \bibinfo {author} {\bibfnamefont {K.}~\bibnamefont {Penc}},\ and\ \bibinfo {author} {\bibfnamefont {R.}~\bibnamefont {Ganesh}},\ }\bibfield  {title} {\bibinfo {title} {Hall effect of triplons in a dimerized quantum magnet},\ }\href {https://doi.org/10.1038/ncomms7805} {\bibfield  {journal} {\bibinfo  {journal} {Nature Communications}\ }\textbf {\bibinfo {volume} {6}},\ \bibinfo {pages} {6805} (\bibinfo {year} {2015})}\BibitemShut {NoStop}%
\bibitem [{\citenamefont {Malki}\ and\ \citenamefont {Schmidt}(2017)}]{malki2017Magnetic}%
  \BibitemOpen
  \bibfield  {author} {\bibinfo {author} {\bibfnamefont {M.}~\bibnamefont {Malki}}\ and\ \bibinfo {author} {\bibfnamefont {K.~P.}\ \bibnamefont {Schmidt}},\ }\bibfield  {title} {\bibinfo {title} {Magnetic {{Chern}} bands and triplon {{Hall}} effect in an extended {{Shastry-Sutherland}} model},\ }\href {https://doi.org/10.1103/PhysRevB.95.195137} {\bibfield  {journal} {\bibinfo  {journal} {Physical Review B}\ }\textbf {\bibinfo {volume} {95}},\ \bibinfo {pages} {195137} (\bibinfo {year} {2017})}\BibitemShut {NoStop}%
\bibitem [{\citenamefont {McClarty}\ \emph {et~al.}(2017)\citenamefont {McClarty}, \citenamefont {Kr{\"u}ger}, \citenamefont {Guidi}, \citenamefont {Parker}, \citenamefont {Refson}, \citenamefont {Parker}, \citenamefont {Prabhakaran},\ and\ \citenamefont {Coldea}}]{mcclarty2017Topological}%
  \BibitemOpen
  \bibfield  {author} {\bibinfo {author} {\bibfnamefont {P.~A.}\ \bibnamefont {McClarty}}, \bibinfo {author} {\bibfnamefont {F.}~\bibnamefont {Kr{\"u}ger}}, \bibinfo {author} {\bibfnamefont {T.}~\bibnamefont {Guidi}}, \bibinfo {author} {\bibfnamefont {S.~F.}\ \bibnamefont {Parker}}, \bibinfo {author} {\bibfnamefont {K.}~\bibnamefont {Refson}}, \bibinfo {author} {\bibfnamefont {A.~W.}\ \bibnamefont {Parker}}, \bibinfo {author} {\bibfnamefont {D.}~\bibnamefont {Prabhakaran}},\ and\ \bibinfo {author} {\bibfnamefont {R.}~\bibnamefont {Coldea}},\ }\bibfield  {title} {\bibinfo {title} {Topological triplon modes and bound states in a {{Shastry}}--{{Sutherland}} magnet},\ }\href {https://doi.org/10.1038/nphys4117} {\bibfield  {journal} {\bibinfo  {journal} {Nature Physics}\ }\textbf {\bibinfo {volume} {13}},\ \bibinfo {pages} {736} (\bibinfo {year} {2017})}\BibitemShut {NoStop}%
\bibitem [{\citenamefont {Joshi}\ and\ \citenamefont {Schnyder}(2019)}]{joshi2019$mathbbZ_2$}%
  \BibitemOpen
  \bibfield  {author} {\bibinfo {author} {\bibfnamefont {D.~G.}\ \bibnamefont {Joshi}}\ and\ \bibinfo {author} {\bibfnamefont {A.~P.}\ \bibnamefont {Schnyder}},\ }\bibfield  {title} {\bibinfo {title} {${Z}_{2}$ topological quantum paramagnet on a honeycomb bilayer},\ }\href {https://doi.org/10.1103/PhysRevB.100.020407} {\bibfield  {journal} {\bibinfo  {journal} {Physical Review B}\ }\textbf {\bibinfo {volume} {100}},\ \bibinfo {pages} {020407} (\bibinfo {year} {2019})}\BibitemShut {NoStop}%
\bibitem [{\citenamefont {Sun}\ \emph {et~al.}(2021)\citenamefont {Sun}, \citenamefont {Sengupta}, \citenamefont {Nam},\ and\ \citenamefont {Yang}}]{sun2021Negative}%
  \BibitemOpen
  \bibfield  {author} {\bibinfo {author} {\bibfnamefont {H.}~\bibnamefont {Sun}}, \bibinfo {author} {\bibfnamefont {P.}~\bibnamefont {Sengupta}}, \bibinfo {author} {\bibfnamefont {D.}~\bibnamefont {Nam}},\ and\ \bibinfo {author} {\bibfnamefont {B.}~\bibnamefont {Yang}},\ }\bibfield  {title} {\bibinfo {title} {Negative thermal {{Hall}} conductance in a two-dimer {{Shastry-Sutherland}} model with a $\ensuremath{\pi}$-flux {{Dirac}} triplon},\ }\href {https://doi.org/10.1103/PhysRevB.103.L140404} {\bibfield  {journal} {\bibinfo  {journal} {Physical Review B}\ }\textbf {\bibinfo {volume} {103}},\ \bibinfo {pages} {L140404} (\bibinfo {year} {2021})}\BibitemShut {NoStop}%
\bibitem [{\citenamefont {Bhowmick}\ and\ \citenamefont {Sengupta}(2021)}]{bhowmick2021Weyl}%
  \BibitemOpen
  \bibfield  {author} {\bibinfo {author} {\bibfnamefont {D.}~\bibnamefont {Bhowmick}}\ and\ \bibinfo {author} {\bibfnamefont {P.}~\bibnamefont {Sengupta}},\ }\bibfield  {title} {\bibinfo {title} {Weyl triplons in {SrCu}$_{2}$({BO}$_{3}$)$_{2}$},\ }\href {https://doi.org/10.1103/PhysRevB.104.085121} {\bibfield  {journal} {\bibinfo  {journal} {Physical Review B}\ }\textbf {\bibinfo {volume} {104}},\ \bibinfo {pages} {085121} (\bibinfo {year} {2021})}\BibitemShut {NoStop}%
\bibitem [{\citenamefont {Thomasen}\ \emph {et~al.}(2021)\citenamefont {Thomasen}, \citenamefont {Penc}, \citenamefont {Shannon},\ and\ \citenamefont {Romh{\'a}nyi}}]{thomasen2021Fragility}%
  \BibitemOpen
  \bibfield  {author} {\bibinfo {author} {\bibfnamefont {A.}~\bibnamefont {Thomasen}}, \bibinfo {author} {\bibfnamefont {K.}~\bibnamefont {Penc}}, \bibinfo {author} {\bibfnamefont {N.}~\bibnamefont {Shannon}},\ and\ \bibinfo {author} {\bibfnamefont {J.}~\bibnamefont {Romh{\'a}nyi}},\ }\bibfield  {title} {\bibinfo {title} {Fragility of ${\mathcal{z}}_{2}$ topological invariant characterizing triplet excitations in a bilayer kagome magnet},\ }\href {https://doi.org/10.1103/PhysRevB.104.104412} {\bibfield  {journal} {\bibinfo  {journal} {Physical Review B}\ }\textbf {\bibinfo {volume} {104}},\ \bibinfo {pages} {104412} (\bibinfo {year} {2021})}\BibitemShut {NoStop}%
\bibitem [{\citenamefont {Esaki}\ \emph {et~al.}(2024)\citenamefont {Esaki}, \citenamefont {Akagi},\ and\ \citenamefont {Katsura}}]{esaki2024Electric}%
  \BibitemOpen
  \bibfield  {author} {\bibinfo {author} {\bibfnamefont {N.}~\bibnamefont {Esaki}}, \bibinfo {author} {\bibfnamefont {Y.}~\bibnamefont {Akagi}},\ and\ \bibinfo {author} {\bibfnamefont {H.}~\bibnamefont {Katsura}},\ }\bibfield  {title} {\bibinfo {title} {Electric field induced thermal hall effect of triplons in the quantum dimer magnets $\mathit{X}${CuCl}$_{3}$ ($\mathit{X}=\mathrm{Tl},\mathrm{K}$)},\ }\href {https://doi.org/10.1103/PhysRevResearch.6.L032050} {\bibfield  {journal} {\bibinfo  {journal} {Physical Review Research}\ }\textbf {\bibinfo {volume} {6}},\ \bibinfo {pages} {L032050} (\bibinfo {year} {2024})}\BibitemShut {NoStop}%
\bibitem [{\citenamefont {Romh{\'a}nyi}(2019)}]{romhanyi2019Multipolar}%
  \BibitemOpen
  \bibfield  {author} {\bibinfo {author} {\bibfnamefont {J.}~\bibnamefont {Romh{\'a}nyi}},\ }\bibfield  {title} {\bibinfo {title} {Multipolar edge states in the anisotropic kagome antiferromagnet},\ }\href {https://doi.org/10.1103/PhysRevB.99.014408} {\bibfield  {journal} {\bibinfo  {journal} {Physical Review B}\ }\textbf {\bibinfo {volume} {99}},\ \bibinfo {pages} {014408} (\bibinfo {year} {2019})}\BibitemShut {NoStop}%
\bibitem [{\citenamefont {Furukawa}\ and\ \citenamefont {Momoi}(2020)}]{furukawa2020Effects}%
  \BibitemOpen
  \bibfield  {author} {\bibinfo {author} {\bibfnamefont {S.}~\bibnamefont {Furukawa}}\ and\ \bibinfo {author} {\bibfnamefont {T.}~\bibnamefont {Momoi}},\ }\bibfield  {title} {\bibinfo {title} {Effects of {{Dzyaloshinskii}}--{{Moriya Interactions}} in {{Volborthite}}: {{Magnetic Orders}} and {{Thermal Hall Effect}}},\ }\href {https://doi.org/10.7566/JPSJ.89.034711} {\bibfield  {journal} {\bibinfo  {journal} {Journal of the Physical Society of Japan}\ }\textbf {\bibinfo {volume} {89}},\ \bibinfo {pages} {034711} (\bibinfo {year} {2020})}\BibitemShut {NoStop}%
\bibitem [{\citenamefont {Ma}\ \emph {et~al.}(2024)\citenamefont {Ma}, \citenamefont {Wang},\ and\ \citenamefont {Chen}}]{ma2024Upperbranch}%
  \BibitemOpen
  \bibfield  {author} {\bibinfo {author} {\bibfnamefont {B.}~\bibnamefont {Ma}}, \bibinfo {author} {\bibfnamefont {Z.~D.}\ \bibnamefont {Wang}},\ and\ \bibinfo {author} {\bibfnamefont {G.~V.}\ \bibnamefont {Chen}},\ }\bibfield  {title} {\bibinfo {title} {Upper-branch thermal {{Hall}} effect in quantum paramagnets},\ }\href {https://doi.org/10.1103/PhysRevResearch.6.023044} {\bibfield  {journal} {\bibinfo  {journal} {Physical Review Research}\ }\textbf {\bibinfo {volume} {6}},\ \bibinfo {pages} {023044} (\bibinfo {year} {2024})}\BibitemShut {NoStop}%
\bibitem [{\citenamefont {Lu}\ \emph {et~al.}(2024)\citenamefont {Lu}, \citenamefont {Ma}, \citenamefont {Yu},\ and\ \citenamefont {Chen}}]{lu2024Spin}%
  \BibitemOpen
  \bibfield  {author} {\bibinfo {author} {\bibfnamefont {B.}~\bibnamefont {Lu}}, \bibinfo {author} {\bibfnamefont {B.}~\bibnamefont {Ma}}, \bibinfo {author} {\bibfnamefont {Y.}~\bibnamefont {Yu}},\ and\ \bibinfo {author} {\bibfnamefont {G.}~\bibnamefont {Chen}},\ }\bibfield  {title} {\bibinfo {title} {Spin {{Nernst}} effects of linear flavor-waves in quantum paramagnets},\ }\href {https://doi.org/10.1103/PhysRevB.110.235109} {\bibfield  {journal} {\bibinfo  {journal} {Physical Review B}\ }\textbf {\bibinfo {volume} {110}},\ \bibinfo {pages} {235109} (\bibinfo {year} {2024})}\BibitemShut {NoStop}%
\bibitem [{\citenamefont {Zhang}\ \emph {et~al.}(2024)\citenamefont {Zhang}, \citenamefont {Gao},\ and\ \citenamefont {Chen}}]{zhang2024Thermal}%
  \BibitemOpen
  \bibfield  {author} {\bibinfo {author} {\bibfnamefont {X.-T.}\ \bibnamefont {Zhang}}, \bibinfo {author} {\bibfnamefont {Y.~H.}\ \bibnamefont {Gao}},\ and\ \bibinfo {author} {\bibfnamefont {G.}~\bibnamefont {Chen}},\ }\bibfield  {title} {\bibinfo {title} {Thermal {{Hall}} effects in quantum magnets},\ }\href {https://doi.org/10.1016/j.physrep.2024.03.004} {\bibfield  {journal} {\bibinfo  {journal} {Physics Reports}\ }\bibinfo {series} {Thermal {{Hall}} Effects in Quantum Magnets},\ \textbf {\bibinfo {volume} {1070}},\ \bibinfo {pages} {1} (\bibinfo {year} {2024})}\BibitemShut {NoStop}%
\bibitem [{\citenamefont {Suetsugu}\ \emph {et~al.}(2022)\citenamefont {Suetsugu}, \citenamefont {Yokoi}, \citenamefont {Totsuka}, \citenamefont {Ono}, \citenamefont {Tanaka}, \citenamefont {Kasahara}, \citenamefont {Kasahara}, \citenamefont {Chengchao}, \citenamefont {Kageyama},\ and\ \citenamefont {Matsuda}}]{suetsugu2022Intrinsic}%
  \BibitemOpen
  \bibfield  {author} {\bibinfo {author} {\bibfnamefont {S.}~\bibnamefont {Suetsugu}}, \bibinfo {author} {\bibfnamefont {T.}~\bibnamefont {Yokoi}}, \bibinfo {author} {\bibfnamefont {K.}~\bibnamefont {Totsuka}}, \bibinfo {author} {\bibfnamefont {T.}~\bibnamefont {Ono}}, \bibinfo {author} {\bibfnamefont {I.}~\bibnamefont {Tanaka}}, \bibinfo {author} {\bibfnamefont {S.}~\bibnamefont {Kasahara}}, \bibinfo {author} {\bibfnamefont {Y.}~\bibnamefont {Kasahara}}, \bibinfo {author} {\bibfnamefont {Z.}~\bibnamefont {Chengchao}}, \bibinfo {author} {\bibfnamefont {H.}~\bibnamefont {Kageyama}},\ and\ \bibinfo {author} {\bibfnamefont {Y.}~\bibnamefont {Matsuda}},\ }\bibfield  {title} {\bibinfo {title} {Intrinsic suppression of the topological thermal {{Hall}} effect in an exactly solvable quantum magnet},\ }\href {https://doi.org/10.1103/PhysRevB.105.024415} {\bibfield  {journal} {\bibinfo  {journal} {Physical Review B}\ }\textbf {\bibinfo {volume} {105}},\ \bibinfo {pages} {024415} (\bibinfo {year} {2022})}\BibitemShut
  {NoStop}%
\bibitem [{\citenamefont {Kawano}\ and\ \citenamefont {Hotta}(2019)}]{kawano2019Thermal}%
  \BibitemOpen
  \bibfield  {author} {\bibinfo {author} {\bibfnamefont {M.}~\bibnamefont {Kawano}}\ and\ \bibinfo {author} {\bibfnamefont {C.}~\bibnamefont {Hotta}},\ }\bibfield  {title} {\bibinfo {title} {Thermal {{Hall}} effect and topological edge states in a square-lattice antiferromagnet},\ }\href {https://doi.org/10.1103/PhysRevB.99.054422} {\bibfield  {journal} {\bibinfo  {journal} {Physical Review B}\ }\textbf {\bibinfo {volume} {99}},\ \bibinfo {pages} {054422} (\bibinfo {year} {2019})}\BibitemShut {NoStop}%
\bibitem [{\citenamefont {Buzo}\ and\ \citenamefont {Doretto}(2024)}]{buzo2024Thermal}%
  \BibitemOpen
  \bibfield  {author} {\bibinfo {author} {\bibfnamefont {L.~S.}\ \bibnamefont {Buzo}}\ and\ \bibinfo {author} {\bibfnamefont {R.~L.}\ \bibnamefont {Doretto}},\ }\bibfield  {title} {\bibinfo {title} {Thermal {{Hall}} conductivity of a valence bond solid phase in the square lattice ${J}_{1}\text{\ensuremath{-}}{J}_{2}$ antiferromagnet {{Heisenberg}} model with a {{Dzyaloshinskii-Moriya}} interaction},\ }\href {https://doi.org/10.1103/PhysRevB.109.134405} {\bibfield  {journal} {\bibinfo  {journal} {Physical Review B}\ }\textbf {\bibinfo {volume} {109}},\ \bibinfo {pages} {134405} (\bibinfo {year} {2024})}\BibitemShut {NoStop}%
\bibitem [{com()}]{commentfig1}%
  \BibitemOpen
  \href@noop {} {}\bibinfo {note} {Even if $J^{\prime}_h$ is included, its effect can be taken into account simply by replacing $J_h$ with $J_h + J^{\prime}_h$ in the triplon Hamiltonian. For this reason, we focus solely on $J_h$ in the following sections.}\BibitemShut {Stop}%
\bibitem [{\citenamefont {Penfield}(1890)}]{penfield1890spangolite}%
  \BibitemOpen
  \bibfield  {author} {\bibinfo {author} {\bibfnamefont {S.~L.}\ \bibnamefont {Penfield}},\ }\bibfield  {title} {\bibinfo {title} {On spangolite, a new copper mineral},\ }\href {https://doi.org/10.2475/ajs.s3-39.233.370} {\bibfield  {journal} {\bibinfo  {journal} {American Journal of Science}\ }\textbf {\bibinfo {volume} {39}},\ \bibinfo {pages} {370} (\bibinfo {year} {1890})}\BibitemShut {NoStop}%
\bibitem [{\citenamefont {Miers}(1893)}]{miers1893Spangolite}%
  \BibitemOpen
  \bibfield  {author} {\bibinfo {author} {\bibfnamefont {H.~A.}\ \bibnamefont {Miers}},\ }\bibfield  {title} {\bibinfo {title} {Spangolite, a {{Remarkable Cornish Mineral}}},\ }\href {https://doi.org/10.1038/048426b0} {\bibfield  {journal} {\bibinfo  {journal} {Nature}\ }\textbf {\bibinfo {volume} {48}},\ \bibinfo {pages} {426} (\bibinfo {year} {1893})}\BibitemShut {NoStop}%
\bibitem [{\citenamefont {Frondel}(1949)}]{frondel1949crystallography}%
  \BibitemOpen
  \bibfield  {author} {\bibinfo {author} {\bibfnamefont {C.}~\bibnamefont {Frondel}},\ }\bibfield  {title} {\bibinfo {title} {Crystallography of spangolite},\ }\href {http://www.minsocam.org/ammin/AM34/AM34_181.pdf} {\bibfield  {journal} {\bibinfo  {journal} {American Mineralogist: Journal of Earth and Planetary Materials}\ }\textbf {\bibinfo {volume} {34}},\ \bibinfo {pages} {181} (\bibinfo {year} {1949})}\BibitemShut {NoStop}%
\bibitem [{\citenamefont {Hawthorne}\ \emph {et~al.}(1993)\citenamefont {Hawthorne}, \citenamefont {Kimata},\ and\ \citenamefont {Eby}}]{hawthorne1993crystal}%
  \BibitemOpen
  \bibfield  {author} {\bibinfo {author} {\bibfnamefont {F.~C.}\ \bibnamefont {Hawthorne}}, \bibinfo {author} {\bibfnamefont {M.}~\bibnamefont {Kimata}},\ and\ \bibinfo {author} {\bibfnamefont {R.~K.}\ \bibnamefont {Eby}},\ }\bibfield  {title} {\bibinfo {title} {The crystal structure of spangolite, a complex copper sulfate sheet mineral},\ }\href {http://www.minsocam.org/ammin/AM78/AM78_649.pdf} {\bibfield  {journal} {\bibinfo  {journal} {American Mineralogist}\ }\textbf {\bibinfo {volume} {78}},\ \bibinfo {pages} {649} (\bibinfo {year} {1993})}\BibitemShut {NoStop}%
\bibitem [{\citenamefont {Betts}(1995)}]{betts1995new}%
  \BibitemOpen
  \bibfield  {author} {\bibinfo {author} {\bibfnamefont {D.~D.}\ \bibnamefont {Betts}},\ }\bibfield  {title} {\bibinfo {title} {A new two-dimensional lattice of coordination number five},\ }\href {http://hdl.handle.net/10222/35332} {\bibfield  {journal} {\bibinfo  {journal} {Proc. N. S. Inst. Sci.}\ }\textbf {\bibinfo {volume} {40}},\ \bibinfo {pages} {95} (\bibinfo {year} {1995})}\BibitemShut {NoStop}%
\bibitem [{\citenamefont {Olmi}\ \emph {et~al.}(1995)\citenamefont {Olmi}, \citenamefont {Sabelli},\ and\ \citenamefont {{Trosti-Ferroni}}}]{olmi1995crystal}%
  \BibitemOpen
  \bibfield  {author} {\bibinfo {author} {\bibfnamefont {F.}~\bibnamefont {Olmi}}, \bibinfo {author} {\bibfnamefont {C.}~\bibnamefont {Sabelli}},\ and\ \bibinfo {author} {\bibfnamefont {R.}~\bibnamefont {{Trosti-Ferroni}}},\ }\bibfield  {title} {\bibinfo {title} {The crystal structure of sabelliite},\ }\href {https://doi.org/10.1127/ejm/7/6/1331} {\bibfield  {journal} {\bibinfo  {journal} {European Journal of Mineralogy}\ }\textbf {\bibinfo {volume} {7}},\ \bibinfo {pages} {1331} (\bibinfo {year} {1995})}\BibitemShut {NoStop}%
\bibitem [{\citenamefont {Schmalfu{\ss}}\ \emph {et~al.}(2002)\citenamefont {Schmalfu{\ss}}, \citenamefont {Tomczak}, \citenamefont {Schulenburg},\ and\ \citenamefont {Richter}}]{schmalfuss2002spin$frac12$}%
  \BibitemOpen
  \bibfield  {author} {\bibinfo {author} {\bibfnamefont {D.}~\bibnamefont {Schmalfu{\ss}}}, \bibinfo {author} {\bibfnamefont {P.}~\bibnamefont {Tomczak}}, \bibinfo {author} {\bibfnamefont {J.}~\bibnamefont {Schulenburg}},\ and\ \bibinfo {author} {\bibfnamefont {J.}~\bibnamefont {Richter}},\ }\bibfield  {title} {\bibinfo {title} {The spin-$\frac{1}{2}$ {{Heisenberg}} antiferromagnet on a $\frac{1}{7}$-depleted triangular lattice: {{Ground-state}} properties},\ }\href {https://doi.org/10.1103/PhysRevB.65.224405} {\bibfield  {journal} {\bibinfo  {journal} {Physical Review B}\ }\textbf {\bibinfo {volume} {65}},\ \bibinfo {pages} {224405} (\bibinfo {year} {2002})}\BibitemShut {NoStop}%
\bibitem [{\citenamefont {Fennell}\ \emph {et~al.}(2011)\citenamefont {Fennell}, \citenamefont {Piatek}, \citenamefont {Stephenson}, \citenamefont {Nilsen},\ and\ \citenamefont {R{\o}nnow}}]{fennell2011Spangolite}%
  \BibitemOpen
  \bibfield  {author} {\bibinfo {author} {\bibfnamefont {T.}~\bibnamefont {Fennell}}, \bibinfo {author} {\bibfnamefont {J.~O.}\ \bibnamefont {Piatek}}, \bibinfo {author} {\bibfnamefont {R.~A.}\ \bibnamefont {Stephenson}}, \bibinfo {author} {\bibfnamefont {G.~J.}\ \bibnamefont {Nilsen}},\ and\ \bibinfo {author} {\bibfnamefont {H.~M.}\ \bibnamefont {R{\o}nnow}},\ }\bibfield  {title} {\bibinfo {title} {Spangolite: an s = 1/2 maple leaf lattice antiferromagnet?},\ }\href {https://doi.org/10.1088/0953-8984/23/16/164201} {\bibfield  {journal} {\bibinfo  {journal} {Journal of Physics: Condensed Matter}\ }\textbf {\bibinfo {volume} {23}},\ \bibinfo {pages} {164201} (\bibinfo {year} {2011})}\BibitemShut {NoStop}%
\bibitem [{\citenamefont {Kampf}\ \emph {et~al.}(2013)\citenamefont {Kampf}, \citenamefont {Mills}, \citenamefont {Housley},\ and\ \citenamefont {Marty}}]{kampf2013Leadtellurium}%
  \BibitemOpen
  \bibfield  {author} {\bibinfo {author} {\bibfnamefont {A.~R.}\ \bibnamefont {Kampf}}, \bibinfo {author} {\bibfnamefont {S.~J.}\ \bibnamefont {Mills}}, \bibinfo {author} {\bibfnamefont {R.~M.}\ \bibnamefont {Housley}},\ and\ \bibinfo {author} {\bibfnamefont {J.}~\bibnamefont {Marty}},\ }\bibfield  {title} {\bibinfo {title} {Lead-tellurium oxysalts from {{Otto Mountain}} near {{Baker}}, {{California}}: {{VIII}}. {{Fuettererite}}, {Pb}$_3$ {Cu} $^{2+}_{6}$ {Te} $^{6+}$ {O}$_6$ {(OH)}$_7$ {Cl}$_5$, a new mineral with double spangolite-type sheets},\ }\href {https://doi.org/10.2138/am.2013.4218} {\bibfield  {journal} {\bibinfo  {journal} {American Mineralogist}\ }\textbf {\bibinfo {volume} {98}},\ \bibinfo {pages} {506} (\bibinfo {year} {2013})}\BibitemShut {NoStop}%
\bibitem [{\citenamefont {Mills}\ \emph {et~al.}(2014)\citenamefont {Mills}, \citenamefont {Kampf}, \citenamefont {Christy}, \citenamefont {Housley}, \citenamefont {Rossman}, \citenamefont {Reynolds},\ and\ \citenamefont {Marty}}]{mills2014Bluebellite}%
  \BibitemOpen
  \bibfield  {author} {\bibinfo {author} {\bibfnamefont {S.~J.}\ \bibnamefont {Mills}}, \bibinfo {author} {\bibfnamefont {A.~R.}\ \bibnamefont {Kampf}}, \bibinfo {author} {\bibfnamefont {A.~G.}\ \bibnamefont {Christy}}, \bibinfo {author} {\bibfnamefont {R.~M.}\ \bibnamefont {Housley}}, \bibinfo {author} {\bibfnamefont {G.~R.}\ \bibnamefont {Rossman}}, \bibinfo {author} {\bibfnamefont {R.~E.}\ \bibnamefont {Reynolds}},\ and\ \bibinfo {author} {\bibfnamefont {J.}~\bibnamefont {Marty}},\ }\bibfield  {title} {\bibinfo {title} {Bluebellite and mojaveite, two new minerals from the central {{Mojave Desert}}, {{California}}, {{USA}}},\ }\href {https://doi.org/10.1180/minmag.2014.078.5.15} {\bibfield  {journal} {\bibinfo  {journal} {Mineralogical Magazine}\ }\textbf {\bibinfo {volume} {78}},\ \bibinfo {pages} {1325} (\bibinfo {year} {2014})}\BibitemShut {NoStop}%
\bibitem [{\citenamefont {Norman}(2018)}]{norman2018Copper}%
  \BibitemOpen
  \bibfield  {author} {\bibinfo {author} {\bibfnamefont {M.~R.}\ \bibnamefont {Norman}},\ }\bibfield  {title} {\bibinfo {title} {Copper tellurium oxides -- {{A}} playground for magnetism},\ }\href {https://doi.org/10.1016/j.jmmm.2017.11.006} {\bibfield  {journal} {\bibinfo  {journal} {Journal of Magnetism and Magnetic Materials}\ }\textbf {\bibinfo {volume} {452}},\ \bibinfo {pages} {507} (\bibinfo {year} {2018})}\BibitemShut {NoStop}%
\bibitem [{\citenamefont {Inosov}(2018)}]{inosov2018Quantum}%
  \BibitemOpen
  \bibfield  {author} {\bibinfo {author} {\bibfnamefont {D.}~\bibnamefont {Inosov}},\ }\bibfield  {title} {\bibinfo {title} {Quantum magnetism in minerals},\ }\href {https://doi.org/10.1080/00018732.2018.1571986} {\bibfield  {journal} {\bibinfo  {journal} {Advances in Physics}\ }\textbf {\bibinfo {volume} {67}},\ \bibinfo {pages} {149} (\bibinfo {year} {2018})}\BibitemShut {NoStop}%
\bibitem [{\citenamefont {Haraguchi}\ \emph {et~al.}(2018)\citenamefont {Haraguchi}, \citenamefont {Matsuo}, \citenamefont {Kindo},\ and\ \citenamefont {Hiroi}}]{haraguchi2018Frustrated}%
  \BibitemOpen
  \bibfield  {author} {\bibinfo {author} {\bibfnamefont {Y.}~\bibnamefont {Haraguchi}}, \bibinfo {author} {\bibfnamefont {A.}~\bibnamefont {Matsuo}}, \bibinfo {author} {\bibfnamefont {K.}~\bibnamefont {Kindo}},\ and\ \bibinfo {author} {\bibfnamefont {Z.}~\bibnamefont {Hiroi}},\ }\bibfield  {title} {\bibinfo {title} {Frustrated magnetism of the maple-leaf-lattice antiferromagnet {MgMn}$_{3}${O}$_7\cdot 3${H}$_2${O}},\ }\href {https://doi.org/10.1103/PhysRevB.98.064412} {\bibfield  {journal} {\bibinfo  {journal} {Physical Review B}\ }\textbf {\bibinfo {volume} {98}},\ \bibinfo {pages} {064412} (\bibinfo {year} {2018})}\BibitemShut {NoStop}%
\bibitem [{\citenamefont {Venkatesh}\ \emph {et~al.}(2020)\citenamefont {Venkatesh}, \citenamefont {Bandyopadhyay}, \citenamefont {Midya}, \citenamefont {Mahalingam}, \citenamefont {Ganesan},\ and\ \citenamefont {Mandal}}]{venkatesh2020Magnetic}%
  \BibitemOpen
  \bibfield  {author} {\bibinfo {author} {\bibfnamefont {C.}~\bibnamefont {Venkatesh}}, \bibinfo {author} {\bibfnamefont {B.}~\bibnamefont {Bandyopadhyay}}, \bibinfo {author} {\bibfnamefont {A.}~\bibnamefont {Midya}}, \bibinfo {author} {\bibfnamefont {K.}~\bibnamefont {Mahalingam}}, \bibinfo {author} {\bibfnamefont {V.}~\bibnamefont {Ganesan}},\ and\ \bibinfo {author} {\bibfnamefont {P.}~\bibnamefont {Mandal}},\ }\bibfield  {title} {\bibinfo {title} {Magnetic properties of the one-dimensional {$S=\frac{3}{2}$} {{Heisenberg}} antiferromagnetic spin-chain compound {${\mathrm{Na}}_{2}{\mathrm{Mn}}_{3}{\mathrm{O}}_{7}$}},\ }\href {https://doi.org/10.1103/PhysRevB.101.184429} {\bibfield  {journal} {\bibinfo  {journal} {Physical Review B}\ }\textbf {\bibinfo {volume} {101}},\ \bibinfo {pages} {184429} (\bibinfo {year} {2020})}\BibitemShut {NoStop}%
\bibitem [{\citenamefont {Haraguchi}\ \emph {et~al.}(2021)\citenamefont {Haraguchi}, \citenamefont {Matsuo}, \citenamefont {Kindo},\ and\ \citenamefont {Hiroi}}]{haraguchi2021Quantum}%
  \BibitemOpen
  \bibfield  {author} {\bibinfo {author} {\bibfnamefont {Y.}~\bibnamefont {Haraguchi}}, \bibinfo {author} {\bibfnamefont {A.}~\bibnamefont {Matsuo}}, \bibinfo {author} {\bibfnamefont {K.}~\bibnamefont {Kindo}},\ and\ \bibinfo {author} {\bibfnamefont {Z.}~\bibnamefont {Hiroi}},\ }\bibfield  {title} {\bibinfo {title} {Quantum antiferromagnet bluebellite comprising a maple-leaf lattice made of spin-$1/2\phantom{\rule{0.16em}{0ex}}$ {Cu}$^{2+}$ ions},\ }\href {https://doi.org/10.1103/PhysRevB.104.174439} {\bibfield  {journal} {\bibinfo  {journal} {Physical Review B}\ }\textbf {\bibinfo {volume} {104}},\ \bibinfo {pages} {174439} (\bibinfo {year} {2021})}\BibitemShut {NoStop}%
\bibitem [{\citenamefont {Makuta}\ and\ \citenamefont {Hotta}(2021)}]{makuta2021Dimensional}%
  \BibitemOpen
  \bibfield  {author} {\bibinfo {author} {\bibfnamefont {R.}~\bibnamefont {Makuta}}\ and\ \bibinfo {author} {\bibfnamefont {C.}~\bibnamefont {Hotta}},\ }\bibfield  {title} {\bibinfo {title} {Dimensional reduction in quantum spin-$\frac{1}{2}$ system on a $\frac{1}{7}$-depleted triangular lattice},\ }\href {https://doi.org/10.1103/PhysRevB.104.224415} {\bibfield  {journal} {\bibinfo  {journal} {Physical Review B}\ }\textbf {\bibinfo {volume} {104}},\ \bibinfo {pages} {224415} (\bibinfo {year} {2021})}\BibitemShut {NoStop}%
\bibitem [{\citenamefont {Ghosh}\ \emph {et~al.}(2022)\citenamefont {Ghosh}, \citenamefont {M{\"u}ller},\ and\ \citenamefont {Thomale}}]{ghosh2022Another}%
  \BibitemOpen
  \bibfield  {author} {\bibinfo {author} {\bibfnamefont {P.}~\bibnamefont {Ghosh}}, \bibinfo {author} {\bibfnamefont {T.}~\bibnamefont {M{\"u}ller}},\ and\ \bibinfo {author} {\bibfnamefont {R.}~\bibnamefont {Thomale}},\ }\bibfield  {title} {\bibinfo {title} {Another exact ground state of a two-dimensional quantum antiferromagnet},\ }\href {https://doi.org/10.1103/PhysRevB.105.L180412} {\bibfield  {journal} {\bibinfo  {journal} {Physical Review B}\ }\textbf {\bibinfo {volume} {105}},\ \bibinfo {pages} {L180412} (\bibinfo {year} {2022})}\BibitemShut {NoStop}%
\bibitem [{\citenamefont {Saha}\ \emph {et~al.}(2023)\citenamefont {Saha}, \citenamefont {Bera}, \citenamefont {Yusuf},\ and\ \citenamefont {Hoser}}]{saha2023Twodimensional}%
  \BibitemOpen
  \bibfield  {author} {\bibinfo {author} {\bibfnamefont {B.}~\bibnamefont {Saha}}, \bibinfo {author} {\bibfnamefont {A.~K.}\ \bibnamefont {Bera}}, \bibinfo {author} {\bibfnamefont {S.~M.}\ \bibnamefont {Yusuf}},\ and\ \bibinfo {author} {\bibfnamefont {A.}~\bibnamefont {Hoser}},\ }\bibfield  {title} {\bibinfo {title} {Two-dimensional short-range spin-spin correlations in the layered spin-$\frac{3}{2}$ maple leaf lattice antiferromagnet {${\mathrm{Na}}_{2}{\mathrm{Mn}}_{3}{\mathrm{O}}_{7}$} with crystal stacking disorder},\ }\href {https://doi.org/10.1103/PhysRevB.107.064419} {\bibfield  {journal} {\bibinfo  {journal} {Physical Review B}\ }\textbf {\bibinfo {volume} {107}},\ \bibinfo {pages} {064419} (\bibinfo {year} {2023})}\BibitemShut {NoStop}%
\bibitem [{\citenamefont {Gresista}\ \emph {et~al.}(2023)\citenamefont {Gresista}, \citenamefont {Hickey}, \citenamefont {Trebst},\ and\ \citenamefont {Iqbal}}]{gresista2023Candidate}%
  \BibitemOpen
  \bibfield  {author} {\bibinfo {author} {\bibfnamefont {L.}~\bibnamefont {Gresista}}, \bibinfo {author} {\bibfnamefont {C.}~\bibnamefont {Hickey}}, \bibinfo {author} {\bibfnamefont {S.}~\bibnamefont {Trebst}},\ and\ \bibinfo {author} {\bibfnamefont {Y.}~\bibnamefont {Iqbal}},\ }\bibfield  {title} {\bibinfo {title} {Candidate quantum disordered intermediate phase in the {{Heisenberg}} antiferromagnet on the maple-leaf lattice},\ }\href {https://doi.org/10.1103/PhysRevB.108.L241116} {\bibfield  {journal} {\bibinfo  {journal} {Physical Review B}\ }\textbf {\bibinfo {volume} {108}},\ \bibinfo {pages} {L241116} (\bibinfo {year} {2023})}\BibitemShut {NoStop}%
\bibitem [{\citenamefont {Ghosh}\ \emph {et~al.}(2023)\citenamefont {Ghosh}, \citenamefont {Seufert}, \citenamefont {M{\"u}ller}, \citenamefont {Mila},\ and\ \citenamefont {Thomale}}]{ghosh2023Maple}%
  \BibitemOpen
  \bibfield  {author} {\bibinfo {author} {\bibfnamefont {P.}~\bibnamefont {Ghosh}}, \bibinfo {author} {\bibfnamefont {J.}~\bibnamefont {Seufert}}, \bibinfo {author} {\bibfnamefont {T.}~\bibnamefont {M{\"u}ller}}, \bibinfo {author} {\bibfnamefont {F.}~\bibnamefont {Mila}},\ and\ \bibinfo {author} {\bibfnamefont {R.}~\bibnamefont {Thomale}},\ }\bibfield  {title} {\bibinfo {title} {Maple leaf antiferromagnet in a magnetic field},\ }\href {https://doi.org/10.1103/PhysRevB.108.L060406} {\bibfield  {journal} {\bibinfo  {journal} {Physical Review B}\ }\textbf {\bibinfo {volume} {108}},\ \bibinfo {pages} {L060406} (\bibinfo {year} {2023})}\BibitemShut {NoStop}%
\bibitem [{\citenamefont {Schmoll}\ \emph {et~al.}(2024)\citenamefont {Schmoll}, \citenamefont {Jeschke},\ and\ \citenamefont {Iqbal}}]{schmoll2024Tensor}%
  \BibitemOpen
  \bibfield  {author} {\bibinfo {author} {\bibfnamefont {P.}~\bibnamefont {Schmoll}}, \bibinfo {author} {\bibfnamefont {H.~O.}\ \bibnamefont {Jeschke}},\ and\ \bibinfo {author} {\bibfnamefont {Y.}~\bibnamefont {Iqbal}},\ }\href {https://doi.org/10.48550/arXiv.2404.14905} {\bibinfo {title} {Tensor network analysis of the maple-leaf antiferromagnet spangolite}} (\bibinfo {year} {2024}),\ \Eprint {https://arxiv.org/abs/2404.14905} {arXiv:2404.14905 [cond-mat]} \BibitemShut {NoStop}%
\bibitem [{\citenamefont {Ghosh}(2024)}]{ghosh2024Triplon}%
  \BibitemOpen
  \bibfield  {author} {\bibinfo {author} {\bibfnamefont {P.}~\bibnamefont {Ghosh}},\ }\bibfield  {title} {\bibinfo {title} {Triplon analysis of magnetic disorder and order in maple-leaf {{Heisenberg}} magnet},\ }\href {https://doi.org/10.1088/1361-648X/ad69f4} {\bibfield  {journal} {\bibinfo  {journal} {Journal of Physics: Condensed Matter}\ }\textbf {\bibinfo {volume} {36}},\ \bibinfo {pages} {455803} (\bibinfo {year} {2024})}\BibitemShut {NoStop}%
\bibitem [{\citenamefont {Zheng}\ \emph {et~al.}(2007)\citenamefont {Zheng}, \citenamefont {Tong}, \citenamefont {Xue}, \citenamefont {Zhang}, \citenamefont {Chen}, \citenamefont {Grandjean},\ and\ \citenamefont {Long}}]{zheng2007Star}%
  \BibitemOpen
  \bibfield  {author} {\bibinfo {author} {\bibfnamefont {Y.-Z.}\ \bibnamefont {Zheng}}, \bibinfo {author} {\bibfnamefont {M.-L.}\ \bibnamefont {Tong}}, \bibinfo {author} {\bibfnamefont {W.}~\bibnamefont {Xue}}, \bibinfo {author} {\bibfnamefont {W.-X.}\ \bibnamefont {Zhang}}, \bibinfo {author} {\bibfnamefont {X.-M.}\ \bibnamefont {Chen}}, \bibinfo {author} {\bibfnamefont {F.}~\bibnamefont {Grandjean}},\ and\ \bibinfo {author} {\bibfnamefont {G.~J.}\ \bibnamefont {Long}},\ }\bibfield  {title} {\bibinfo {title} {A ``{{Star}}'' {{Antiferromagnet}}: {{A Polymeric Iron}}({{III}}) {{Acetate That Exhibits Both Spin Frustration}} and {{Long-Range Magnetic Ordering}}},\ }\href {https://doi.org/10.1002/anie.200701954} {\bibfield  {journal} {\bibinfo  {journal} {Angewandte Chemie International Edition}\ }\textbf {\bibinfo {volume} {46}},\ \bibinfo {pages} {6076} (\bibinfo {year} {2007})}\BibitemShut {NoStop}%
\bibitem [{\citenamefont {Sorolla}\ \emph {et~al.}(2020)\citenamefont {Sorolla}, \citenamefont {Wang}, \citenamefont {Koo}, \citenamefont {Whangbo},\ and\ \citenamefont {Jacobson}}]{sorolla2020Synthesis}%
  \BibitemOpen
  \bibfield  {author} {\bibinfo {author} {\bibfnamefont {M.~I.}\ \bibnamefont {Sorolla}}, \bibinfo {author} {\bibfnamefont {X.}~\bibnamefont {Wang}}, \bibinfo {author} {\bibfnamefont {H.-J.}\ \bibnamefont {Koo}}, \bibinfo {author} {\bibfnamefont {M.-H.}\ \bibnamefont {Whangbo}},\ and\ \bibinfo {author} {\bibfnamefont {A.~J.}\ \bibnamefont {Jacobson}},\ }\bibfield  {title} {\bibinfo {title} {Synthesis of the {{Elusive S}} = 1/2 {{Star Structure}}: {{A Possible Quantum Spin Liquid Candidate}}},\ }\href {https://doi.org/10.1021/jacs.0c00901} {\bibfield  {journal} {\bibinfo  {journal} {Journal of the American Chemical Society}\ }\textbf {\bibinfo {volume} {142}},\ \bibinfo {pages} {5013} (\bibinfo {year} {2020})}\BibitemShut {NoStop}%
\bibitem [{\citenamefont {{d'Ornellas}}\ and\ \citenamefont {Knolle}(2024)}]{dornellas2024KitaevHeisenberg}%
  \BibitemOpen
  \bibfield  {author} {\bibinfo {author} {\bibfnamefont {P.}~\bibnamefont {{d'Ornellas}}}\ and\ \bibinfo {author} {\bibfnamefont {J.}~\bibnamefont {Knolle}},\ }\bibfield  {title} {\bibinfo {title} {Kitaev-{{Heisenberg}} model on the star lattice: {{From}} chiral {{Majorana}} fermions to chiral triplons},\ }\href {https://doi.org/10.1103/PhysRevB.109.094421} {\bibfield  {journal} {\bibinfo  {journal} {Physical Review B}\ }\textbf {\bibinfo {volume} {109}},\ \bibinfo {pages} {094421} (\bibinfo {year} {2024})}\BibitemShut {NoStop}%
\bibitem [{\citenamefont {Ishikawa}\ \emph {et~al.}(2024)\citenamefont {Ishikawa}, \citenamefont {Ishii}, \citenamefont {Yajima}, \citenamefont {Matsuda}, \citenamefont {Kindo}, \citenamefont {Shimizu}, \citenamefont {Rousochatzakis}, \citenamefont {R{\"o}{\ss}ler},\ and\ \citenamefont {Janson}}]{ishikawa2024Geometric}%
  \BibitemOpen
  \bibfield  {author} {\bibinfo {author} {\bibfnamefont {H.}~\bibnamefont {Ishikawa}}, \bibinfo {author} {\bibfnamefont {Y.}~\bibnamefont {Ishii}}, \bibinfo {author} {\bibfnamefont {T.}~\bibnamefont {Yajima}}, \bibinfo {author} {\bibfnamefont {Y.~H.}\ \bibnamefont {Matsuda}}, \bibinfo {author} {\bibfnamefont {K.}~\bibnamefont {Kindo}}, \bibinfo {author} {\bibfnamefont {Y.}~\bibnamefont {Shimizu}}, \bibinfo {author} {\bibfnamefont {I.}~\bibnamefont {Rousochatzakis}}, \bibinfo {author} {\bibfnamefont {U.~K.}\ \bibnamefont {R{\"o}{\ss}ler}},\ and\ \bibinfo {author} {\bibfnamefont {O.}~\bibnamefont {Janson}},\ }\bibfield  {title} {\bibinfo {title} {Geometric frustration and {{Dzyaloshinskii-Moriya}} interactions in a quantum star lattice hybrid copper sulfate},\ }\href {https://doi.org/10.1103/PhysRevB.109.L180401} {\bibfield  {journal} {\bibinfo  {journal} {Physical Review B}\ }\textbf {\bibinfo {volume} {109}},\ \bibinfo {pages} {L180401} (\bibinfo {year} {2024})}\BibitemShut {NoStop}%
\bibitem [{\citenamefont {Moriya}(1960)}]{moriya1960anisotropic}%
  \BibitemOpen
  \bibfield  {author} {\bibinfo {author} {\bibfnamefont {T.}~\bibnamefont {Moriya}},\ }\bibfield  {title} {\bibinfo {title} {Anisotropic superexchange interaction and weak ferromagnetism},\ }\href {https://journals.aps.org/pr/abstract/10.1103/PhysRev.120.91} {\bibfield  {journal} {\bibinfo  {journal} {{Physical Review}}\ }\textbf {\bibinfo {volume} {120}},\ \bibinfo {pages} {91} (\bibinfo {year} {1960})}\BibitemShut {NoStop}%
\bibitem [{Note1()}]{Note1}%
  \BibitemOpen
  \bibinfo {note} {An exact dimer ground state is realized only when $J_t = J_h$ and without the DM interaction \cite {ghosh2022Another}. However, the ground state is well approximated by the dimer ground state even when $J_t \gg J_h$ and with the DM interaction. See Appendix \ref {Appendix:ED_vs_Triplon} for details.}\BibitemShut {Stop}%
\bibitem [{\citenamefont {Sachdev}\ and\ \citenamefont {Bhatt}(1990)}]{sachdev1990Bondoperator}%
  \BibitemOpen
  \bibfield  {author} {\bibinfo {author} {\bibfnamefont {S.}~\bibnamefont {Sachdev}}\ and\ \bibinfo {author} {\bibfnamefont {R.~N.}\ \bibnamefont {Bhatt}},\ }\bibfield  {title} {\bibinfo {title} {Bond-operator representation of quantum spins: {{Mean-field}} theory of frustrated quantum {{Heisenberg}} antiferromagnets},\ }\href {https://doi.org/10.1103/PhysRevB.41.9323} {\bibfield  {journal} {\bibinfo  {journal} {Physical Review B}\ }\textbf {\bibinfo {volume} {41}},\ \bibinfo {pages} {9323} (\bibinfo {year} {1990})}\BibitemShut {NoStop}%
\bibitem [{Note2()}]{Note2}%
  \BibitemOpen
  \bibinfo {note} {We note that the constant term ${\protect \mathcal {H}^{(0)}}$ is not the ground state energy. The appropriate expression of the ground state energy is given in Eq. (\ref {Eq:Correction_GS}).}\BibitemShut {Stop}%
\bibitem [{\citenamefont {Colpa}(1978)}]{colpa1978Diagonalization}%
  \BibitemOpen
  \bibfield  {author} {\bibinfo {author} {\bibfnamefont {J.~H.~P.}\ \bibnamefont {Colpa}},\ }\bibfield  {title} {\bibinfo {title} {Diagonalization of the quadratic boson hamiltonian},\ }\href {https://doi.org/10.1016/0378-4371(78)90160-7} {\bibfield  {journal} {\bibinfo  {journal} {Physica A: Statistical Mechanics and its Applications}\ }\textbf {\bibinfo {volume} {93}},\ \bibinfo {pages} {327} (\bibinfo {year} {1978})}\BibitemShut {NoStop}%
\bibitem [{\citenamefont {Mook}\ \emph {et~al.}(2014{\natexlab{b}})\citenamefont {Mook}, \citenamefont {Henk},\ and\ \citenamefont {Mertig}}]{mook2014Edge}%
  \BibitemOpen
  \bibfield  {author} {\bibinfo {author} {\bibfnamefont {A.}~\bibnamefont {Mook}}, \bibinfo {author} {\bibfnamefont {J.}~\bibnamefont {Henk}},\ and\ \bibinfo {author} {\bibfnamefont {I.}~\bibnamefont {Mertig}},\ }\bibfield  {title} {\bibinfo {title} {Edge states in topological magnon insulators},\ }\href {https://doi.org/10.1103/PhysRevB.90.024412} {\bibfield  {journal} {\bibinfo  {journal} {Physical Review B}\ }\textbf {\bibinfo {volume} {90}},\ \bibinfo {pages} {024412} (\bibinfo {year} {2014}{\natexlab{b}})}\BibitemShut {NoStop}%
\bibitem [{\citenamefont {Lee}\ \emph {et~al.}(2015)\citenamefont {Lee}, \citenamefont {Han},\ and\ \citenamefont {Lee}}]{lee2015Thermal}%
  \BibitemOpen
  \bibfield  {author} {\bibinfo {author} {\bibfnamefont {H.}~\bibnamefont {Lee}}, \bibinfo {author} {\bibfnamefont {J.~H.}\ \bibnamefont {Han}},\ and\ \bibinfo {author} {\bibfnamefont {P.~A.}\ \bibnamefont {Lee}},\ }\bibfield  {title} {\bibinfo {title} {Thermal {{Hall}} effect of spins in a paramagnet},\ }\href {https://doi.org/10.1103/PhysRevB.91.125413} {\bibfield  {journal} {\bibinfo  {journal} {Physical Review B}\ }\textbf {\bibinfo {volume} {91}},\ \bibinfo {pages} {125413} (\bibinfo {year} {2015})}\BibitemShut {NoStop}%
\bibitem [{Note3()}]{Note3}%
  \BibitemOpen
  \bibinfo {note} {In the $2\times 2$ spin subspace, the U(1) rotation is represented by $e^{-i\phi \sigma _z} \simeq \cos \phi ~I_{2\times 2} - i\sin \phi ~\sigma _z$, where $\phi $ is a rotation angle. Thus, the commutation relation (\ref {Eq:U1_sym}) ensures the U(1) symmetry in the spin space.}\BibitemShut {Stop}%
\bibitem [{\citenamefont {Kondo}\ \emph {et~al.}(2019)\citenamefont {Kondo}, \citenamefont {Akagi},\ and\ \citenamefont {Katsura}}]{kondo2019$mathbbZ_2$}%
  \BibitemOpen
  \bibfield  {author} {\bibinfo {author} {\bibfnamefont {H.}~\bibnamefont {Kondo}}, \bibinfo {author} {\bibfnamefont {Y.}~\bibnamefont {Akagi}},\ and\ \bibinfo {author} {\bibfnamefont {H.}~\bibnamefont {Katsura}},\ }\bibfield  {title} {\bibinfo {title} {${\mathbb{z}}_{2}$ topological invariant for magnon spin {{Hall}} systems},\ }\href {https://doi.org/10.1103/PhysRevB.99.041110} {\bibfield  {journal} {\bibinfo  {journal} {Physical Review B}\ }\textbf {\bibinfo {volume} {99}},\ \bibinfo {pages} {041110} (\bibinfo {year} {2019})}\BibitemShut {NoStop}%
\bibitem [{Note4()}]{Note4}%
  \BibitemOpen
  \bibinfo {note} {We used the method in Ref. \cite {fukui2005Chern}}\BibitemShut {NoStop}%
\bibitem [{\citenamefont {Hirschberger}\ \emph {et~al.}(2015)\citenamefont {Hirschberger}, \citenamefont {Chisnell}, \citenamefont {Lee},\ and\ \citenamefont {Ong}}]{hirschberger2015Thermal}%
  \BibitemOpen
  \bibfield  {author} {\bibinfo {author} {\bibfnamefont {M.}~\bibnamefont {Hirschberger}}, \bibinfo {author} {\bibfnamefont {R.}~\bibnamefont {Chisnell}}, \bibinfo {author} {\bibfnamefont {Y.~S.}\ \bibnamefont {Lee}},\ and\ \bibinfo {author} {\bibfnamefont {N.~P.}\ \bibnamefont {Ong}},\ }\bibfield  {title} {\bibinfo {title} {Thermal {{Hall Effect}} of {{Spin Excitations}} in a {{Kagome Magnet}}},\ }\href {https://doi.org/10.1103/PhysRevLett.115.106603} {\bibfield  {journal} {\bibinfo  {journal} {Physical Review Letters}\ }\textbf {\bibinfo {volume} {115}},\ \bibinfo {pages} {106603} (\bibinfo {year} {2015})}\BibitemShut {NoStop}%
\bibitem [{\citenamefont {Koyama}\ and\ \citenamefont {Nasu}(2023{\natexlab{a}})}]{koyama2023Flavorwave}%
  \BibitemOpen
  \bibfield  {author} {\bibinfo {author} {\bibfnamefont {S.}~\bibnamefont {Koyama}}\ and\ \bibinfo {author} {\bibfnamefont {J.}~\bibnamefont {Nasu}},\ }\bibfield  {title} {\bibinfo {title} {Flavor-wave theory with quasiparticle damping at finite temperatures: {{Application}} to chiral edge modes in the {{Kitaev}} model},\ }\href {https://doi.org/10.1103/PhysRevB.108.235162} {\bibfield  {journal} {\bibinfo  {journal} {Physical Review B}\ }\textbf {\bibinfo {volume} {108}},\ \bibinfo {pages} {235162} (\bibinfo {year} {2023}{\natexlab{a}})}\BibitemShut {NoStop}%
\bibitem [{\citenamefont {Koyama}\ and\ \citenamefont {Nasu}(2023{\natexlab{b}})}]{koyama2023Nonlinear}%
  \BibitemOpen
  \bibfield  {author} {\bibinfo {author} {\bibfnamefont {S.}~\bibnamefont {Koyama}}\ and\ \bibinfo {author} {\bibfnamefont {J.}~\bibnamefont {Nasu}},\ }\bibfield  {title} {\bibinfo {title} {Nonlinear {{Spin-wave Theory}} for {{Dynamical Spin Correlations}} in the {{Kitaev Model}} at {{Finite Temperatures}}},\ }\href {https://doi.org/10.3938/NPSM.73.1123} {\bibfield  {journal} {\bibinfo  {journal} {New Physics: Sae Mulli}\ }\textbf {\bibinfo {volume} {73}},\ \bibinfo {pages} {1123} (\bibinfo {year} {2023}{\natexlab{b}})}\BibitemShut {NoStop}%
\bibitem [{\citenamefont {Choi}\ \emph {et~al.}(2023)\citenamefont {Choi}, \citenamefont {Yang}, \citenamefont {Park},\ and\ \citenamefont {Park}}]{choi2023Sizable}%
  \BibitemOpen
  \bibfield  {author} {\bibinfo {author} {\bibfnamefont {Y.}~\bibnamefont {Choi}}, \bibinfo {author} {\bibfnamefont {H.}~\bibnamefont {Yang}}, \bibinfo {author} {\bibfnamefont {J.}~\bibnamefont {Park}},\ and\ \bibinfo {author} {\bibfnamefont {J.-G.}\ \bibnamefont {Park}},\ }\bibfield  {title} {\bibinfo {title} {Sizable suppression of magnon {{Hall}} effect by magnon damping in {${\mathrm{Cr}}_{2}{\mathrm{Ge}}_{2}{\mathrm{Te}}_{6}$}},\ }\href {https://doi.org/10.1103/PhysRevB.107.184434} {\bibfield  {journal} {\bibinfo  {journal} {Physical Review B}\ }\textbf {\bibinfo {volume} {107}},\ \bibinfo {pages} {184434} (\bibinfo {year} {2023})}\BibitemShut {NoStop}%
\bibitem [{\citenamefont {Koyama}\ and\ \citenamefont {Nasu}(2024{\natexlab{a}})}]{koyama2024Thermal}%
  \BibitemOpen
  \bibfield  {author} {\bibinfo {author} {\bibfnamefont {S.}~\bibnamefont {Koyama}}\ and\ \bibinfo {author} {\bibfnamefont {J.}~\bibnamefont {Nasu}},\ }\bibfield  {title} {\bibinfo {title} {Thermal {{Hall}} effect incorporating magnon damping in localized spin systems},\ }\href {https://doi.org/10.1103/PhysRevB.109.174442} {\bibfield  {journal} {\bibinfo  {journal} {Physical Review B}\ }\textbf {\bibinfo {volume} {109}},\ \bibinfo {pages} {174442} (\bibinfo {year} {2024}{\natexlab{a}})}\BibitemShut {NoStop}%
\bibitem [{\citenamefont {Koyama}\ and\ \citenamefont {Nasu}(2024{\natexlab{b}})}]{koyama2024Impact}%
  \BibitemOpen
  \bibfield  {author} {\bibinfo {author} {\bibfnamefont {S.}~\bibnamefont {Koyama}}\ and\ \bibinfo {author} {\bibfnamefont {J.}~\bibnamefont {Nasu}},\ }\href {https://doi.org/10.48550/arXiv.2411.10065} {\bibinfo {title} {Impact of triplon damping on thermal {{Hall}} conductivity in {{Shastry-Sutherland}} model}} (\bibinfo {year} {2024}{\natexlab{b}}),\ \Eprint {https://arxiv.org/abs/2411.10065} {arXiv:2411.10065 [cond-mat]} \BibitemShut {NoStop}%
\bibitem [{\citenamefont {Chatzichrysafis}\ and\ \citenamefont {Mook}(2025)}]{Chatzichrysafis2025Thermal}%
  \BibitemOpen
  \bibfield  {author} {\bibinfo {author} {\bibfnamefont {D.}~\bibnamefont {Chatzichrysafis}}\ and\ \bibinfo {author} {\bibfnamefont {A.}~\bibnamefont {Mook}},\ }\bibfield  {title} {\bibinfo {title} {Thermal hall effect of magnons from many-body skew scattering},\ }\href {https://doi.org/10.1103/PhysRevB.111.134405} {\bibfield  {journal} {\bibinfo  {journal} {Phys. Rev. B}\ }\textbf {\bibinfo {volume} {111}},\ \bibinfo {pages} {134405} (\bibinfo {year} {2025})}\BibitemShut {NoStop}%
\bibitem [{\citenamefont {Chen}\ and\ \citenamefont {Wu}(2024)}]{chen2024Multiflavor}%
  \BibitemOpen
  \bibfield  {author} {\bibinfo {author} {\bibfnamefont {G.~V.}\ \bibnamefont {Chen}}\ and\ \bibinfo {author} {\bibfnamefont {C.}~\bibnamefont {Wu}},\ }\bibfield  {title} {\bibinfo {title} {Multiflavor {{Mott}} insulators in quantum materials and ultracold atoms},\ }\href {https://doi.org/10.1038/s41535-023-00614-2} {\bibfield  {journal} {\bibinfo  {journal} {npj Quantum Materials}\ }\textbf {\bibinfo {volume} {9}},\ \bibinfo {pages} {1} (\bibinfo {year} {2024})}\BibitemShut {NoStop}%
\bibitem [{\citenamefont {Mila}(2024)}]{mila2024Mott}%
  \BibitemOpen
  \bibfield  {author} {\bibinfo {author} {\bibfnamefont {F.}~\bibnamefont {Mila}},\ }\bibfield  {title} {\bibinfo {title} {Mott physics in the multiflavored age},\ }\href {https://doi.org/10.1038/s41535-024-00639-1} {\bibfield  {journal} {\bibinfo  {journal} {npj Quantum Materials}\ }\textbf {\bibinfo {volume} {9}},\ \bibinfo {pages} {1} (\bibinfo {year} {2024})}\BibitemShut {NoStop}%
\bibitem [{\citenamefont {IUCr}(2010)}]{ITCVolE}%
  \BibitemOpen
  \bibfield  {author} {\bibinfo {author} {\bibnamefont {IUCr}},\ }\href {https://onlinelibrary.wiley.com/iucr/itc/Eb/contents/fullindex.html} {\emph {\bibinfo {title} {Subperiodic groups}}},\ \bibinfo {edition} {2nd}\ ed.,\ edited by\ \bibinfo {editor} {\bibfnamefont {V.}~\bibnamefont {Kopsky}}\ and\ \bibinfo {editor} {\bibfnamefont {D.~B.}\ \bibnamefont {Litvin}},\ \bibinfo {series} {International Tables for Crystallography}, Vol.\ \bibinfo {volume} {E: Subperiodic groups}\ (\bibinfo  {publisher} {Kluwer Academic Publishers},\ \bibinfo {address} {Dordrecht, Boston, London},\ \bibinfo {year} {2010})\BibitemShut {NoStop}%
\bibitem [{\citenamefont {Sze}\ \emph {et~al.}(2025)\citenamefont {Sze}, \citenamefont {Xi},\ and\ \citenamefont {Zhu}}]{SZE2025e01009}%
  \BibitemOpen
  \bibfield  {author} {\bibinfo {author} {\bibfnamefont {W.~H.~R.}\ \bibnamefont {Sze}}, \bibinfo {author} {\bibfnamefont {B.}~\bibnamefont {Xi}},\ and\ \bibinfo {author} {\bibfnamefont {J.}~\bibnamefont {Zhu}},\ }\bibfield  {title} {\bibinfo {title} {Key difference of input data organization to the predictions of symmetry information and layer number for quasi-2d films from band structure},\ }\href {https://doi.org/https://doi.org/10.1016/j.cocom.2025.e01009} {\bibfield  {journal} {\bibinfo  {journal} {Computational Condensed Matter}\ }\textbf {\bibinfo {volume} {42}},\ \bibinfo {pages} {e01009} (\bibinfo {year} {2025})}\BibitemShut {NoStop}%
\bibitem [{\citenamefont {Neumann}\ \emph {et~al.}(2020)\citenamefont {Neumann}, \citenamefont {Mook}, \citenamefont {Henk},\ and\ \citenamefont {Mertig}}]{neumann2020Orbital}%
  \BibitemOpen
  \bibfield  {author} {\bibinfo {author} {\bibfnamefont {R.~R.}\ \bibnamefont {Neumann}}, \bibinfo {author} {\bibfnamefont {A.}~\bibnamefont {Mook}}, \bibinfo {author} {\bibfnamefont {J.}~\bibnamefont {Henk}},\ and\ \bibinfo {author} {\bibfnamefont {I.}~\bibnamefont {Mertig}},\ }\bibfield  {title} {\bibinfo {title} {Orbital {{Magnetic Moment}} of {{Magnons}}},\ }\href {https://doi.org/10.1103/PhysRevLett.125.117209} {\bibfield  {journal} {\bibinfo  {journal} {Physical Review Letters}\ }\textbf {\bibinfo {volume} {125}},\ \bibinfo {pages} {117209} (\bibinfo {year} {2020})}\BibitemShut {NoStop}%
\bibitem [{\citenamefont {Fukui}\ \emph {et~al.}(2005)\citenamefont {Fukui}, \citenamefont {Hatsugai},\ and\ \citenamefont {Suzuki}}]{fukui2005Chern}%
  \BibitemOpen
  \bibfield  {author} {\bibinfo {author} {\bibfnamefont {T.}~\bibnamefont {Fukui}}, \bibinfo {author} {\bibfnamefont {Y.}~\bibnamefont {Hatsugai}},\ and\ \bibinfo {author} {\bibfnamefont {H.}~\bibnamefont {Suzuki}},\ }\bibfield  {title} {\bibinfo {title} {Chern {{Numbers}} in {{Discretized Brillouin Zone}}: {{Efficient Method}} of {{Computing}} ({{Spin}}) {{Hall Conductances}}},\ }\href {https://doi.org/10.1143/JPSJ.74.1674} {\bibfield  {journal} {\bibinfo  {journal} {Journal of the Physical Society of Japan}\ }\textbf {\bibinfo {volume} {74}},\ \bibinfo {pages} {1674} (\bibinfo {year} {2005})}\BibitemShut {NoStop}%
\end{thebibliography}%

\end{document}